 \documentclass[journal]{IEEEtran}

\usepackage{bbm,dsfont,mathrsfs,fixmath}
\usepackage{stmaryrd}
\usepackage{amsmath,amstext,amsfonts,amssymb}
\usepackage{epsfig,float}
\usepackage{graphics,dblfloatfix}
\usepackage{cite}
\usepackage{psfrag}
\usepackage{subfigure}
\usepackage{url}
\usepackage{upgreek}
\usepackage{shadow,color,pifont,times,rotate}

\newcommand{\mb}{\mathbf}

\newcommand{\mc}{\mathcal}

\def\esp{{\mathbb E}}  
\DeclareMathOperator{\Tr}{Tr}
\newcommand{\mkv}{-\!\!\!\!\minuso\!\!\!\!-}
 \DeclareMathAlphabet{\mathpzc}{OT1}{pzc}{m}{it}
\DeclareRobustCommand{\prob}[1][P]{\ensuremath {\mathbb{#1}}}

\newtheorem{theorem}{Theorem}
\newtheorem{lemma}{Lemma}
\newtheorem{proposition}{Proposition}
\newtheorem{corollary}{Corollary}
\newtheorem{remark}{Remark}
\newtheorem{definition}{Definition}

\usepackage{enumerate}    
\ifCLASSINFOpdf
		\usepackage{epstopdf}
\else
\fi

\usepackage{ifthen}

\begin{document}
\title{Mixed Noisy Network Coding and \\ Cooperative Unicasting in Wireless  Networks}
\author{
Arash~Behboodi,~\IEEEmembership{Member,~IEEE,}
 and Pablo~Piantanida,~\IEEEmembership{Member,~IEEE,} 
\thanks{The work of P. Piantanida is partially supported by the ANR grant (FIREFLIES) INTB 0302 01.
The material in this paper was presented in part at the 5th International Symposium on Communications, Control, and Signal Processing, May 2012, and the 2012 and  2013 IEEE International Symposium  on Information Theory.}
\thanks{Arash Behboodi is with the Telecommunication Network Groups, Technische Universit\"{a}t Berlin, Einsteinufer 25, FT 5 10587 Berlin, Germany, Email: arash.behboodi@tu-berlin.de.

Pablo Piantanida is with the Department of Telecommunications, SUPELEC, 91192 Gif-sur-Yvette, France, Email: pablo.piantanida@supelec.fr.}
}
\maketitle

\begin{abstract}
The problem of communicating a single message to a destination in presence of multiple relay nodes, referred  to as cooperative unicast network, is considered.  First, we introduce ``Mixed Noisy Network Coding" (MNNC) scheme which  generalizes ``Noisy Network Coding" (NNC) where relays are allowed to decode-and-forward (DF) messages while all of them (without exception) transmit noisy descriptions of their observations. These descriptions are exploited at the destination and the DF relays aim to decode the transmitted messages while creating full cooperation among the nodes. Moreover, the destination and the DF relays can independently select the set of descriptions to be decoded or treated as interference. This  concept is further extended to multi-hopping scenarios, referred to as ``Layered MNNC" (LMNNC), where DF relays are organized into disjoint groups representing one hop in the network. For cooperative unicast additive white Gaussian noise (AWGN) networks we show that --provided DF relays are 
properly chosen-- MNNC improves over all previously established constant gaps to the cut-set bound. 
Secondly, we consider the composite cooperative unicast network where the channel parameters are randomly drawn  before communication starts and remain fixed during the transmission. Each draw is assumed to be unknown at the source and fully known at the destination but  only partly known at the relays. We introduce through MNNC scheme the concept of ``Selective Coding Strategy" (SCS) that enables relays to decide dynamically  whether,  in addition to communicate noisy descriptions, is possible to decode and forward messages. It is demonstrated through slow-fading AWGN relay networks that  SCS clearly outperforms conventional coding schemes. 
\end{abstract}



\section{Introduction}

Cooperation in multi-terminal networks is becoming the essential part of modern communication systems, e.g., wireless mobile systems, device-to-device (D2D) communications, network coding, sensor and ad-hoc networks. The increasing development of these networks during recent years has revitalized the interest in understanding the most basic information-theoretic setups such as broadcast, interference and relay networks. A convenient wireless model for such scenarios, as has been widely adopted in the literature, is slow-fading one where accurate channel state information (CSI) may be available to the receivers but not to transmitters, and only partial CSI is available to intermediate nodes. In these cases, classical Shannon capacities are typically zero due to the non-zero probability of channels experiencing an arbitrarily deep fade, so performance is instead quantified in terms of maximum achievable rates subject to a constraint on the  tolerated error probability (see~\cite{720551} and references therein).

The term ``cooperation strategy" stands for the procedure used to forward information from source to destination in relay networks. In selecting a cooperative scheme for wireless scenarios, several factors must be considered in order to preserve the capability of relay nodes to deal with the physical and statistical nature of their channel disturbances. The main cooperation strategies have been first introduced  by Cover-El Gamal~\cite{Cover1979} for the relay channel. Although these coding schemes were not shown to achieve capacity of the additive white Gaussian noise (AWGN) channel, Decode-and-Forward (DF) scheme has been shown to be well suitable for situations where the source-to-relay channel is stronger than the others channels while Compress-and-Forward (CF) scheme is preferable for situations where the relay-to-destination channel is the strongest link. Essentially, a relay using DF scheme forwards information based on a hard estimate of the encoder's message  whereas CF scheme is based on a soft 
estimate.  Lately El Gamal-Mohseni-Zahedi~\cite{Elgamal2006} developed an alternative version of CF scheme (not based on Wyner-Ziv coding and sequential decoding at the destination) which achieves the same rate that CF scheme~\cite{Cover1979}. In fact, both CF schemes can perform within a constant gap to the information-theoretic capacity of the AWGN relay channel, regardless of channel parameters~\cite{Avestimehr2011}. 

More recently, there has been a growing interest in cooperative networks with multiple relays and several attempts were made to develop cooperation strategies, e.g., for multiple access and broadcast relay channels (see~\cite{Kramer2005, Liang2007B, 6341082} and references therein). The capacity of degraded  unicast cooperative networks is derived in~\cite{Kramer2005} by using a sequential DF scheme while the capacity of a class of modulo-sum relay channels is found in~\cite{4787628} by using a CF based scheme. Graphical multicast networks were studied in~\cite{Ahlswede2000} where the ``max-flow min-cut theorem'' for network information flow was presented for the point-to-point communication network. Deterministic networks with no interference were studied in~\cite{Ratnaker2006} whereas the capacity of wireless erasure multicast networks was determined in~\cite{Dana2006}, and  the scaling behavior of cooperative multicasting in wireless networks was studied in~\cite{1638558}. 

\subsection{Related Work}

 An approximation approach to general networks via deterministic channels was introduced by Avestimehr-Diggavi-Tse~\cite{Avestimehr2011}. This approach yields a novel   improvement over CF scheme, referred to as ``Quantize-Map-and-Forward" (QMF), which achieves performance within constant gap of capacity for unicast AWGN networks with arbitrary number of relays. This important feature  guarantees the uniformity in the channel coefficients and hence the fading statistics. Relay nodes quantize their received signals at noise level, map them randomly to Gaussian codewords and forward them to the others nodes. The fundamental difference between CF and QMF schemes relies on the delay and CSI aspects. The standard CF scheme~\cite{Cover1979} requires successive decoding at the destination and forward channel knowledge at the relays while QMF uses joint decoding of descriptions and messages with only CSI at the destination. As a matter of fact, this approach has played a key role in the development of several 
further results on cooperative wireless networks.

In~\cite{6034734}, Nazer-Gastpar propose an ingenious coding scheme, referred to as compute-and-forward, which aims at allowing the relays to decode and send noiseless functions --linear combinations-- of the transmitted messages. By combining all these descriptions, the destination determines the original messages being sent. Indeed, due to the additive nature of the channel, each relay receives a linear combination of the lattice codewords~\cite{6034734} in addition to some additive noise, which have the property that any integer linear combination  is still a codeword~\cite{6200862, 6522158}. The relays then decode the linear combination of the codewords and thus a noiseless function of the messages. Nevertheless, the lattice property requires a integer linear combination of codewords to guarantee that it is still a codeword, however the linear combination induced by wireless channels have arbitrary real (or complex) channel gains. In order to overcome this difficulty, the authors in \cite{6034734} 
propose to scale the received channel output so that the received signal is close to an integer linear combination. The tightness of this approximation relies on the scaling factor which introduces a tradeoff between closeness of approximation and noise amplification.

Recent work~\cite{Lim2011} by Lim~\emph{et al.} generalizes QMF approach to arbitrary memoryless multicast networks by introducing the notion of ``Noisy Network Coding" (NNC) scheme, which implies the previous inner bounds in~\cite{Dana2006,Avestimehr2011}. As a matter of fact,  Yassaee~\emph{et al.}  in \cite{Yassaee2008,Yassaee2009,Yassaee2011} independently introduced for the first time the idea of NNC and derived the same achievable rate regions. In \cite{Lim2011}, relay nodes based on NNC scheme send the same --long-- message over many blocks of equal length --repetitive encoding-- and the descriptions at the relays do not require binning while their indices are non-uniquely decoded at the destination. While the same result was obtained by using short messages in \cite{Yassaee2008,Yassaee2009,Yassaee2011}. 

The achievable region from NNC scheme is shown to be tight for specific cases, e.g., deterministic and erasure networks, and in particular, it achieves within constant gap of capacity for multicast AWGN relay networks.
Further progress was made in~\cite{Wu2013} where authors showed that the gain in NNC comes from backward decoding and delaying the decoding procedure. The use of different message coding opens up the possibility of combining DF and  NNC scheme. This approach, referred to as ``Short-Message Noisy Network Coding" (SNNC),  was taken in~\cite{6217861,Behboodi2012A} and~\cite{Kramer2011,Hou2012}. Transmission is performed over $(B+L)$ blocks, where $B$ denotes  the number of blocks in which a new message is being transmitted and $L$ denotes the number of blocks in which the previous messages are repeated according to a specific pattern. Both $(B,L)$ are required to be large enough in~\cite{Wu2013} while only $B$ needs to be large enough in~\cite{6217861}, and the destination uses backward decoding. In this case,  relays are divided into two sets, the relays in the first set use NNC scheme while those in the second set use DF scheme. 

The previously mentioned works have neglected two aspects of cooperative unicast networks. First, all relay nodes are capable of collaborating with each other to increase their chances of decoding the source message, similarly as done in compute-and-forward~\cite{6034734}, and second, the destination can  benefit from  noisy descriptions of all nodes which also includes DF relays.  Actually, NNC and SNNC schemes have since then been exploited in various ways, e.g., multi-level DF schemes for DF relays are investigated in~\cite{DBLP:journals/corr/abs-1209-4889,Du2012,Hou2013} where an aware source exploits the existence of a hierarchy of the relays based on their channel quality.  

\subsection{Contribution and Outline}

In this paper, we investigate coding strategies for cooperative unicasting in wireless networks. This problem consists of a source that wishes to communicate a single message to a destination in presence of multiple relay nodes. The focus is on wireless configurations where without CSI, the source cannot any longer agree with the relays to jointly select an adequate cooperative strategy for each specific draw of the network parameters. Traditional approaches to deal with this scenario falls into composite models for networks~\cite{Effros2010} which, unlike compound models~\cite{Wolfowitz1960}, channel uncertainty is addressed by introducing a probability distribution (PD) from which the current channel index (or vector of channels parameters) is drawn but remains fixed during the communication. Composite cooperative AWGN networks have been studied beforehand via the notion of capacity versus outage (see~\cite{Katz2009, Behboodi2011A,1638558} among other references). This setup prevents, in general, the 
source use of any hierarchical multi-level scheme~\cite{DBLP:journals/corr/abs-1209-4889,Du2012} to enhance cooperation among the nodes. Notice that without CSI at the source such approach would clearly result in performance degradation.

We shall follow an approach similar to that of compute-and-forward~\cite{6034734} to the study of simultaneous coding strategies~\cite{Behboodi2010A} that are capable of enabling all nodes to decide --depending on their instantaneous channel measurements-- whether would be possible to decode-and-forward messages (e.g. the amount of available noisy descriptions provides enough information) and which nodes should cooperate with each other by decoding noisy descriptions of observations (or noisy functions of the transmitted messages). In the first part, we introduce ``Mixed Noisy Network Coding'' (MNNC)  for memoryless unicast networks with perfect CSI at all nodes while in the second part, we focus on composite cooperative unicast networks where the channel parameters are assumed to be unknown at the source and fully known at the destination, but  only partly known at the intermediate nodes. We introduce through MNNC scheme the concept of ``Selective Coding Strategy" (SCS) that enables all relays to 
dynamically select, based on on the available CSI, both the cooperative strategy and the nodes to benefit from cooperation. 

Our main contributions are summarized below:
\begin{enumerate}
\item We introduce MNNC scheme where part of the nodes are allowed to decode-and-forward (DF) messages while all of them (without exception) transmit noisy descriptions of their observations (cf. Theorem~\ref{thm:MNNC} and Corollary~\ref{coro:MNNC}, section~\ref{SectionII-2}). Moreover, these descriptions can be exploited at the destination and the DF relays ends to decode the intended message --based on offset coding-- while creating full cooperation among the nodes. A general achievable rate  is derived from this cooperative strategy which can be viewed as a generalization of several existing results in the literature (cf. section~\ref{SectionII-3}). 

It is worth mentioning  that this coding strategy differs from NNC~\cite{Lim2011} in at least two important aspects: (i) the use of short-messages coding~\cite{Yassaee2009,Wu2013, Kramer2011,Kramer2011A,6217861} where transmission expands over $(B+L)$ blocks and  the relays retransmit the compression index of  block $(B+2)$ during the last $(L-2)$ blocks while backward decoding is used at the destination, and (ii) the relays use simultaneous cooperative strategies~\cite{Behboodi2011A} by decoding and forwarding messages in addition to communicate noisy descriptions of their observations. 

\item  We further extend the concept of MNNC to multi-hopping scenarios that we refer to as ``Layered MNNC" (LMNNC) where DF relays are organized into disjoint groups representing one hop in the network (cf. section~\ref{SectionII-4}). Furthermore, LMNNC performs at least as good as MNNC and improves upon the existent results~\cite[Theorem 2]{Hou2013}.

\item For cooperative unicast AWGN networks  we show that, provided DF relays are properly chosen,  MNNC improves over all previously established constant gaps to the cut-set bound~\cite{Avestimehr2011,Lim2011} (cf. Proposition~\ref{prop-constant-gap}, section~\ref{SectionIII-2}). As a matter of fact, the presence of mixed cooperation strategies~\cite{Hou2012, Behboodi2012A} introduces considerable difficulty when attempting to compare such scheme to the original NNC scheme~\cite{Lim2011}. This issue is solved by the proposed MNNC scheme where all nodes are enable to simultaneously decode and forward messages and transmit noisy descriptions of their observations. 

\item We finally study composite cooperative unicast networks where the channel parameters are randomly drawn from a probability distribution (cf. section~\ref{SectionIV-0}). Each random  draw is assumed to be unknown at the source and fully known at the destination, but  only partly known at the relay nodes. We introduce through MNNC the concept of SCS that  enables relays to decide dynamically  (e.g. based on their channel measurements) whether would be possible to decode and forward messages in addition to communication of noisy descriptions to the destination and other nodes. It is demonstrated through the asymptotic average error probability of the slow-fading AWGN relay channel that  SCS clearly outperforms conventional cooperative schemes  (cf. section~\ref{SectionV}). 
\end{enumerate} 

Within the framework of wireless networks, the results of this paper are therefore useful to analyze the relationship between simultaneous cooperative strategies and the use of available CSI at nodes to dynamically select the coding strategy, transmission rate and the asymptotic error probability. A connection is established between the asymptotic error probability and the outage probability. As a matter of fact,  assuming codes of sufficiently long block lengths, outage probability dominates from above and below the asymptotic error probability.  Although our results are specific to cooperative unicasting, we believe that the framework is enough general to be useful more broadly  in the analysis of user cooperation for multi-source/multicast networks.
 
\subsection*{Notations}
The vector notation $\underline{x}$ stands for the collection of $n$ samples $(x_1,\dots,x_n)$ while upper-case letters $X^n$ are used to denote a vector of random variables (RVs) $(X_1,\dots,X_n)$. The random channels parameters are denoted by $\uptheta$ and any specific draw is denoted by $\theta$. Let $\mc{N}\triangleq \{1,\dots  ,N\}$ denote a set of indices, then for any subset  $\mc{S}\subseteq\mc{N}$ the vector of RVs $X_{\mc{S}}$ stands for the collection $(X_i)_{i\in \mc{S}}$; and similarly $X_{\mc{S}^c}=(X_i)_{i\in \mc{N} -\mc{S} }$ where ``$-$" is understood as setminus. The indicator function for the event $\mc{A}$ is denoted by $\mathbf{1}[\mc A]$.  Differential entropy is denoted by $h(\cdot)$, and mutual information by $I(\cdot\,;\,\cdot)$ while $\mathcal{C}(x)\triangleq \frac{1}{2}\log\left(1+x \right)$. Then, the entropy of $X_{\mc{S}}$ is defined by $H(X_{\mc{S}})=H\left( (X_i)_{i\in \mc{S}} \right)$ and similarly with mutual information. Let $X$, $Y$ and $Z$ be three RVs on some alphabets 
with probability distribution~$p$. If $p(x|yz)=p(x|y)$ for each $x,y,z$, then they form a Markov chain, which is denoted by $X\mkv Y\mkv Z$. Finally we denote {\it strong $\epsilon$-typical}  and {\it conditional strong $\epsilon$-typical } sets by $\mc{A}^n_\epsilon(X)$ and $\mc{A}^n_\epsilon(Y|X)$, respectively  (see \cite{cover-book} for details).  Logarithms are taken in base $2$ and denoted by $\log(\cdot)$. 
\begin{figure} [t]
\centering  
\includegraphics [width=.5 \textwidth] {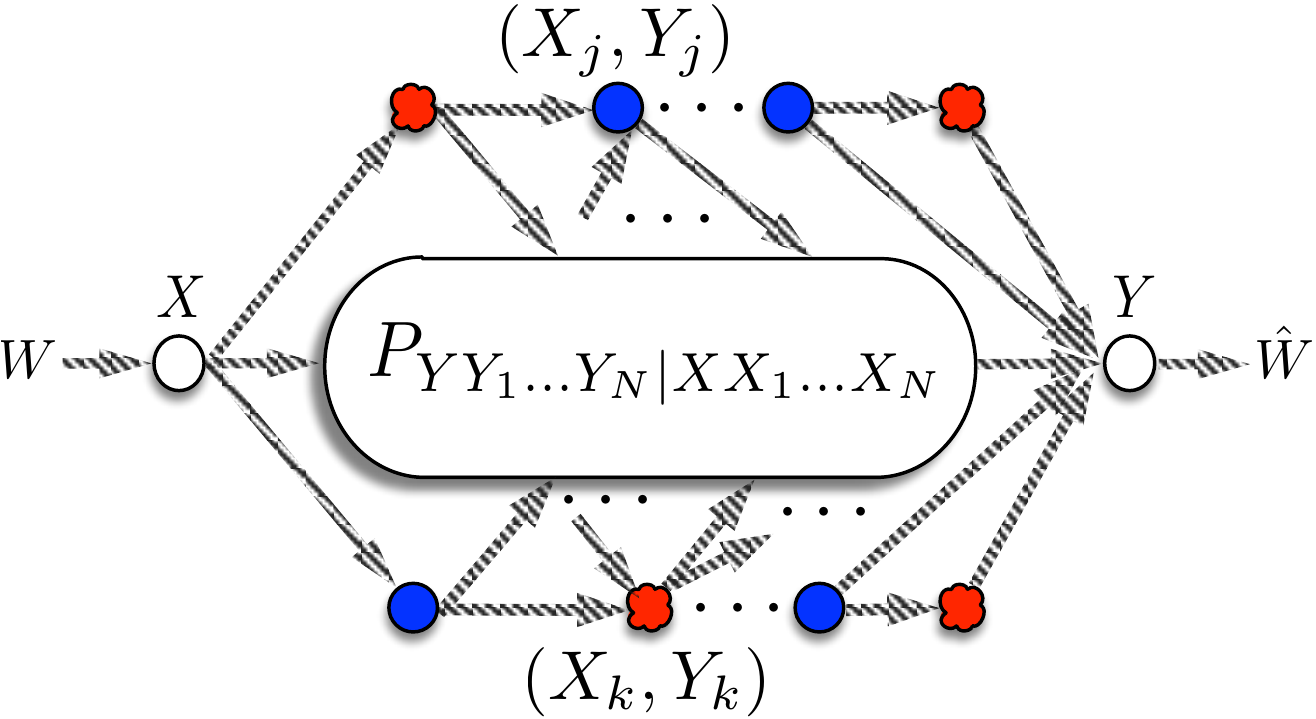}    
\caption{Cooperative unicast network with relay nodes $(\underbar{X}_j,\underbar{Y}_j)$, for all $j\in\mc{N}$, source input $\underbar{X}$ and destination $\underbar{Y}$.}
\label{fig:UMRC}
\end{figure}
 

\section{Mixed Noisy Network Coding}\label{SectionII-0}
In this section, we introduce the problem of communicating a single message to a destination in presence of multiple relay nodes. Through this section, we shall assume that all channel parameters involved in the network are known to all terminals. We introduce a novel cooperation scheme referred to as ``Mixed Noisy Network Coding" (MNNC)  that yields a rather general achievable rate expression improving over several existing results in the literature which can be viewed as particular cases. Then, we further extend this concept to multi-hopping scenarios, referred  to as ``Layered MNNC" (LMNNC), where DF relays are organized into disjoint groups representing one hop in the network.

\subsection{Definition of the Cooperative Unicast Network}\label{SectionII-1}

The cooperative unicast network as depicted in Fig.~\ref{fig:UMRC} is defined by a conditional probability distribution (PD) characterizing the destination output $Y$ and the observations at each of the relay nodes $Y_\mathcal{N}\triangleq (Y_{1},Y_{2},\dots ,Y_{N})$ given the source input $X$ and all relay inputs $X_\mathcal{N}\triangleq (X_{1},X_{2},\dots ,X_{N})$, and
$$
\mathbb{P}_{Y Y_\mathcal{N} |X X_\mathcal{N} }:\,  \mathcal{X}\times \mathcal{X}_1\times\dots  \times \mathcal{X}_N \longmapsto  \mathcal{Y} \times \mathcal{Y}_1\times\dots  \times \mathcal{Y}_N\, ,
$$
where $\mc{N}=\{1,\dots,N\}$. This network is assumed to be memoryless and without channel feedback, and all alphabets are assumed to be finite. We let  $\mathbb{P}_{Y^n Y_\mathcal{N}^n |X^n X_\mathcal{N}^n }$ denote the PD of the $n$-memoryless extension.
 \vspace{1mm} 
\begin{definition}[code and achievability] \label{def-code-MNNC}
A code-$\mc{C}(n,M_n,\epsilon_n)$ for the cooperative unicast network consists of the following mappings:
\begin{itemize}
\item An encoder mapping:  
$$
\varphi:\mc{M}_{n}=\{1,\dots,M_n\} \longmapsto \mathcal{X}^n \, ,
$$ 
\item A decoder mapping: $ \phi:\mathcal{Y}^n \longmapsto \mc{M}_{n}$\, ,
\item A sequence of relay functions: 
$$
\left\{ f^{(k)}_{i} :\mathcal{Y}_k^{i-1}  \longmapsto \mathcal{X}_{k} \right \}_{i=1}^n\, ,
$$
for $k\in\mc{N}$, and average error probability given by
\begin{equation*}
\epsilon_n \triangleq \frac{1}{M_n} \sum\limits_{w=1}^{M_n} \Pr\big[\phi(Y^n) \neq w \big]\, .
\label{errorprob_def-MNNC}   
\end{equation*}    
\end{itemize}

A positive rate $R$ is said to be achievable for the cooperative unicast network if there exists a code-$\mc{C}(n,M_n,\epsilon_n)$ defined as above such that
\begin{equation*}
\liminf\limits_{n\rightarrow \infty}\frac{1}{n} \log M_n \geq R 
\end{equation*} 
and  
\begin{equation*}
 \limsup\limits_{n\rightarrow \infty}\,\epsilon_n=0\,.
\label{errorprob_def-MR}   
\end{equation*}  
  The supremum of all achievable rates is the capacity of the unicast multi-relay network.
\end{definition}

\subsection{Mixed Noisy Network Coding}\label{SectionII-2}

\begin{figure} [t]
\centering  
\includegraphics [width=.45 \textwidth] {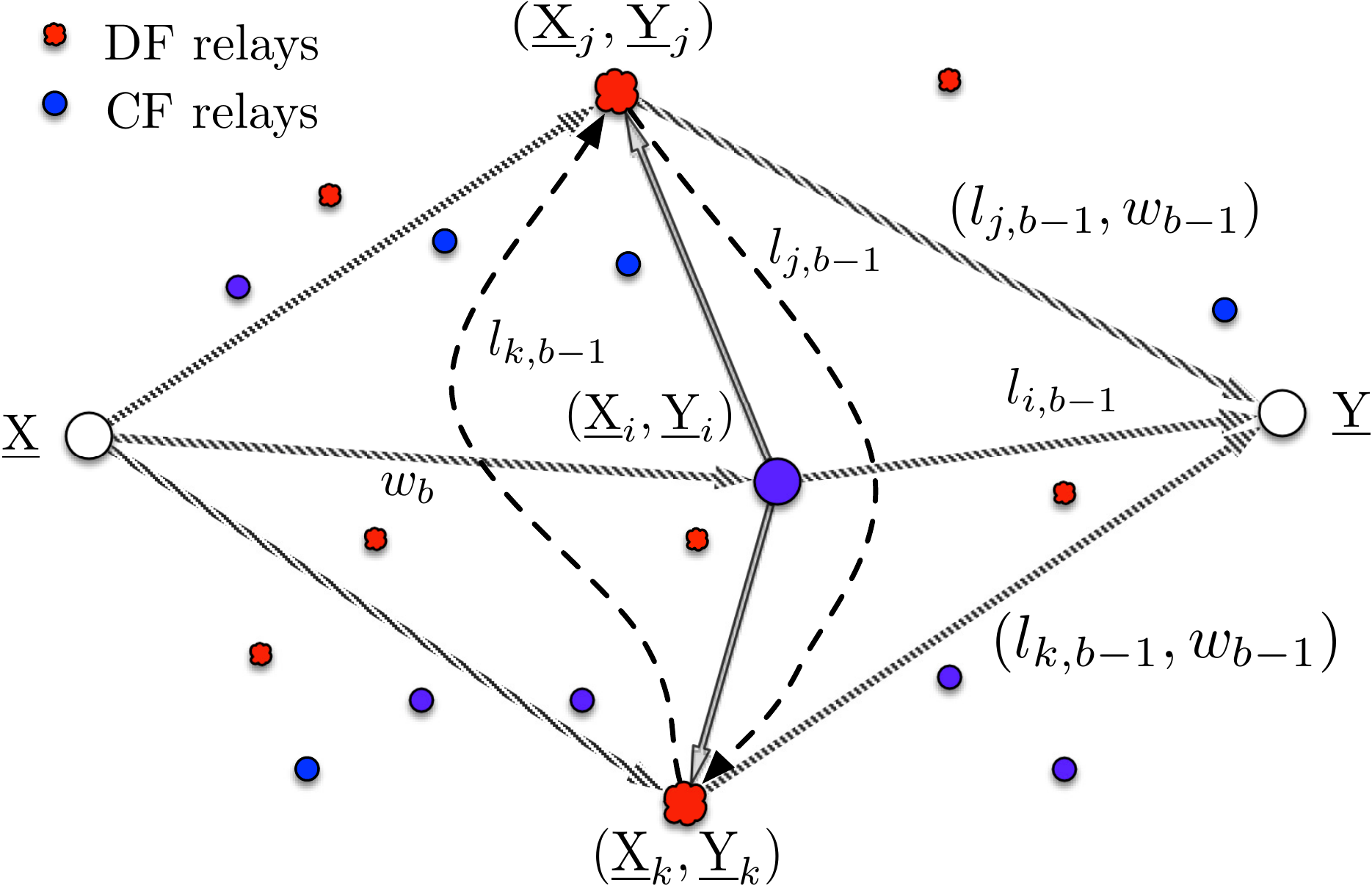}    
\caption{Mixed Noisy Network Coding (MNNC).}
\label{fig:MNNC}
\end{figure}

The key ingredients behind MNNC scheme rely on the following ideas.

\begin{enumerate}
\item Simultaneous use of different cooperative strategies among the relay nodes: Relay nodes are divided  into two disjoint groups of relays denoted by $(\mc{V},\mc{V}^c)$ satisfying $\mc{V} \cup \mc{V}^c=\mc{N}$. As shown in Fig.~\ref{fig:MNNC}, relays in the group $\mc{V}^c$ with nodes indices $(j,k)$ are simultaneously employing partly DF and CF scheme as cooperation strategy while relays in $\mc{V}$, denoted by index $i$, simply employ CF scheme. 
\item Full cooperation among all nodes in the network: Each relay in the network is transmitting the compressed version of its observation and thus all DF relays take advantage of the compression indices of the other relays in the network, which clearly improves the decoding of source messages. In this setting,  transmission takes place via \emph{block-Markov} coding in $B+L$ blocks each of length $n$, where DF relays for each block $b\leq B+2$ forward the source message of the $(b-2)$-th block. We remark that this is slightly different from the conventional DF scheme~\cite{Cover1979}, where the relay forwards the message of the previous block. Indeed, the compression index of the $b$-th block is transmitted only in block $b+1$ and thus DF relays have to wait until the end of block $b+1$ to decode the compression index as well as the message of block $b$. Hence, they can forward the message of block $b$ only after block $b+1$. 
\item Selection of descriptions to be decoded at both destination and  relays: The destination selects the help of only the best subset $\mc{T}$ of nodes among all possible relays $\mc{N}$. Similarly, the $k$-th DF relay in the set
 $\mc{V}^c$ is allowed to exploit only the help of a selected subset of relays, which are denoted by $\mc{T}_k$. 
\end{enumerate}
The next theorem provides the corresponding achievable rate for this scheme.

\vspace{1mm}
\begin{theorem} [Mixed Noisy Network Coding (MNNC)]
All rates $R$ satisfying the following inequality are achievable for the cooperative unicast network:
\begin{IEEEeqnarray}{rCl}
R &\leq& \max_{P\in\mc{P}}
\max_{\mc{V}\subseteq\mc{N}}\, \min\,\bigg(\max_{\mc{T}\in\Upsilon(\mc{N})}\,\min_{\mc{V}^c \subseteq\mc{S}\subseteq\mc{T}} R_{\mc{T}}(\mc{S})\,,\,\nonumber\\
\IEEEeqnarraymulticol{3}{r}{\min_{k\in\mc{V}^c}\,\max_{\mc{T}_k\in\Upsilon_k(\mc{N})}\,\min_{\mc{S}\subseteq\mc{T}_k} R^{(k)}_{\mc{T}_k}(\mc{S})\bigg)\ ,}	
\label{MNNC}
\end{IEEEeqnarray}
where
\begin{IEEEeqnarray*}{rCl}
R_{\mc{T}}(\mc{S})&\triangleq& I(XX_{\mc{S}};\hat{Y}_{\mc{S}^c}Y|X_{\mc{S}^c}Q)\nonumber\\
\IEEEeqnarraymulticol{3}{c}{ -I(\hat{Y}_{\mc{S}};{Y}_{\mc{S}}|XX_{\mc{T}}\hat{Y}_{\mc{S}^c}YQ)\,,}\label{src-rate}\\
R^{(k)}_{\mc{T}_k}(\mc{S}) &\triangleq&
I(X;\hat{Y}_{\mc{T}_k}Y_k|VX_kX_{\mc{T}_k}Q)\nonumber\\
\IEEEeqnarraymulticol{3}{c}{ +I(X_{\mc{S}};Y_k|VX_kX_{\mc{S}^c}Q)   -I(\hat{Y}_{\mc{S}};Y_{\mc{S}}|VX_kX_{\mc{T}_k}\hat{Y}_{\mc{S}^c}Y_kQ)\ ,}\label{relay-rate}
\end{IEEEeqnarray*}
with $\mc{S}^c\triangleq \mc{T}-\mc{S}$ in \eqref{src-rate} and $\mc{S}^c\triangleq \mc{T}_k-\mc{S}$ in \eqref{relay-rate}. Moreover, $\mc{T}\subseteq \mc{N}$, $\mc{T}_k\subseteq \mc{N}-\{k\}$ and $\mc{V}^c=\mc{N}-\mc{V}$, and $\Upsilon(\mc{N})$ and $\Upsilon_k(\mc{N})$ are defined as follows:
\begin{IEEEeqnarray*}{rCl}
 \Upsilon(\mc{N})& \triangleq &\big\{\mc{T}\subseteq\mc{N}: \,Q_{\mc{T}}(\mc{S})\geq 0\,\,\forall \, \mc{S}\subseteq\mc{T} \big\}\ ,\\
\Upsilon_k(\mc{N})& \triangleq &\big\{\mc{T}\subseteq\mc{N}-\{k\}:  \,Q^{(k)}_{\mc{T}}(\mc{S})\geq 0\, \,\forall \,\mc{S}\subseteq\mc{T}\big\}\ ,
\label{MNNCcondition}
\end{IEEEeqnarray*}
where $Q_{\mc{T}}(\mc{S})$ and  $Q^{(k)}_{\mc{T}}(\mc{S})$ are used to denote: 
\begin{IEEEeqnarray*}{rcl}
Q_{\mc{T}}(\mc{S})&\triangleq &I(X_{\mc{S}};\hat{Y}_{\mc{S}^c}Y|VXX_{\mc{S}^c}Q)\nonumber\\
&&-I(\hat{Y}_{\mc{S}};{Y}_{\mc{S}}|VXX_{\mc{T}}\hat{Y}_{\mc{S}^c}YQ)\ ,\\
Q^{(k)}_{\mc{T}}(\mc{S})& \triangleq &I(X_{\mc{S}};Y_k|VX_kX_{\mc{S}^c}Q)\nonumber\\
&&-I(\hat{Y}_{\mc{S}};Y_{\mc{S}}|VXX_kX_{\mc{T}}\hat{Y}_{\mc{S}^c}Y_kQ)\ .
\end{IEEEeqnarray*}
The set of all admissible input PDs $\mc{P}$ is defined by
\begin{IEEEeqnarray*}{lCl}
\mathcal{P} &\triangleq&   \displaystyle\Big\{P_{QVXX_{\mc{N}}Y_{\mc{N}}\hat{Y}_{\mc{N}}Y}= P_{QVX} \mathbb{P}_{YY_{\mc{N}}|XX_{\mc{N}}}\nonumber\\
&&\prod_{j\in \mc{V}^c}P_{X_{j}|VQ}P_{\hat{Y}_j|VX_jY_jQ}
\prod_{j\in \mc{V}} P_{X_j|Q}P_{\hat{Y}_j|X_jY_jQ}\Big\}\ . 
\label{eq:MNNC-APD}
\end{IEEEeqnarray*}
\label{thm:MNNC}
\end{theorem}
\begin{IEEEproof}
The proof of this theorem is relegated to Appendix~\ref{proof:MNNC}.
\end{IEEEproof}

\begin{remark}
By following similar arguments to those used in~\cite{Kramer2011}, it is not difficult to check that the optimization of the term $R_{\mc{T}}(\mc{S})$ in expression \eqref{MNNC} can be taken over $\mc{T}\subseteq\mc{N}$ instead of a subset $\mc{T}\in\Upsilon(\mc{N})$. To this end, it is enough to show that for every $\mc{T}\subseteq\mc{N}$, if there is a subset $\mc{A}\subseteq\mc{T}$ such that $Q_{\mc{T}}(\mc{A})<0$, then there must be another subset $\mc{T}^\prime\subset\mc{T}\subseteq\mc{N}$ such that the region with respect to $\mc{T}^\prime$ is increased. Therefore, for each $\mc{S}^\prime\subseteq\mc{T}^\prime$ there is a unique $\mc{S}\subseteq\mc{T}$ such that $R_{\mc{T}}(\mc{S})\leq R_{\mc{T}^\prime}(\mc{S}^\prime)$ which yields the desired inequality. Interestingly, it appears that this new set is obtained by removing  from the set $\mc{T}$ the relays that are present in $\mc{A}$. Thus, for all $\mc{A}\subseteq\mc{S}\subseteq\mc{T}$, it holds that
\begin{IEEEeqnarray}{rCl}
R_{\mc{T}}(\mc{S})=R_{\mc{T}\cap\mc{A}^c}(\mc{S}\cap\mc{A}^c)+Q_{\mc{T}}(\mc{A})\ ,
\label{Qnegative}
\end{IEEEeqnarray}
for each $\mc{S}\subseteq\mc{T}\cap\mc{A}^c$. Hence, for every set $\mc{T}\subseteq\mc{N}$, if  there is a set $\mc{A}\subseteq\mc{T}$ such that $Q_{\mc{T}}(\mc{A})<0$, then it can be seen from \eqref{Qnegative} that 
\begin{IEEEeqnarray}{rCl}
R_{\mc{T}}(\mc{S}\cup\mc{A})<R_{\mc{T}\cap\mc{A}^c}(\mc{S})\ ,
\end{IEEEeqnarray}
which implies the final rate is increased by replacing $\mc{T}$ with $\mc{T}\cap\mc{A}^c$. For instance, for each $\mc{T}\subseteq\mc{N}$ and $\mc{T}\in{\Upsilon(\mc{N})}^c$, we can find a subset $\mc{T}^\prime \subset\mc{T}\subseteq\mc{N}$ that is not necessarily in $\Upsilon^c(\mc{N})$ such that the region with respect to $\mc{T}^\prime$ is enlarged and this proves the claim.

 A direct consequence of the above observation is that, for every $\mc{T}$ and $\mc{A}\subseteq\mc{T}$ such that $Q_{\mc{T}}(\mc{A})<0$, it is enough to ignore --not looking at their description indices-- the relay nodes in $\mc{A}$. Thus, the next achievable rate simply follows from this replacement:
 \begin{IEEEeqnarray}{rCl}
R &\leq& \max_{P\in\mc{P}}
\max_{\mc{V}\subseteq\mc{N}}
\, \min\bigg(\max_{\mc{T}\subseteq\mc{N}}\, \min_{\mc{V}^c\subseteq\mc{S}\subseteq\mc{T}} R_{\mc{T}}(\mc{S})\,,\,\nonumber\\
\IEEEeqnarraymulticol{3}{r}{\min_{k\in\mc{V}^c}\max_{\mc{T}_k\in\Upsilon_k(\mc{N})}\min_{\mc{S}\subseteq\mc{T}_k} R^{(k)}_{\mc{T}_k}(\mc{S})\bigg)\ .} 
\label{MNNC-new}
\end{IEEEeqnarray}
\label{remark-MNNC}
\end{remark}

It is worth mentioning here that by setting $\mc{V}=\mc{N}$, the rate expression in Theorem~\ref{thm:MNNC} reduces to that of SNNC~\cite{Wu2013,Kramer2011}, which was shown to be equivalent to NNC first derived in~\cite{Lim2011}. Thus, Theorem~\ref{thm:MNNC} can be seen as a generalization that includes the previous results based on NNC schemes while it also provides a potentially larger rate (e.g. it achieves the capacity of degraded ``Relay Channels" which is not the case of NNC region). 

We also note that since DF relays require the use of ``forward decoding", the rate $R^{(k)}_{\mc{T}_k}(\mc{S})$ is clearly expected to be smaller compared to the situation where all relays are allowed to use ``backward decoding". The reason for this, as was also pointed out in~\cite{Wu2013}, is that the gain of NNC is due to delaying decoding until the last block. However, postponing decoding to the last block would not be possible for those relays cooperating via DF scheme which brings the rate loss we mentioned. In order to better explore this rate loss, let us assume the unicast relay network where all relays are forced to use CF scheme, but the destination decodes based on ``forward decoding" --instead of ``backward decoding"--, i.e., the same decoding method as DF relays in Theorem~\ref{thm:MNNC}. As a consequence of Theorem~\ref{thm:MNNC}, we can obtain the next corollary that provides an achievable rate based on ``forward decoding".\vspace{1mm}

\begin{corollary}  [Forward decoding NNC]\label{coro:MNNC}
Assuming that all nodes are forced to use ``forward decoding", then all rates $R$ satisfying the following inequality are achievable:
\begin{IEEEeqnarray}{rCl}
R \leq \max_{P\in\mc{P}}\,
\max_{\mc{T}\subseteq\Upsilon(\mc{N})}\, \min_{\mc{S}\subseteq\mc{T}} \,R^{\textrm{FD}}_{\mc{T}}(\mc{S})\ ,
\label{NNC-FD}
\end{IEEEeqnarray}
where
\begin{IEEEeqnarray}{rCl}
R^{\textrm{FD}}_{\mc{T}}(\mc{S})&\triangleq &I(X;\hat{Y}_{\mc{T}}Y|X_{\mc{T}}Q)+I(X_{\mc{S}};Y|X_{\mc{S}^c}Q)\nonumber\\
\IEEEeqnarraymulticol{3}{r}{ -I(\hat{Y}_{\mc{S}};{Y}_{\mc{S}}|X_{\mc{T}}\hat{Y}_{\mc{S}^c}YQ) }
\end{IEEEeqnarray}
with $\mc{S}^c\triangleq \mc{T}-\mc{S}$ and  $\Upsilon(\mc{N})$ defined by
\begin{IEEEeqnarray}{rCl}
\Upsilon(\mc{N})&\triangleq& \big\{\mc{T}\subseteq\mc{N}: \, I(X_{\mc{S}};Y|X_{\mc{S}^c}Q)\nonumber\\
&&-I(\hat{Y}_{\mc{S}};Y_{\mc{S}}|XX_{\mc{T}}\hat{Y}_{\mc{S}^c}YQ) \geq 0\,\,\forall\ \mc{S}\subseteq\mc{T}\big\}\ .\,\,\,\,\,\,\,
\end{IEEEeqnarray}
\end{corollary}

The first observation from the above rate is that ``forward decoding" at the destination does not perform  in general as good as NNC. Nevertheless, it potentially improves on the use of ``binning" and other ``forward decoding" techniques~\cite{Wu2013}. Particularly, the condition which determines the optimization region in~\cite{Wu2013}, i.e.,
\begin{equation}
I(X_{\mc{S}};Y|X_{\mc{S}^c}Q)-I(\hat{Y}_{\mc{S}};Y_{\mc{S}}|X_{\mc{T}}\hat{Y}_{\mc{S}^c}YQ)\geq 0\,,
\end{equation}
creates a smaller optimization region than $\Upsilon(\mc{N})$ because
\begin{equation}
I(\hat{Y}_{\mc{S}};Y_{\mc{S}}|X_{\mc{T}}\hat{Y}_{\mc{S}^c}YQ)\geq I(\hat{Y}_{\mc{S}};Y_{\mc{S}}|XX_{\mc{T}}\hat{Y}_{\mc{S}^c}YQ)\,.
\end{equation}
Actually, the use of ``joint-forward decoding" without ``binning" performs potentially better respect to ``joint-forward decoding" with ``binning".\vspace{1mm}

\begin{remark}
The reason for the sub-optimality of the current forward-decoding scheme is two fold and can be easily understood from the proof. The DF relays decode the compression indices of others relays and also the source message by using the typical sets~\eqref{DFrelaydecoding}. The destination, however, does the same but using the typical set given by~\eqref{Destdecoding}. By comparing these decoding rules, it is not difficulty to see that the destination  decodes jointly the compression index and the source message using a typical set involving: the description $\hat{Y}_{\mc{T}}$, the compression index of the other relays and the source codeword $X$. Whereas these terms are absent from the second decoding rule of DF relays in~\eqref{DFrelaydecoding}. Indeed, DF relays decode jointly  the source message and compression index of other relays of block $b$ based on the consecutive blocks $b$ and $b+1$. In the decoding block $b+1$, DF relays cannot dispose of any information about the fresh compression index and source'
s message of the same block. Intuitively, these relays cannot benefit fully from the presence of the descriptions $\hat{Y}_{\mc{T}}$ and $X$. In contrast with this, when backward decoding is used decoding  is performed  in a single block and full cooperation can be obtained. 

This problem may  be partially fixed by using layering coding~\cite{Yassaee2011}. In this case, the set of DF relays is partitioned into a set of layers $\{\mc{L}_1,\dots,\mc{L}_k\}$, representing the order in which the compression indices are decoded. Therefore, the decoding process is delayed at a given layer for few blocks, and then by proper choice of layers one can achieve each corner point of the original region. This is possible because DF relays at the layer $b$ are able to use $\hat{Y}_{\mc{L}_{i}}$ for all blocks $i\leq b-1$ while decoding the compression index. The drawback, however, is that this would require new layering and thus a new coding to achieve each of the corner points. Specifically, in communication scenarios where the current channel is unknown at the source is beneficial to have an oblivious coding schemes for which the source code does not require to be aware of the cooperation strategy. As a matter of fact, the proper choice of layering at the source to achieve a certain corner 
point is mostly dependent on the channels  parameters. As we will discuss later, the same problem occurs in multi-hopping setups  where the optimal choice and number of hops strongly depends on the channels parameters.

Although this layering coding can improve on the results of Corollary~\ref{coro:MNNC}, still there are some problems that cannot be completely solved. In our setting, the single source does not observe any output and thus, it can only cooperate with DF relays by helping them in decoding the compression indices. As a matter of fact, by looking at the proof together with expression~\eqref{eq:FWDdecoding}, it is easy to check that the message index of source at block $b+1$ must be unknown to the other relays when decoding the messages of block $b$. As a consequence of this, the source codewords cannot be fully exploited in the decoding rule. On one hand, since the source does not have any channel observation and hence cannot cooperate by transmitting its compression index while on the other, the source is the only node having new messages and therefore, its fresh information must be superimposed on the last layer of the relays. Otherwise, the relays would need to know the message beforehand which is not 
possible. 
\end{remark}

We consider now another scenario. Let us assume that each node in the network, including the destination and all relays, decides to use the help of all nodes. In other words, we set $\mc{T}\triangleq \mc{N}$ and $\mc{T}_k\triangleq \mc{N}-\{k\}$, then the following corollary easily follows from Theorem~\ref{thm:MNNC}.
\vspace{1mm}

\begin{corollary}[Fully cooperative MNNC]
Assuming that all nodes cooperate each other in the unicast network, then all rates $R$ satisfying the following inequality are achievable:
\begin{IEEEeqnarray}{rCl}
R &\leq & \max_{P\in\mc{P}}
\max_{\mc{V}\subseteq\mc{N}}
\min\bigg( \min_{\mc{V}^c\subseteq\mc{S}\subseteq\mc{N}} R_{\mc{N}}(\mc{S})\,,\,\nonumber\\
\IEEEeqnarraymulticol{3}{r}{\min_{k\in\mc{V}^c}\min_{\mc{S}\subseteq\mc{N}-\{k\}} R^{(k)}_{\mc{N}}(\mc{S})
\bigg)\ ,}
\label{MNNC-allrelay}
\end{IEEEeqnarray}
where
\begin{IEEEeqnarray}{rCl}
R_{\mc{N}}(\mc{S})&\triangleq &I(XX_{\mc{S}};\hat{Y}_{\mc{S}^c}Y|X_{\mc{S}^c}Q)\nonumber\\
&&-I(\hat{Y}_{\mc{S}};{Y}_{\mc{S}}|XX_{\mc{N}}\hat{Y}_{\mc{S}^c}YQ)\,, \label{src-rate-allrelay}\\
R^{(k)}_{\mc{N}}(\mc{S})&\triangleq &
I(X;\hat{Y}_{\mc{N}-\{k\}}Y_k|VX_{\mc{N}}Q)\nonumber\\
&&+I(X_{\mc{S}};Y_k|VX_kX_{\mc{S}^c}Q)\nonumber\\
&&{-I(\hat{Y}_{\mc{S}};Y_{\mc{S}}|VX_{\mc{N}}\hat{Y}_{\mc{S}^c}Y_kQ)}  \label{relay-rate-allrelay}
\end{IEEEeqnarray}
with $\mc{S}^c\triangleq \mc{N}-\mc{S}$ in expression \eqref{src-rate-allrelay},  $\mc{S}^c\triangleq \mc{N}-(\mc{S}\cup\{k\})$  in expression \eqref{relay-rate-allrelay} and $\mc{V}^c \triangleq \mc{N}-\mc{V}$ satisfying the constraints:
\begin{IEEEeqnarray}{rCl}
I(X_{\mc{S}}&&;\hat{Y}_{\mc{S}^c}Y|VXX_{\mc{S}^c}Q)\nonumber\\
&&-I(\hat{Y}_{\mc{S}};{Y}_{\mc{S}}|VXX_{\mc{N}}\hat{Y}_{\mc{S}^c}YQ)\geq 0\,,  \forall  \mc{S}\subseteq\mc{N}\,,\\
I(X_{\mc{S}}&&;Y_k|VX_kX_{\mc{S}^c}Q)\nonumber\\
&&-I(\hat{Y}_{\mc{S}};Y_{\mc{S}}|VXX_{\mc{N}}\hat{Y}_{\mc{S}^c}Y_kQ)\geq 0\,,\nonumber\\
&&\forall  \mc{S}\subseteq\mc{N}-\{k\}\,, \,\forall \, k\in\mc{V}^c\,.\,\,\,\,
\label{MNNCcondition-allrelay}
\end{IEEEeqnarray}
The set of all admissible input distributions $\mc{P}$ is again defined  by \eqref{eq:MNNC-APD}.
\label{col:MNNC}
\end{corollary}

\subsection{Variations of Mixed Noisy Network Coding}\label{SectionII-3}

For the rest of this section, we shall focus on some variations of MNNC scheme. 
\begin{figure} [t]
\centering  
\includegraphics [width=.45 \textwidth] {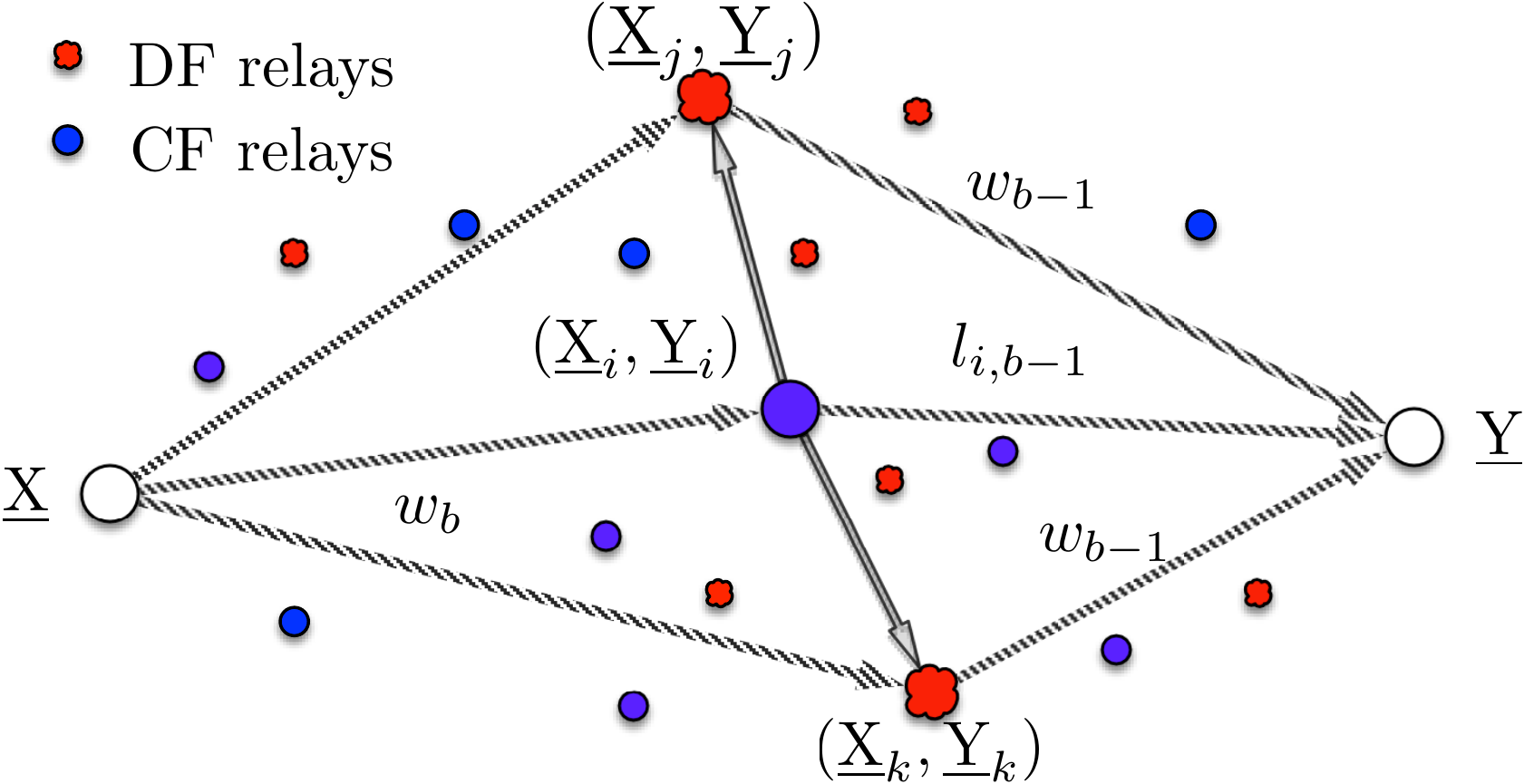}    
\caption{Non-cooperative Mixed Noisy Network Coding (MNNC) scheme.}
\label{fig:NC-MNNC}
\end{figure}
The cooperative strategy that yields Theorem \ref{thm:MNNC} allows all DF relays to cooperate with each other and with CF relays via the exchanges of their  compression indices. It appears clear that, decoding of the source message at any node becomes dependent on all other relays which in many cases will lead to rather  complex optimization problems. In order to further simplify the coding scheme and thus to avoid such complex dependences,  DF relays can be constrained to decode messages as stand alone terminals, i.e., without the use of additional description indices, as it is shown in~Fig. \ref{fig:NC-MNNC}. Moreover, this simplification can also avoid decoding delay at all DF relays, which was necessary for decoding based on the compression indices of other relays. The next theorem provides the achievable rate of this simplified coding scheme that is referred to as ``non-cooperative MNNC'', which was independently derived in~\cite{Hou2013}. 

\begin{theorem} [Non-cooperative MNNC]
Assuming that no cooperation is allowed among the relay nodes, then all rates $R$ satisfying the following inequality are achievable:
\begin{IEEEeqnarray}{rCl}
 R &\leq & \max_{P\in\mc{P}}\,\max_{\mc{V}\subseteq\mc{N}\,,\,\mc{T}\in\Upsilon(\mc{V})}\min\bigg(\min_{\mc{S}\subseteq\mc{T}} R_{\mc{T}}(\mc{S})\, ,\nonumber\\
 \IEEEeqnarraymulticol{3}{r}{\min_{i\in \mc{V}^c}I(X;Y_i|X_{\mc{V}^c}Q)\bigg)\ ,}
\label{NC-MNNC}
\end{IEEEeqnarray}
where
\begin{IEEEeqnarray}{rCl}
R_{\mc{T}}(\mc{S})& \triangleq & I(XX_{\mc{V}^c}X_{\mc{S}};\hat{Y}_{\mc{S}^c}Y|X_{\mc{S}^c}Q)\nonumber\\
 \IEEEeqnarraymulticol{3}{r}{- I(Y_{\mc{S}};\hat{Y}_{\mc{S}}|XX_{\mc{T}\cup \mc{V}^c}\hat{Y}_{\mc{S}^c}YQ) }
\end{IEEEeqnarray}
with $\mc{T}\subseteq \mc{V}\subseteq \mc{N}$, and $\mc{V}^c\triangleq\mc{N}-\mc{V}$, and $\mc{S}^c\triangleq \mc{T}-\mc{S}$. For notation convenience,  we use $\min_{\emptyset}(\cdot) \triangleq +\infty$. Moreover, $\Upsilon(\mc{V})$ is defined by
\begin{IEEEeqnarray}{rCl}
 \Upsilon(\mc{V})  &\triangleq&  \big\{\mc{T}\subseteq\mc{V}\,\big| \,I(X_{\mc{S}};\hat{Y}_{\mc{S}^c}Y|XX_{\mc{S}^c\cup\mc{V}^c}Q)\nonumber\\
 &&-I(Y_{\mc{S}};\hat{Y}_{\mc{S}}|XX_{\mc{T}\cup \mc{V}^c}\hat{Y}_{\mc{S}^c}YQ)\geq 0 \,\,\, \forall\,\,\,  \mc{S}\subseteq\mc{T} \big\}\ .\,\,\, \,\,\,  \,\,\, 
\label{NC-MNNCcondition}
\end{IEEEeqnarray}
The set of all admissible input PDs $\mc{P}$ is defined as follows:
\begin{IEEEeqnarray}{rCl}
\mathcal{P} & \triangleq & \Big\{P_{QXX_{\mc{N}}Y_{\mc{N}}\hat{Y}_{\mc{N}}Y}= \nonumber\\
 \IEEEeqnarraymulticol{3}{r}{P_{Q} P_{XX_{\mc{V}^c}|Q} \mathbb{P}_{YY_{\mc{N}}|XX_{\mc{N}}}\prod_{j\in \mc{V}} P_{X_j|Q}P_{\hat{Y}_j|X_jY_jQ}\Big\}\ .}
 \end{IEEEeqnarray}
\label{thm:NC-MNNC}
\end{theorem}
\begin{IEEEproof}
The proof of this theorem is provided in Appendix~\ref{proof:NC-MNNC}.
\end{IEEEproof}

As we have previously stated in Theorem \ref{thm:MNNC}, all DF relay nodes  forward simultaneously the source message and their  description indices. However, we can reduce the complexity of relaying functions  by forcing partial cooperation, i.e., all DF relays in $\mc{V}^c$ to use only the help of CF relays in $\mc{V}$. This simplified scheme yields the following corollary, which is a special case of Theorem~\ref{thm:MNNC}.\vspace{1mm}

\begin{corollary} [Partially cooperative MNNC]
Assuming that only partial cooperation is allowed among the relays, then all rates $R$ satisfying the following inequality are achievable:
\begin{IEEEeqnarray}{rCl}
R \leq \max_{P\in\mc{P}}
\max_{\mc{V}\subseteq\mc{N}}\, &&\min\bigg(\max_{\mc{T}\in\Upsilon(\mc{V})}\,\min_{\mc{S}\subseteq\mc{T}} R_{\mc{T}}(\mc{S})\, , \nonumber\\
&&\min_{k\in\mc{V}^c}\max_{\mc{T}_k\in\Upsilon_k(\mc{V})}\,\min_{\mc{S}\subseteq\mc{T}_k} R^{(k)}_{\mc{T}_k}(\mc{S})\bigg)\,,
\label{thm:PC-MNNC}
\end{IEEEeqnarray}
where
\begin{IEEEeqnarray}{rCl}
R_{\mc{T}}(\mc{S}) & \triangleq & I(XX_{\mc{V}^c}X_{\mc{S}};\hat{Y}_{\mc{S}^c}Y|X_{\mc{S}^c}Q)\nonumber\\
 \IEEEeqnarraymulticol{3}{C}{-I(Y_{\mc{S}};\hat{Y}_{\mc{S}}|XX_{\mc{T}\cup \mc{V}^c}\hat{Y}_{\mc{S}^c}YQ)\,,} \\
R^{(k)}_{\mc{T}_k}(\mc{S}) &\triangleq & I(X;\hat{Y}_{\mc{T}_k}Y_k|X_{\mc{V}^c}X_{\mc{T}_k}Q)+I(X_{\mc{S}};Y_k|X_{\mc{V}^c\cup\mc{S}^c}Q)\nonumber\\
 \IEEEeqnarraymulticol{3}{C}{-I(\hat{Y}_{\mc{S}};{Y}_{\mc{S}}|X_{\mc{V}^c\cup\mc{T}_k}\hat{Y}_{\mc{S}^c}Y_kQ)} \label{DFrelaycondition}
\end{IEEEeqnarray}
with sets  $\mc{T}$, and $\mc{T}_k\subseteq \mc{V}\subseteq \mc{N}$, and $\mc{S}^c\triangleq \mc{T}-\mc{S}$, and  $\mc{S}^c \triangleq \mc{T}_k-\mc{S}$, and $\mc{V}^c\triangleq \mc{N}-\mc{V}$. Moreover, $\Upsilon(\mc{V})$ and $\Upsilon_k(\mc{V})$ are defined by
\begin{IEEEeqnarray}{rCl}
 &&\Upsilon(\mc{V})\triangleq  \big\{\mc{T}\subseteq\mc{V}\,\big| \, I(X_{\mc{S}};\hat{Y}_{\mc{S}^c}Y|XX_{\mc{S}^c\cup\mc{V}^c}Q)\nonumber\\
 \IEEEeqnarraymulticol{3}{C}{ -I(Y_{\mc{S}};\hat{Y}_{\mc{S}}|XX_{\mc{T}\cup \mc{V}^c}\hat{Y}_{\mc{S}^c}YQ)\geq 0,  \,\, \,\,\forall\, \mc{S}\subseteq\mc{T}\big\}\,, }\\
 &&\Upsilon_k(\mc{V})\triangleq\big\{\mc{T}\subseteq\mc{V}\,\big| \, I(X_{\mc{S}};Y_k|X_{\mc{V}^c\cup\mc{S}^c}Q)\nonumber\\
 \IEEEeqnarraymulticol{3}{C}{-I(\hat{Y}_{\mc{S}};Y_{\mc{S}}|XX_{\mc{V}^c\cup\mc{T}}\hat{Y}_{\mc{S}^c}Y_kQ)\geq 0,  \,\, \,\,\forall\,\mc{S}\subseteq\mc{T}  \big\}\ .}
\label{PC-MNNCcondition}
\end{IEEEeqnarray}
The set of all admissible input PDs $\mc{P}$ is defined by \eqref{eq:MNNC-APD}.
 \label{cor:PC-MNNC}
\end{corollary}
\subsection{Layered Mixed Noisy Network Coding}\label{SectionII-4}

In this section, we extend the previous results to multi-hopping scenarios. As before, the relays are divided into two disjoint subsets $\mc{V}$ and $\mc{V}^c$. Nodes in $\mc{V}^c$ are purely dedicated to use DF scheme. Whereas the main difference relies on the fact that DF relays are organized into disjoint groups where each of them represents one hop in the network. Each of these groups are referred to as ``layer". Theses layeres form an ordered $T-$tuple, denoted by $\big\langle \mc{L}_j\,:\,j=[1:T]\big\rangle$, where $T$ represents the number of hops present in the scheme. We denote the set of all such ordered partitions of an arbitrary set $\mc{X}$ by $\Pi_o(\mc{X})$, i.e.,  the set $\Pi_o(\mc{X})$ contains all possible hops and layers for all relays. As it is the case for MNNC, decoding at DF relays is delayed by one block in order to benefit from the compression indices of other CF relays. Moreover, decoding at DF relays is  sequentially performed, such that DF relays at a higher layer, with higher 
indices, start to decode sooner than the other lower layers, i.e. with lower indices. In  this case, DF relays at each layer can enjoy the help of higher relays which have already decoded the message.  We next introduce notation needed for the rest of this  section:
$$
\mc{L}_{\leq d}\triangleq\bigcup_{t\leq d}\,\mc{L}_t \,\,\text{		and 	} \,\,\mc{L}_{> d}\triangleq\bigcup_{t> d}\,\mc{L}_t \ ,
$$
where  $\mc{L}_{\leq T}=\mc{V}^c$. For any sequence of RVs $\{X_i\}$ with $i=\{1,\dots ,n\}$, we define 
$$
V_{\leq m}\triangleq (V_1,V_2,\dots, V_m)\ ,
$$
and similarly define $x_{\mc{L}}\triangleq \left(x_{k}\right)_{k\in\mc{L}}$.

We next present an achievable rate for Layered Mixed Noisy Network Coding  (LMNNC).\vspace{1mm}

\begin{theorem} [Layered Mixed Noisy Network Coding]
All rates $R$ satisfying the following inequality are achievable:
\begin{IEEEeqnarray}{rCl}
R \leq & & \max_{P\in\mc{P}}
\max_{\mc{V}\subseteq\mc{N}}\, \max_{\big(\mc{L}_j,j\in[1:T]\big)\subseteq\Pi_o(\mc{V}^c)}
\, \min   \nonumber\\
& &\left(\max_{\mc{T}\in\Upsilon(\mc{V})}\, \min_{\mc{S}\subseteq\mc{T}} R_{\mc{T}}(\mc{S})\,, \right. \nonumber\\
& &
\left.\min_{i\in[1:T]}\, \min_{k\in\mc{L}_i}\,\max_{\mc{T}_k\in\Upsilon_k(\mc{V})}\,\min_{\mc{S}\subseteq\mc{T}_k} R^{(k)}_{\mc{T}_k}(\mc{L}_i,\mc{S})\,\,\right)\ ,
\label{LMNNC}
\end{IEEEeqnarray}
where
\begin{IEEEeqnarray}{rCl}
R_{\mc{T}}(\mc{S})&\triangleq& 
I(XX_{\mc{V}^c\cup\mc{S}};\hat{Y}_{\mc{S}^c}Y|X_{\mc{S}^c}Q)\nonumber\\
& &-I(\hat{Y}_{\mc{S}};{Y}_{\mc{S}}|XX_{\mc{V}^c\cup\mc{T}}\hat{Y}_{\mc{S}^c}YQ)
\,,\label{src-rateL}\\
R^{(k)}_{\mc{T}_k}(\mc{L}_t,\mc{S}) &\triangleq&
I(XV_{>t}X_{\mc{L}_{>t}};Y_k|V_{\leq t}X_{\mc{L}_{\leq t}}Q)\nonumber\\
& &+I(X_{\mc{S}};Y_k|V_{\leq T}X_{\mc{V}^c\cup\mc{S}^c}Q)  \nonumber\\
& &{-I(\hat{Y}_{\mc{T}_k};Y_{\mc{S}}|V_{\leq T}X_{\mc{V}^c\cup\mc{T}_k}Y_kQ)}
\label{relay-rateL}
\end{IEEEeqnarray}
with $\mc{S}^c\triangleq \mc{T}-\mc{S}$ in \eqref{src-rateL} and $\mc{S}^c\triangleq \mc{T}_k-\mc{S}$ in \eqref{relay-rateL}. Moreover, $\mc{T},\mc{T}_k\subseteq \mc{N}$, and $\mc{V}^c=\mc{N}-\mc{V}$, and $\Upsilon(\mc{V})$ and $\Upsilon_k(\mc{V})$ are defined as follows:
\begin{IEEEeqnarray}{rCl}
 \Upsilon(\mc{V})& \triangleq &\big\{\mc{T}\subseteq\mc{V}: \,Q_{\mc{T}}(\mc{S})\geq 0\,\,\forall \, \mc{S}\subseteq\mc{T} \big\}\,,\\
\Upsilon_k(\mc{V})& \triangleq &\big\{\mc{T}\subseteq\mc{V}:  \,Q^{(k)}_{\mc{T}}(\mc{S})\geq 0\, \,\forall \,\mc{S}\subseteq\mc{T}\big\}\,,
\label{LMNNCcondition}
\end{IEEEeqnarray}
where $Q_{\mc{T}}(\mc{S})$ and  $Q^{(k)}_{\mc{T}}(\mc{S})$ are used to denote: 
\begin{IEEEeqnarray}{rCl}
Q_{\mc{T}}(\mc{S})&\triangleq &
I(X_{\mc{S}};\hat{Y}_{\mc{S}^c}Y|V_{\leq T}XX_{\mc{V}^c}X_{\mc{S}^c}Q)\nonumber\\
&&-I(\hat{Y}_{\mc{S}};{Y}_{\mc{S}}|V_{\leq T}XX_{\mc{V}^c}X_{\mc{T}}\hat{Y}_{\mc{S}^c}YQ)\ ,
\\
Q^{(k)}_{\mc{T}}(\mc{S})& \triangleq &
I(X_{\mc{S}};Y_k|V_{\leq T}X_{\mc{V}^c}X_{\mc{S}^c}Q)\nonumber\\
&&-I(\hat{Y}_{\mc{S}};Y_{\mc{S}}|V_{\leq T}XX_{\mc{V}^c}X_{\mc{T}_k}\hat{Y}_{\mc{S}^c}Y_kQ)\ .
\end{IEEEeqnarray}
The set of all admissible input distributions $\mc{P}$ is defined by
\begin{IEEEeqnarray}{lCl}
\mathcal{P} &\triangleq& \displaystyle\Big\{P_{QV_1\dots V_TXX_{\mc{N}}Y_{\mc{N}}\hat{Y}_{\mc{V}}Y}= P_{Q}\prod_{t=1}^T P_{V_t|V^{t-1}Q}P_{X_t|V_tQ}\nonumber\\
&&\mathbb{P}_{YY_{\mc{N}}|XX_{\mc{N}}Q}
\prod_{j\in \mc{V}}P_{X_{j}|Q}P_{\hat{Y}_j|X_jY_j}\Big\}\,.  
\label{eq:LMNNC-APD}
\end{IEEEeqnarray}
\label{thm:LMNNC}
\end{theorem}
\begin{IEEEproof}
The proof of this theorem is provided in Appendix~\ref{proof:LMNNC}.
\end{IEEEproof}

We first remark that, by comparing the decoding condition~\eqref{relay-rateL} with~\eqref{DFrelaycondition}, the contribution of having different layers brings the term $(V_{>t},X_{\mc{L}_{>t}})$ in the mutual information, which corresponds to the help of higher layers from DF relays, i.e., the  $t$-th  layers shared at the source. In other words, the source superimposes the fresh information over the layers $(V_1,\dots,V_T)$. Moreover, the rate region presented in Theorem~\ref{thm:LMNNC}  performs at least as good as partially cooperative MNNC while by exploiting the help of CF relays it improves upon the existent results in~\cite[Theorem 2]{Hou2013}. Furthermore, this multi-hopping scheme achieves the capacity of some networks, e.g., line networks where relays can be ordered in a way that the observation of lower layer nodes is a {\it physically degraded} version of that of higher nodes. 

In networks with random parameters, e.g., wireless networks, it is hard to assume a fixed hierarchy between relays for all channel draws and thus the  degradedness assumption does not usually hold. Even the optimal number of hops $T$ depends on the specific channels realizations. Hence the source cannot adaptively change the number of hops $T$ and  set a priori coding based on a hierarchy. Nevertheless, this problem can be partially addressed through the adaptive use of $(V_1,\dots,V_T)$ where after the source transmission, DF relays  can choose a set of layers by looking at their channels and superimpose the information over the corresponding layers, generating a conditional codebook. But the number of hops must be selected in advance.

\section{Capacity of Cooperative Unicast AWGN  Networks within a Constant Gap}\label{SectionIII-0}

In this section, we study the characterization of the capacity of  cooperative unicast additive Gaussian noise (AWGN) networks within a constant gap with respect to the cut-set bound. In particular, we show that MNNC under certain conditions  --provided that DF relays are chosen properly-- can achieve a tighter ``constant gap" than the standard  NNC.

\subsection{Single-Relay AWGN  Channel}\label{SectionIII-1}
We first review the constant gap of DF rate for the single AWGN relay channel while CF constant gap follows along the same lines as shown in~\cite{Lim2011}.  Consider the AWGN relay channel defined by the channel outputs:
\begin{IEEEeqnarray}{lCl}
Y& =& g_3\,X+ g_2 \,X_1+\mathpzc{V}_1\ , \\
Y_1& = & g_1 \,X+{\mathpzc{V}}_2\ ,
\label{eq:1}
\end{IEEEeqnarray}
where the inputs are constrained to satisfy the average power $\esp[X^2]\leq P$, $\esp[X_1^2]\leq P$, and the Gaussian noises ${\mathpzc{V}}_1$ and ${\mathpzc{V}}_2$ are zero-mean of equal variance $N$; the channel gains  $(g_1,g_2,g_3)$ are assumed to take arbitrary real values. Lower bounds on the capacity of this channel are well-known from literature~\cite{Cover1979}. The lower bound given by DF rate can be written as follows 
\begin{IEEEeqnarray}{rcl}
R_{\textrm{DF}}  \triangleq &&\max_{\beta\in[0,1]}\min    \left\{\mathcal{C}\left( \frac{g_1^2 \beta P}{N}\right) \ , \right. \nonumber\\
&& \left. \,\mathcal{C}\left(\frac{g_3^2 P+g_2^2 P+2\sqrt{\overline{\beta}g_2^2g_3^2}P}{N}\right)\right\}\ ,
\label{eq:singleRC-DF}
\end{IEEEeqnarray}
and the cut-set bound (CB) reads as 
\begin{align}
C_{\textrm{CB}}\triangleq & \max_{\beta\in[0,1]}\min\left\{ \mathcal{C}\left(\frac{ g_1^2 \beta P+g_3^2 \beta P}{N}\right)\, ,\right.\nonumber\\
&\left.\,\mathcal{C}\left(\frac{g_3^2 P+g_2^2 P+2\sqrt{\overline{\beta}g_2^2 g_3^2}P}{N}\right)
\right\}\,,
\label{eq:singleRC-CB}
\end{align}
where  $\beta$ denotes the correlation coefficient. Observe that the second term in \eqref{eq:singleRC-DF}  appears unchnaged in \eqref{eq:singleRC-CB}, and let us assume that $\beta^\star$ is the value maximizing the CB in~\eqref{eq:singleRC-CB} that is also chosen to evaluate the achievable rate in \eqref{eq:singleRC-DF}.  Hence, only the difference between the first two terms affects the gap that is bounded as follows:
\begin{IEEEeqnarray}{rCl}
\Delta(C_{\textrm{CB}},R_{\textrm{DF}}) &\triangleq&  C_{\textrm{CB}} - R_{\textrm{DF}} \nonumber\\
& \leq & \mathcal{C}\left(\frac{g_1^2 \beta^\star P+g_3^2 \beta^\star P}{N}\right)-\mathcal{C}\left(\frac{g_1^2 \beta^\star P}{N}\right)\nonumber\\
&=& \frac{1}{2}\log\left(\frac{N+\beta^\star g_1^2 P+\beta^\star g_3^2 P}{N+\beta^\star g_1^2 P}\right)\nonumber\\
&=& \mathcal{C}\left(\frac{g_3^2}{g_1^2}\frac{\beta P}{\frac{N}{g_1^2}+\beta P}\right) \leq\mathcal{C}\left(\frac{g_3^2}{g_1^2}\right)\ .
\end{IEEEeqnarray}
From our previous analysis we remark that --unlike the conventional NNC-- the gap for DF rate cannot be made independent of the channel gains. For instance, if  the channel gain source-to-relay is enough strong with respect to that of source-to-destination, the capacity gap can be made arbitrarily small. Furthermore, we may expect that in general, as will be the case later, the performances of DF based schemes are heavily related to channel conditions and therefore cannot be evaluated independently. Also, it can be seen that when the quality of source-to-destination channel is better than the quality of source-to-relay channel, i.e., $g_3>g_1$, then direct transmission and thus CF scheme perform better than DF scheme. 

\subsection{Cooperative Unicast AWGN Networks} \label{SectionIII-2}

Consider a cooperative unicast AWGN network composed of $N$ relay nodes, a single source and single destination node, which  yields  in total to $N+2$ nodes. The relays are indexed as before with index belonging to the set $\mc{N}\triangleq \{1,\dots  ,N\}$ but for simplicity, we will also associate the source with index $0$ and the destination with index $N+1$, i.e., $X_0 \triangleq X$ and $Y_{N+1} \triangleq Y$. Thus, there is a bijection from the set of nodes to the set $\{0,1,\dots  ,N,N+1\}$. The transmitters' set is denoted by $\mc{M} \triangleq  \{0,1,\dots  ,N\}$ and the receivers' set is denoted by $\mc{D}\triangleq \{1,\dots  ,N,N+1\}$. The channel gain from node $i$ to node $j$ is denoted by $\{g_{ij}\}$, and $\mathpzc{V}_{j}$ denotes the noise at node $j$, which is assumed to follow a Gaussian distribution of zero-mean and unit variance. The channel outputs at the different nodes  are given by
\begin{equation*}
\left\{
 \begin{array}{lcl}
 {Y}_{\mc{D}}&=&G(\mc{D},\mc{T}){X}_{\mc{M}}+{\mathpzc{V}}_{\mc{D}}\,, \\
 \hat{{Y}}_{\mc{N}}&= &{Y}_{\mc{N}}+\hat{\mathpzc{V}}_{\mc{N}}\,,
\end{array}
\right.
\end{equation*}
where $ {Y}_{\mc{D}} = [Y_1 \, Y_2 \dots Y_N \, Y]^T$, $ {Y}_{\mc{N}} = [Y_1 \, Y_2 \dots Y_N]^T$, $ {X}_{\mc{M}} = [X_0 \, X_1 \dots X_N]^T$ and $  \mathpzc{V}_{\mc{D}} = [ \mathpzc{V}_1 \,  \mathpzc{V}_2 \dots  \mathpzc{V}_{N+1}]^T$ and $G(\mc{D},\mc{T})$ denotes the channel matrix with the corresponding channel gains, where we use the definition $g_{ii}\triangleq 0$ for all $i\in\mc{D}$;  and $\hat{\mathpzc{V}}_{\mc{N}}$ denotes the compression noise vector that is chosen to follow the same statistic as the channel noise, i.e., Gaussian distribution of zero-mean and unit variance. All through this section, the notation $G(\mc{S}_1,\mc{S}_2)$ is used to indicate the set of channel gains $\big\{g_{ij}\,|\, i\in\mc{S}_1,j\in\mc{S}_2\big\}$. We simply use $G$  when the respective sub-matrices can be understood implicitly. All channel inputs are constrained to satisfy average power $\esp[X^2_i]\leq P$, for all $i\in\mc{M}$. The covariance matrix of any subset $X_{\mc{S}}$ of channel inputs is denoted by ${\
Sigma}{(\mc{S})}=\left[P\,\rho_{ij}\right]$ for all $i,j\in\mc{S}$ with corresponding correlation coefficients $\rho_{ij}$  between the input 
components $(X_i,X_j)$. Similarly, we have ${\Sigma}{(\mc{S}_1,\mc{S}_2)}=\left[P\,\rho_{ij}\right]$ for all $i\in\mc{S}_1$ and $j\in\mc{S}_2$. Also $\mathbf{I}$ denotes the identity matrix. 

We first recall the capacity within a constant gap which has been derived in~\cite{Lim2011} based on NNC scheme.\vspace{1mm}

\begin{proposition}[Constant Gap of NNC~\cite{Lim2011}]
A constant gap between NNC rate and cut-set bound for the AWGN network with $N$ relays is given by
\begin{IEEEeqnarray}{rCl}
\Delta^*(C_{\textrm{CB}},R_{\textrm{NNC}})  &\triangleq & 0.63({N+2})\, .
\label{CGAPII-NNC}
\end{IEEEeqnarray}
\end{proposition} 
We proceed to evaluate the achievable rate given in Theorem~\ref{thm:MNNC} from which we shall derive a novel constant gap to the capacity. Let us assume that $\mc{T}=\mc{N}$, which implies that the destination decodes the compression indices of all relays. Channel inputs are chosen to be Gaussian random variables of zero-mean and unit variance satisfying the corresponding average power constraints. The set $\mc{V}^c$ denotes the index set of all relays using DF scheme and $\mc{V}$ those using CF scheme. Based on these setups, we need to evaluate:
\begin{align}
 R_{\textrm{MNNC}} \triangleq 
\sup\limits_{P\in \mathcal{P}} \,\min&\left(\min_{\mc{V}^c\subseteq\mc{S}\subseteq\mc{N}}\, R_{\mc{N}}(\mc{S})\,,\, \right.\nonumber\\
&\left.\min_{k\in\mc{V}^c}\,\max_{\mc{T}_k\in\Upsilon_k(\mc{N})}\,\min_{\mc{S}\subseteq\mc{T}_k} \, R^{(k)}_{\mc{T}_k}(\mc{S})\right)
\label{eq:MNNC-Evaluation}
\end{align}
where
\begin{IEEEeqnarray}{rcl}
R_{\mc{N}}(\mc{S})&\triangleq &I(XX_{\mc{S}};\hat{Y}_{\mc{S}^c}Y|X_{\mc{S}^c})\nonumber\\
&&-I(\hat{Y}_{\mc{S}};{Y}_{\mc{S}}|XX_{\mc{N}}\hat{Y}_{\mc{S}^c}Y)\,, \\
R^{(k)}_{\mc{T}_k}(\mc{S})&\triangleq &
I(X;\hat{Y}_{\mc{T}_k}Y_k|VX_kX_{\mc{T}_k}Q)\nonumber\\
 \IEEEeqnarraymulticol{3}{l}{+I(X_{\mc{S}};Y_k|VX_kX_{\mc{S}^c}) -I(\hat{Y}_{\mc{S}};Y_{\mc{S}}|VX_kX_{\mc{T}_k}\hat{Y}_{\mc{S}^c}Y_k)\,. }
\end{IEEEeqnarray}
In order to evaluate expression \eqref{eq:MNNC-Evaluation} and thus compute the gap from capacity based on MNNC, we first need to evaluate the cut-set bound in an more convenient manner. 

\begin{figure*} [!b]
\hrule
\begin{IEEEeqnarray}{lCl}
  C_{\textrm{CB}}&\triangleq& 
\max\limits_{\Sigma{(\cdot)}} \,\min\limits_{\mc{V}^c \subseteq \mc{S}\subseteq\mc{N}} \,\frac{1}{2}
\log\left| \mathbf{I}{(\mc{S}^{c}\cup\{N+1\})}+\frac{1}{2}G\left[
\begin{array} {cc}
\Sigma{( \{0\} \cup \mc{V}^c ) } & 0 \\
0 &  P \mathbf{I}{( {\mc{S}}\cap{\mc{V}})}
\end{array}
\right]G^T
\right|\nonumber\\
 \IEEEeqnarraymulticol{3}{r}{+ \frac{1+\min\{|\mc{S}^c|,|\mc{S}|\}}{2}\log\left(4\,\max(1,|{\mc{S}}\cap{\mc{V}}|)\right)\,,}
\label{CBbndfinal}
\end{IEEEeqnarray} 
\end{figure*}
\vspace{1mm}
\begin{lemma}[Cut-set bound]
The capacity of the unicast cooperative AWGN network is upper bounded by equation \eqref{CBbndfinal}, for an arbitrary set of nodes $\mc{V}\subseteq \mc{N}$, where the maximum is taken over all covariance matrices $\Sigma{(\cdot)}$ satisfying the corresponding inputs constraints. Indeed, the set  $\mc{V}^c\triangleq \mc{N} - \mc{V}$ can be seen as the set of relays already having or decoding the source message. 
\end{lemma}

\begin{IEEEproof}
Let $A$ and $B$ be two positive semidefinite matrices such that:
\begin{equation}
 A=\left[
\begin{array} {cc}
A_{11} & A_{12} \\
A_{21} & A_{22}
\end{array}
\right]\succeq 0\,\, \text{  and  }\,\,
B=\left[
\begin{array} {cc}
A_{11} & 0 \\
0 & A_{22}
\end{array}
\right]\succeq 0\ ,  \,\,\, 
\label{eq:identity}
\end{equation}
{ then  $ 2B \succeq A$ .} It is enough to check that the matrix $A^{(-)}\triangleq \left[
\begin{array} {cc}
A_{11} & -A_{12} \\
-A_{21} & A_{22}
\end{array}
\right] $ 
is positive semidefinite and hence $A^{(-)}+A=2B$. On the other hand, the cut-set bound is as follows:

\begin{equation}
C_{\textrm{CB}} \triangleq \max\limits_{P\in \mathcal{P}} \,\min\limits_{\mc{S}\subseteq\mc{N}}\, I(XX_{\mc{S}};{Y}_{\mc{S}^c}Y|X_{\mc{S}^c})\,.
\label{eq:cut-set}
\end{equation}

For convenience, we define the sets $\mc{S}^{c*}\triangleq \mc{S}^{c}\cup\{N+1\}$, $\mc{V}^{c*}\triangleq \mc{V}^{c}\cup\{0\}$ and $ {\mc{S}}_{\text{CF}}\triangleq {\mc{S}}\cap{\mc{V}}$. The following inequalities hold true:
\begin{align}
&I(XX_{\mc{S}}; {Y}_{\mc{S}^c}Y|X_{\mc{S}^c})\nonumber \\
&\leq h( {Y}_{\mc{S}^c} {Y})-h({Y}_{\mc{S}^c}{Y}|XX_{\mc{N}})\\
&=\frac{1}{2} \log\left|   \mathbf{I}(\mc{S}^{c*})   +G\left[
\begin{array} {cc}
\Sigma{(\mc{V}^{c*})} & \Sigma{(\mc{V}^{c*},\mc{S}_{\text{CF}})} \\
 \Sigma{(\mc{S}_{\text{CF}},\mc{V}^{c*})} & \Sigma{(\mc{S}_{\text{CF}})}
\end{array}
\right]G^T
\right|\nonumber \\
&\leq \frac{1}{2}	 \log\left| 2\mathbf{I}(\mc{S}^{c*}) +2G\left[
\begin{array} {cc}
\Sigma{(\mc{V}^{c*})} & 0 \\
0 &  \Sigma{(\mc{S}_{\text{CF}})}
\end{array}
\right]G^T\right| \label{eq:gap-point(a)}\\
&\leq \frac{1}{2} \log\left|  2\mathbf{I}(\mc{S}^{c*})+2G\left[
\begin{array} {cc}
\Sigma{(\mc{V}^{c*})} & 0 \\
0 &  \Tr(\mc{S}_{\text{CF}})\mathbf{I}(\mc{S}_{\text{CF}})
\end{array}
\right]G^T
\right| \label{eq:gap-point(b)}\\
&{\leq}
\frac{1}{2}
\log\Big| 
2\mathbf{I}(\mc{S}^{c*})+  \nonumber\\
&\left. 2\max(1,|\mc{S}_{\text{CF}}|)G\left[
\begin{array} {cc}
\Sigma{(\mc{V}^{c*})} & 0 \\
0 &  P\mathbf{I}(\mc{S}_{\text{CF}})
\end{array}
\right]G^T
\right|\label{CBform1}\\
&\leq
\frac{1}{2}
\log\left| 
\mathbf{I}(\mc{S}^{c*})+\frac{1}{2}G\left[
\begin{array} {cc}
\Sigma{(\mc{V}^{c*})} & 0 \\
0 &  P\mathbf{I}(\mc{S}_{\text{CF}})
\end{array}
\right]G^T
\right|\nonumber\\
&+ \frac{|\mc{S}^c|+1}{2}\log\big(4\max(1,|\mc{S}_{\text{CF}}|)\big)
\label{eq:gap-point(c)}
\end{align}
where \eqref{eq:gap-point(a)} follows from the identity \eqref{eq:identity} and \eqref{eq:gap-point(b)}  follows by noting that $\Tr(\mc{S}_{\text{CF}})\mathbf{I}(\mc{S}_{\text{CF}}) \succeq \Sigma{(\mc{S}_{\text{CF}})}$. For $|\mc{S}_{\text{CF}}|\neq 0$, \eqref{CBform1} and~\eqref{eq:gap-point(c)} follow from 
$\Tr(\mc{S}_{\text{CF}})=P|\mc{S}_{\text{CF}}|$ and basic matrix operations. For $|\mc{S}_{\text{CF}}|=0$, one can bound~\eqref{eq:gap-point(b)} directly by $\frac{|\mc{S}^c|+1}{2}\log(4)$. By rewriting expression \eqref{eq:gap-point(c)}, it is not difficult to check that \eqref{eq:gap-point(c)} implies \eqref{CBbndfinal} which concludes the proof.
\end{IEEEproof}

\begin{proposition}[Constant Gap of MNNC]\label{prop-constant-gap}
Provided source-to-relays channels allow decoding at all relay nodes in $\mc{V}^c\triangleq  \mc{N} - \mc{V}$ for some set $\mc{V}\subseteq \mc{N}$, capacity can be stated within a constant gap from MNNC rate satisfying
\begin{IEEEeqnarray}{lCl}
\Delta(C_{\textrm{CB}},R_{\textrm{MNNC}}) &\triangleq &\max_{\mc{V}^c \subseteq \mc{S} \subseteq \mc{N}} \left[ \frac{|\mc{S}|}{2} \right. \nonumber\\
 \IEEEeqnarraymulticol{3}{r}{\left. +\frac{1+\min\{|\mc{S}|, |\mc{S}^c| \}}{2} \log \left(4\max(1,|\mc{V}| - |\mc{S}^c|) \right)\right]\ .}\label{eq:final-gap-MNNC}
\end{IEEEeqnarray}

Furthermore, if all source-to-relay channels are \emph{good} enough to select DF then the constant gap verifies:
\begin{IEEEeqnarray}{lCl}
\Delta(C_{\textrm{CB}},R_{\textrm{MNNC}}) & \leq & 0.5N+0.7 <   \Delta(C_{\textrm{CB}},R_{\textrm{NNC}})\,,
\end{IEEEeqnarray}
which leads to a strictly tighter gap than that of NNC scheme~\cite{Lim2011}.
\end{proposition}

It should be emphasized that the gains in terms of ``constant gap''  provided by MNNC  with respect to NNC scheme are obtained by taking  $\mc{T}_k$ as empty sets. However, the original rate $R^{(k)}_{\mc{T}_k}$ in Theorem~\ref{thm:MNNC} can be maximized over all general (non necessarily empty) sets $\mc{T}_k$ which may improve  the final rate.  Although we have assumed that the gap incurred by restricting this maximization is not significant (at least in terms of the notion of constant gap), the main outcome of our analysis is that the constant gap can be improved,  provided the relays using DF scheme are adequately chosen.

\begin{IEEEproof}
Consider the first term of MNNC rate given by expression~\eqref{eq:MNNC-Evaluation}. This can be lower bounded, for any set $\mc{S}\subseteq \mc{N}$, as follows:
\begin{IEEEeqnarray}{lCl}
R_{\mc{N}}(\mc{S}) &=& I(XX_{\mc{S}};\hat{Y}_{\mc{S}^c}Y|X_{\mc{S}^c})-I(\hat{Y}_{\mc{S}};{Y}_{\mc{S}}|XX_{\mc{T}}\hat{Y}_{\mc{S}^c}Y)\nonumber\\
&\geq &I(XX_{\mc{S}};\hat{Y}_{\mc{S}^c}Y|X_{\mc{S}^c})- I(\hat{Y}_{\mc{S}};{Y}_{\mc{S}}|XX_{\mc{N}})\nonumber\\
&= &I(XX_{\mc{S}};\hat{Y}_{\mc{S}^c}Y|X_{\mc{S}^c})- \frac{|\mc{S}|}{2}\,.
\end{IEEEeqnarray} 
By convenience, we select the sets $\mc{S}^{c*}\triangleq \mc{S}^{c}\cup\{N+1\}$, $\mc{V}^{c*}\triangleq \mc{V}^{c}\cup\{0\}$ and $ {\mc{S}}_{\text{CF}}\triangleq {\mc{S}}\cap{\mc{V}}$. Since all channel inputs among the nodes in $\mc{V}^c$ are not necessarily independent, we have that
\begin{IEEEeqnarray}{lCl}
&&I(XX_{\mc{S}};\hat{Y}_{\mc{S}^c}Y|X_{\mc{S}^c}) \geq I(XX_{\mc{S}};\hat{Y}_{\mc{S}^c}\hat{Y}|X_{\mc{S}^c}) \nonumber\\
&&=h(\hat{Y}_{\mc{S}^c}\hat{Y}|X_{\mc{S}^c})-h(\hat{Y}_{\mc{S}^c}\hat{Y}|XX_{\mc{N}})\nonumber\\
&&=\frac{1}{2}
\log\left| \mathbf{I}(\mc{S}^{c*})+\frac{1}{2}G\left[
\begin{array} {cc}
\Sigma{(\mc{V}^{c*})} & 0 \\
0 & P\mathbf{I}({\mc{S}}_{\text{CF}})
\end{array}
\right]G^T
\right|\,,\nonumber
\end{IEEEeqnarray}
where the covariance matrix $\Sigma{(\mc{V}^{c*})}$ is the one that maximizes the cut-set bound in \eqref{CBbndfinal}. Hence, the maximum rate $R_{\mc{N}}(\mc{S})$ is lower bounded by 
\begin{align}
R_{\mc{N}}(\mc{S}) &\geq 
\frac{1}{2}
\log\left| \mathbf{I}(\mc{S}^{c*})+\frac{1}{2}G\left[
\begin{array} {cc}
\Sigma{(\mc{V}^{c*})} & 0 \\
0 & P\mathbf{I}({\mc{S}}_{\text{CF}})
\end{array}
\right]G^T\right|\nonumber\\
&-\frac{|\mc{S}|}{2}\,.\label{eq:gapMNNC-final}
\end{align}
Finally, from \eqref{eq:gapMNNC-final} the gap between MNNC rate \eqref{eq:MNNC-Evaluation} and the cut-set bound \eqref{CBbndfinal} is bounded by
\begin{IEEEeqnarray}{rCl}
\Delta_1&\triangleq& \max_{\mc{V}^c\subseteq\mc{S}\subseteq\mc{N}}\left[ \frac{|\mc{S}|}{2}\right. \nonumber\\
&&\left. + \frac{1+\min\{|\mc{S}^c|,|\mc{S}|\}}{2} \log\left(4\max(1,|{\mc{S}}\cap{\mc{V}}|)\right)\right]\ .\,\,\,\,
\label{CGAP1-MNNC}
\end{IEEEeqnarray}

\begin{figure*} [b]
\hrule
\begin{IEEEeqnarray}{ll}
\Delta_2 &\triangleq \max_{k\in\mc{V}^c}\frac{1}{2}
\log\left(
\frac{
\left| I(\{2\})+G\Sigma{(\mc{M}-\{k\})G^T}\right|\left[\sum\limits_{i\in\mc{V}} g_{ik}^2 P+\sum\limits_{i\in\mc{V}^c} g_{ik}^2(1-\rho_{ik}^2)P+1\right]}
{
g_{0k}^2(1-\rho_{0k}^2)P+\sum\limits_{i\in\mc{V}} g_{ik}^2P+\sum\limits_{i\in\mc{V}^c} g_{ik}^2(1-\rho_{ik}^2)P+1
}
\right)\,.\,\,\,\,
\label{CGAP2-MNNC}
\end{IEEEeqnarray}
\end{figure*}

The remanning part of the MNNC rate, which is related to all relays using DF scheme, can be bounded as follows. We remark that the channel output $Y$ is absent in the rate expression  while it is present in the cut-set bound. Hence, any bound on the gap between the achievable rate and the cut-set bound --no matter how tight it is-- will depend on the channel gains between the output $Y$ and all inputs. For sake of simplicity, we shall assume that each DF relay is decoding the source message without looking at the compression indices of others relays, which yields  $\mc{T}_k=\emptyset$. Then, the rate $R^{(k)}_{\mc{T}_k}$ is simply reduced to $R^{(k)}_{\textrm{DF}}=I(X;Y_k|VX_{k})$ and it can be bounded using the same steps as before. The outputs $Y_k$ can be re-written as
\begin{IEEEeqnarray*}{rCl}
Y_k&=&g_{0k}X+G(\{k\},\mc{N}){X}_{\mc{N}}+ \mathpzc{V}_k \\
&=& g_{0k}X+\sum_{i\in\mc{V}} g_{ik}X_i+\sum_{i\in\mc{V}^c} g_{ik}X_i+ \mathpzc{V}_k\,,
\end{IEEEeqnarray*}
where the relays in the set $\mc{V}$ use CF scheme. In order to evaluate the conditional entropy $h(Y_k|VX_{k})$, we can use the standard linear decomposition $X_i=\tilde{X}_i+\alpha_{i}V$, based on independent descriptions $\tilde{X}_i$ and $V$ that satisfy the power constraints. It is easy to check that
\begin{IEEEeqnarray*}{lll}
h(Y_k|VX_{k})&=&  \frac{1}{2}\log(2\pi e) \Big[g_{0k}^2(1-\rho_{0k}^2)P \nonumber\\
&+& \left. \sum_{i\in\mc{V}} g_{ik}^2 P+\sum_{i\in\mc{V}^c} g_{ik}^2(1-\rho_{ik}^2)P+1 \right]\ ,\,\,\,\,
\end{IEEEeqnarray*}
and similarly
\begin{IEEEeqnarray*}{lll}
&&h(Y_k|VXX_{k})= \nonumber\\
&& \frac{1}{2}\log(2\pi e)\left[  \sum_{i\in\mc{V}} g_{ik}^2P+\sum_{i\in\mc{V}^c} g_{ik}^2(1-\rho_{ik}^2)P+1
\right]\ . 
\end{IEEEeqnarray*}
From which the mutual information is obtained as
\begin{IEEEeqnarray*}{lll}
&&I(X;Y_k|VX_{k})=\nonumber\\
&&\frac{1}{2}\log\left(1+\frac{g_{0k}^2(1-\rho_{0k}^2)P}
{
\sum\limits_{i\in\mc{V}} g_{ik}^2P+\sum\limits_{i\in\mc{V}^c} g_{ik}^2(1-\rho_{ik}^2)P+1
}
\right)\ .
\end{IEEEeqnarray*}
By evaluating the the cut-set bound, we get:
\begin{IEEEeqnarray}{rCl}
I(XX_{\mc{N}-\{k\}}; {Y}_{k}Y|X_k) &=&\nonumber\\
 \IEEEeqnarraymulticol{3}{l}{
\frac{1}{2}
\log\left| I(\{2\})+G\Sigma{(\mc{M}-\{k\})}G^T
\right|}\ ,
\end{IEEEeqnarray}
where
\begin{align}
G&=G(\{k,N+1\},\mc{M}-\{k\})=\nonumber\\
&\left[
\begin{array} {cccc}
g_{0k} & g_{1k} & \dots & g_{Nk} \\
g_{0(N+1)} & g_{1(N+1)} & \dots & g_{N(N+1)}
\end{array}
\right]\,,
\end{align}

the gap between the achievable rate in this case and the cut-set bound can be bounded by \eqref{CGAP2-MNNC}.
As it was expected, this gap does depend on channel gains. From expressions~\eqref{CGAP1-MNNC} and~\eqref{CGAP2-MNNC}, the final gap reads:
\begin{equation}
\Delta(C_{\textrm{CB}},R_{\textrm{MNNC}})\triangleq \max\left( \Delta_1,\Delta_2\right)\ .
\label{eq:totalgap}
\end{equation}
Nevertheless, assuming that all relays using DF  scheme are chosen appropriately, the first term $\Delta_1$ of the gap in \eqref{eq:totalgap}, which is independent of the channel gains is expected to be dominant. For instance, the question that arises here is what is the most appropriate condition to select the set of DF relays. A simple way to deal with this issue is to pick all relay nodes whose channel gains $\{g_{0k}\}$ are enough large compared to those of the other nodes. This selection rule only requires the optimization of the first term in~\eqref{eq:totalgap} which yields the desired gap in~\eqref{eq:final-gap-MNNC}.  
\end{IEEEproof}

Therefore, it is of interest to understand how the gap of MNNC is compared with that of NNC~\cite{Lim2011}. As it was previously discussed, there is no definitive answer for general cases. However, if a considerable  amount of relays are well positioned to perform DF scheme, then the constant gap can be strictly improved. Notice the interest behind~\eqref{eq:final-gap-MNNC} is to emphasize that MNNC can improve the gap to the capacity. The improvement of  this  gap is two fold. Firstly, the gap decreases logarithmically as the number of DF relays increases. Moreover, the expression of constant gap is optimized over all sets  $\mc{S}$ such that $\mc{V}^c \subseteq \mc{S} \subseteq \mc{N}$. This leads to a reduced optimization space which eventually can lead to a better gap. Secondly, all this is conditioned by whether the DF relays affect the gap or not. For instance, if all relays are  good enough then the encoder chooses $\mc{V}^c = \mc{N}$ and thus the optimization set is reduced to a single element $\mc{N}
$, and the constant  gap becomes $0.5N+0.7$ which is strictly better than~\eqref{CGAPII-NNC}. More precisely, the gap can be bounded by $0.5N+C(k)$, where the value of $C(k)$ is independent of $N$ and $k$ is the number of CF relays provided that $N-k$ relays perform DF scheme without degrading the rate. Once again, it should be emphasized that the constant gap results are dependent on the assumption of \textit{good} source-to-relay channel gains.


\section{Cooperative Unicasting in Wireless Networks}\label{SectionIV-0}

In this section, we consider a direct application of MNNC to the problem of cooperative unicasting in wireless networks where a single source wishes to communicate with a destination in presence of multiple relay nodes. A composite unicast network is assumed where the channel parameters are randomly drawn and this draw is assumed to be unknown at the source, fully known at the destination and only partly known at the relay nodes. We exploit MNNC scheme to introduce a novel transmission scheme that enables the relays to select --based on their channel measurements-- the best cooperative strategy. Bounds on the asymptotic average error probability of this class of networks are derived.

\subsection{Composite Relay Channels}\label{SectionIV-1}

\begin{figure} [t]
\centering  
\includegraphics [width=.35 \textwidth] {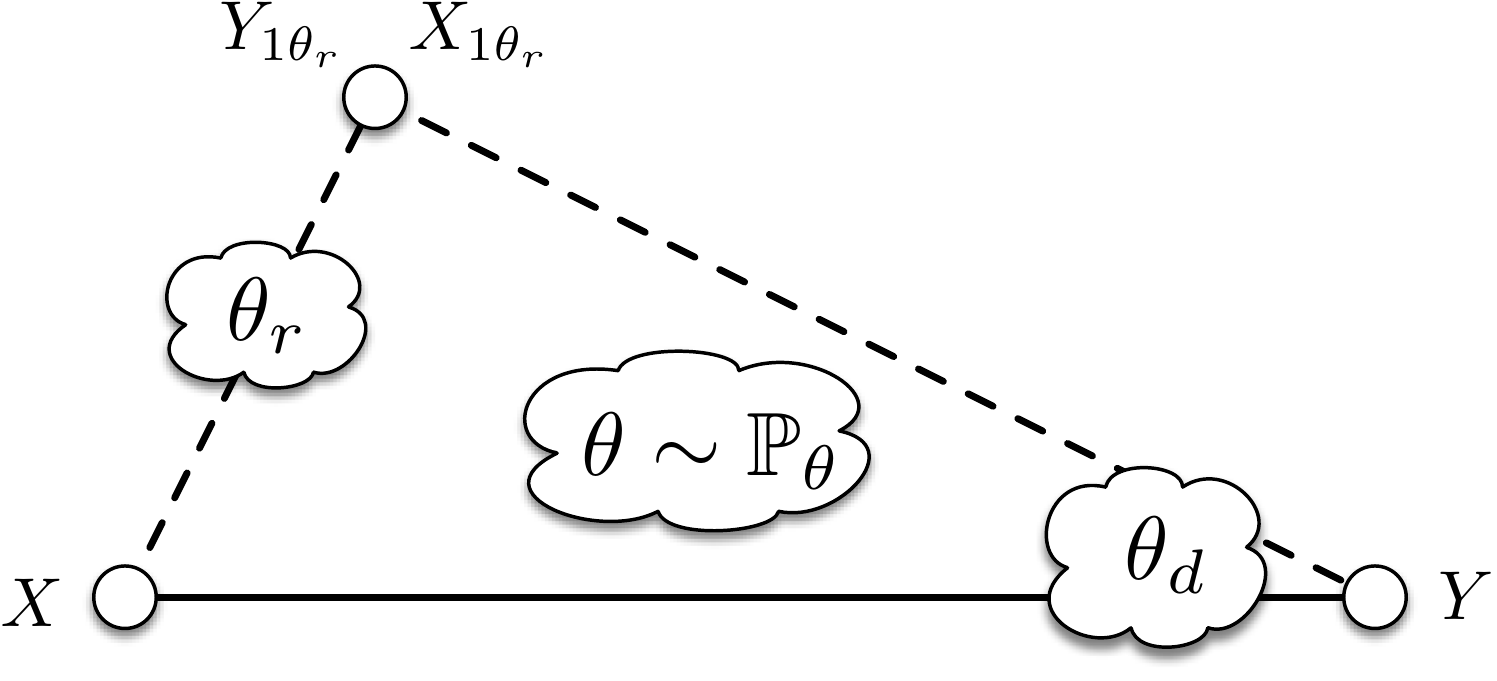}    
\caption{Composite Relay Channel.}
\label{fig:singlerelay}
\end{figure}

Consider the composite relay channel described in Fig.~\ref{fig:singlerelay}, where the probability distribution characterizing the channel is indexed with  parameters $\theta$. This channel can be defined as a set of  memoryless probability distributions:
\begin{IEEEeqnarray*}{lCl}
\mc{W}_\Theta& = &
\left\{P_{Y^{n}Y_{1}^{n}|X^{n} X^{n}_{1};\theta}(\underline{y},\underline{y}_1|\underline{x}, \underline{x}_1;\theta)\,\Big |\, \underline{x}\in \mathcal{X}^n,\right.\nonumber\\
 \IEEEeqnarraymulticol{3}{r}{\left. \underline{x_1}\in \mathcal{X}_1^n, \, \underline{y}_1\in \mathcal{Y}_1^n,\,  \underline{y}\in \mathcal{Y}^n,\, \theta\in\Theta \Big\}^{\infty}_{n=1}\right.}\ .
\end{IEEEeqnarray*}
This family of channels corresponds to the definition of the compound relay channel for which the channel is chosen in an arbitrary manner but remains fixed during the communication. Whereas, the composite relay channel introduces a probability measure $\prob_\uptheta$ on $\Theta$ to handle the channel selection and thus an index $\theta$ is present with probability $\prob_\uptheta(\theta)$ but also remains fix during the communication. The index  $\theta$ represents vectors of parameters $\theta=(\theta_d,\theta_r)$ with $(\theta_d,\theta_r) \in \Theta$, where $\theta_r\in\Theta_r$ denotes all parameters affecting the relay output and $\theta_d\in\Theta_d$ are the remaining parameters involved in the communication. More precisely, the marginal distributions read as:
\begin{IEEEeqnarray*}{ll}
 P_{Y_{1}^{n}|X^{n} X^{n}_{1};\uptheta}&=P_{Y_{1}^{n}|X^{n} X^{n}_{1};\uptheta_r}\ ,\\
P_{Y^{n}|X^{n} X^{n}_{1};\uptheta}&=P_{Y^{n}|X^{n} X^{n}_{1};\uptheta_d}\ .
\end{IEEEeqnarray*}
The specific draw $\theta=(\theta_d,\theta_r)$ is assumed to be unknown at the source and fully known at the destination while the relay only knows $\theta_r $. The notion of capacity-versus-outage shall be used to characterize the performance of this channel\footnote{Notice that non-zero rate cannot always be guaranteed for the compound model which means that the weak capacity of many compound models would be zero.}. 
\vspace{1mm}
\begin{definition}[code and achievability] \label{def-code-singlerelay}
A code-$\mc{C}(n,M_n,$ $\epsilon_{n,\uptheta})$ for the composite relay channel with $(\mc{W}_\Theta,\prob_\uptheta)$ defined as before consists of:
\begin{itemize}
\item An encoder mapping  $\left \{\varphi:\mc{M}_{n} \longmapsto \mathcal{X}^n  \right\}$, 
\item A decoder mapping $\left\{ \phi:\mathcal{Y}^n\times\Theta \longmapsto \mc{M}_{n}\right \}$,
\item A set of relay functions $\left\{ f_{i} :\mathcal{Y}_1^{i-1}\times\Theta_r  \longmapsto \mathcal{X}_{1} \right \}_{i=1}^n$, 
\end{itemize}
for some set of uniformly distributed messages $W\in\mc{M}_{n}=\big\{ 1,\dots, M_{n} \big\}$. For each $\uptheta$, the average error probability is defined as:
$$
\epsilon_{n,\uptheta}=\Pr\left\{\phi(Y^n,{\uptheta}) \neq W\big | \uptheta\right\}\ .
$$
An error probability $0\leq \epsilon<1$  is said to be $r$-achievable, or the rate $r$ is said to be $\epsilon$-achievable, if there exists a code-$\mc{C}(n,M_n,r)$ with rate satisfying
\begin{equation*}
\liminf\limits_{n\rightarrow \infty}\frac{1}{n} \log M_n \geq r 
\label{rate-condition}
\end{equation*} 
and average error probability 
\begin{equation*}
 \limsup\limits_{n\rightarrow \infty}\, \esp_{\uptheta}\left[\Pr\left\{\phi(Y^n,{\uptheta}) \neq W\big | \uptheta\right\}\right]\leq \epsilon\ .
\label{errorprob_def}   
\end{equation*}    
The infimum of all $r$-achievable error probabilities $\bar{\epsilon}(r)$ is defined as 
\begin{equation*}
\bar{\epsilon}(r)\triangleq \inf\left\{0\leq \epsilon<1\,:\, \textrm{$\epsilon$ is $r$-achievable}\right\}\ .
\label{errorprob_def2}   
\end{equation*}    
\end{definition}

\begin{remark}
It is important to remark that the reliability function of the composite relay channel may be defined in different ways. If the expectation of the error probability is chosen as the reliability function \eqref{errorprob_def}, then the definition remains the same as that of averaged channels in \cite{Ahlswede1968,Han2003}. The notion of $\epsilon$-achievability stays the same as the previous definition where the supremum of all $\epsilon$-achievable rates is refereed to as  $\epsilon$-capacity of the averaged channel
\begin{equation*}
C_{\epsilon}\triangleq \sup\left\{r\geq  0 \,:\, \textrm{$r$ is $\epsilon$-achievable}\right\}\ .
\label{epscapacity-def}   
\end{equation*}     
Indeed, composite channels provide more general models since the reliability function unlike averaged channels is not uniquely determined. 
\end{remark}

We aim at characterizing the smallest possible average error probability \eqref{errorprob_def} as a function of the coding rate $r$. In wireless scenarios, the notion of outage probability is extensively used to characterize the average error probability. To properly define this notion, let us assume that the decoder is equipped with an outage identification function~\cite{Effros2010}:
\begin{equation*}
I:\Theta \longmapsto\{0,1\}
\label{eq:funI}
\end{equation*}    
such that  $I(\theta)$ equal to one indicates that the decoder is able to recover  the message, i.e.,
\begin{equation*}
\lim_{n\to\infty}\Pr\left\{\phi(Y^n,{\theta}) \neq W\big | \theta\right\}=0\ ,
\end{equation*}    
otherwise if $I(\theta)$ is zero, the decoder does not decode the message and declares an outage event. The outage probability is then defined by
\begin{equation*}
P_{\text{out}}\triangleq \Pr\{I(\uptheta)=0\}\ .
\end{equation*}    
Hence, for any code with outage probability $P_{\text{out}}$ we know that if $I(\theta)$ is equal to one the error probability tends to zero as $n$ goes to infinity and thus \eqref{errorprob_def}  can be upper bounded by the outage probability, i.e., 
\begin{equation*}
\limsup\limits_{n\rightarrow \infty}\, \esp_{\uptheta}\left[\Pr\left\{\phi(Y^n,{\uptheta}) \neq W\big | \uptheta\right\}\right]\leq P_{\text{out}}\ .
\end{equation*}    
On the other hand, for $I(\theta)=0$ the error probability can be only bounded away from zero. Let us now assume that the decoder is provided with the full error genie aided  function:
\begin{equation*}
J:\Theta \longmapsto\{0,1\} \ ,
\label{eq:funJ}
\end{equation*}  
where $J(\theta)$ equal to zero indicates that the error probability tends to one, i.e., 
\begin{equation*}
\limsup_{n\to\infty}\Pr\left\{\phi(Y^n,{\theta}) \neq W\big | \theta\right\}=1\ ,
\end{equation*}    
and the value one is assigned to indicate that the error probability tends not to one, but neither necessarily to zero. Thus, we have that  the average error probability is bounded away from  the probability that $J(\theta)$ equals zero,  yielding the lower bound:
\begin{equation*}
\liminf\limits_{n\rightarrow \infty}\, \esp_{\uptheta}\left[\Pr\left\{\phi(Y^n,{\uptheta}) \neq W\big | \uptheta\right\}\right] \geq \Pr\{J(\uptheta)=0\}\ .
\end{equation*}    

In a recent work~\cite{Behboodi2011A}, it has been shown that any rate bigger than the cut-set bound will produce an error probability tending to one. This implies that all relay channels for which the cut-set bound is tight satisfy the strong converse property. Furthermore, it also implies that the next genie aided function is a full error identification function:
\begin{equation*}
J(\theta)\triangleq \mathbf{1}[r>C_{\textrm{CB}}(\theta)]\,,
\end{equation*}    
where $C_{\textrm{CB}}(\theta)$ is the cut-set bound indexed by $\theta$,
$$
C_{\textrm{CB}}(\theta) \triangleq \max_{p(x,x_1)} \min\left\{I_\theta(X;YY_{1}|X_{1}),I_\theta(XX_{1};Y)\right\}.
$$ 
By using the previous inequalities, the average error probability $ \bar{\epsilon}(r)$ can be bounded as follows 
\begin{equation*}
\prob_{\uptheta }\{{r}>C_{\textrm{CB}}(\uptheta)\} \leq \bar{\epsilon}(r) \leq P_{\text{out}}(r)
\label{general-bounds}   
\end{equation*}  
and $P_{\text{out}}(r)$ is the outage probability of a given coding strategy (e.g.  DF and CF schemes).  The use of DF scheme yields an outage probability given by
\begin{IEEEeqnarray}{lCl}
P_{\textrm{out}}^{\textrm{DF}}(r) \triangleq 
\min_{p(x,x_1)} \prob_{\uptheta}&& \big[ r>\min\{I_{\uptheta_r}(X;Y_{1}|X_{1}), \nonumber\\
&& I_{\uptheta}(XX_{1};Y)\} \big]\ ,
\label{DFoutage-2}   
\end{IEEEeqnarray} 
where $I_{\uptheta}$ denotes the mutual information for a given $\uptheta$. Notice that since the source is unaware of $\theta=(\theta_r,\theta_d)$, and $p(x,x_1)$ must be known at both source and relay end, then $p(x_1)$ cannot be independently optimized on $\theta_r$ to minimize the outage probability. 

\begin{figure*}[t]
\centering{
\subfigure[Selective Coding Strategy (SCS).]{
             \includegraphics [width=.45 \textwidth] {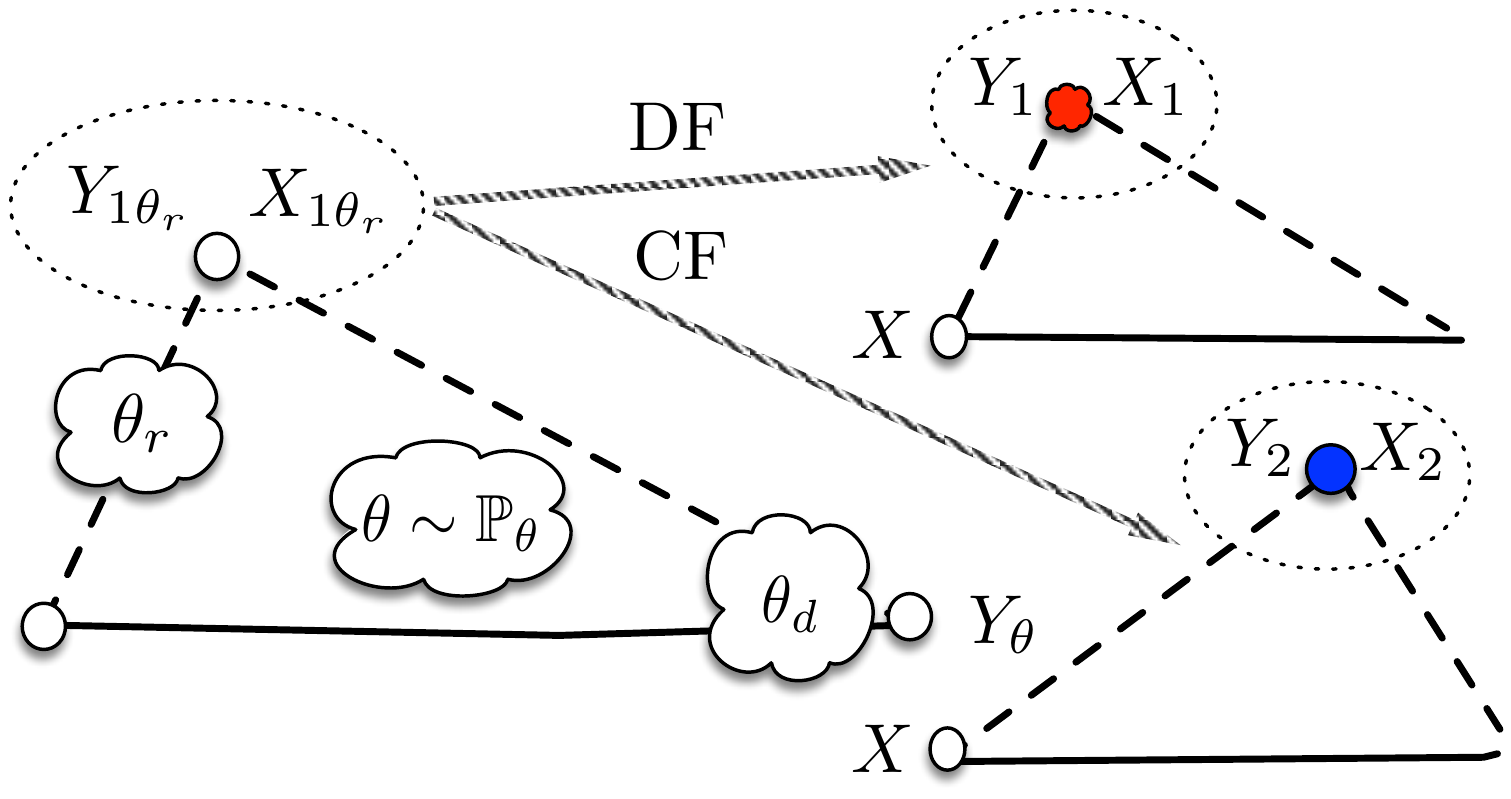} 
         \label{fig:scs-singlerelay}
 }\hspace{.5mm}
\subfigure[Two-relay network.]{  
    \includegraphics [width=.4 \textwidth] {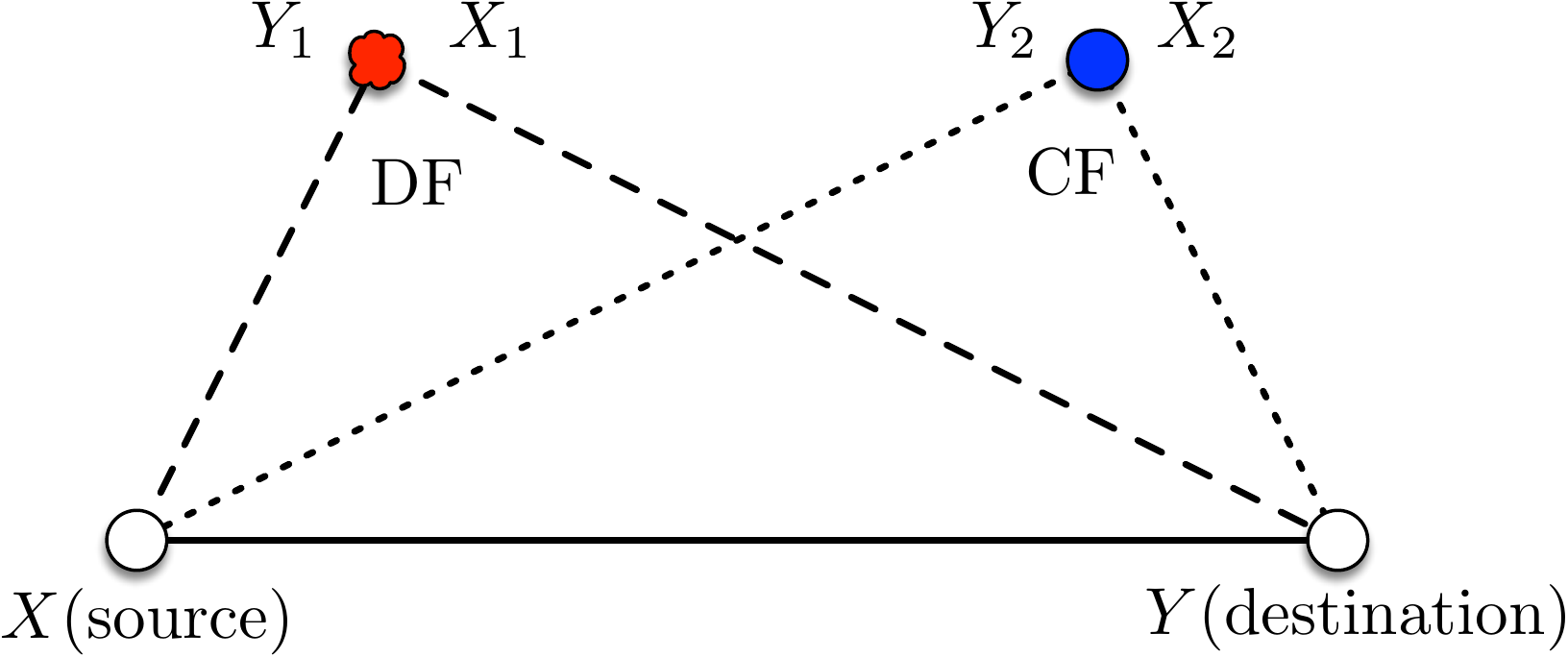}   
    \label{fig:tworelay}
 }
\label{myfigure}
\caption{Selective Coding Strategy (SCS) and Two-relay network.}}
\end{figure*}
Consider now the case of CF, for which the source does not need to know $p(x_1)$, so the relay can choose $p(x_1)$ to minimize the outage probability conditioned on each value $\theta_r$. This requires two steps of optimization, the outage probability of CF scheme~\cite{Lim2011} reads as: 
\begin{IEEEeqnarray}{rCl}
P_{\textrm{out}}^{\textrm{CF}}(r)&\triangleq& \min_{p(x,q)}\,
\prob[E]_{\uptheta_r}\Big[\min_{p(x_1|q)p(\hat{y}_1|x_1,y_1,q)}\nonumber\\
\IEEEeqnarraymulticol{3}{r}{\prob_{\uptheta|\uptheta_r}\big[ r> \min\{ I_{\uptheta}(X;\hat{Y}_1Y|X_{1}Q),I_{\uptheta}(XX_{1};Y|Q)-}\nonumber\\
\IEEEeqnarraymulticol{3}{r}{ I_{\uptheta}(Y_{1};\hat{Y}_1|XX_{1}YQ)\} \big | \uptheta_r\big]
\Big]\,.}
\label{CFoutage}   
\end{IEEEeqnarray}    
Moreover, from \eqref{DFoutage-2}  and \eqref{CFoutage} the selection of the best strategy minimizing the outage probability provides the tightest bound on  the error probability, 
\begin{equation}
P_{\textrm{out}}(r)\triangleq  \min \left \{ P_{\textrm{out}}^{\textrm{DF}}(r), P_{\textrm{out}}^{\textrm{CF}}(r)\right\}\,. 
\label{usual-upperbounds}
\end{equation}

The central question that arises here is whether the achievable error probability~\eqref{usual-upperbounds} can be improved by some kind of smart coding strategy in which the relay selects  instantaneously the best strategy between DF or CF, according to its channel measurement $\theta_r$. To this purpose, the source code should be capable of being used simultaneously with DF and CF schemes, as shown in Fig.~\ref{fig:scs-singlerelay}. Nevertheless, the source is not aware of the channel measurement at the relay and hence it is not able to know which strategy is going to be selected at the relay. This is an example of simultaneous relay channel with two possible situation, DF relay and CF relay. As it is discussed previously in~\cite{Behboodi2013J}, this problem can be studied using an equivalent model. The source must consider the two-relay network where one relay node employs DF scheme while the other one uses CF scheme, as shown in Fig.~\ref{fig:tworelay}. Now the source code should be designed to account  
for  both relays. Also in the equivalent model, the relays cannot collaborate since only one of them is present at once. This model sheds light on the proper code design for composite setting due to the simultaneous presence of relays with heterogeneous cooperative strategy.

The next corollary is a special case of Theorem~\ref{thm:NC-MNNC} for a two-relay network where the DF relay decodes directly the source message without the help of the other relay. \vspace{1mm}

\begin{corollary}[Two-relay network] A lower bound on the capacity of the two-relay network is given by all rates satisfying 
\begin{IEEEeqnarray}{rCl}
R&\leq &\max\limits_{P\in\mathcal{P}}
\min\bigg\{I(X;Y_1\vert X_1Q)\,,\,\max\Big\{ I(XX_1;Y|Q),\nonumber\\
\IEEEeqnarraymulticol{3}{r}{\min\big[I(XX_1;\hat{Y}_2Y\vert X_2Q),I(XX_1X_2;Y|Q)}\nonumber\\
\IEEEeqnarraymulticol{3}{r}{-I(Y_2;\hat{Y}_2\vert YXX_1X_2Q)\big]\Big\}\bigg\}}
\label{CFconditionA}
\end{IEEEeqnarray}
and the set of all admissible input PDs $\mathcal{P}$ is defined as 
\begin{IEEEeqnarray}{rCl}
\mathcal{P} \triangleq  
&&\left \{P_{QX_2X_1XYY_1Y_2\hat{Y}_2}= P_{Q} P_{X_2|Q}P_{XX_1|Q} \right.\nonumber\\
\IEEEeqnarraymulticol{3}{r}{\left.\times\prob_{YY_1Y_2|XX_1X_2}P_{\hat{Y}_2|X_2Y_2Q}\right\}}\ .  
\label{eq:II-2}
\end{IEEEeqnarray}
\label{thm:tworelay}
\end{corollary}
The maximum in \eqref{CFconditionA} determines whether the second relay that uses CF scheme is increasing the rate or would be better to treat its transmission as interference. It is not difficult to check that the second relay increases the rate provided the following condition is satisfied:
\begin{equation}
I(X_2;Y|XX_1Q)\geq I(Y_2;\hat{Y}_2\vert YXX_1X_2Q)\,.
\label{CFcondition}
\end{equation}
The last two terms in \eqref{CFconditionA} represent the condition of successful decoding at the destination while the first term is the condition of successful decoding of $X_1$ at the first relay. By comparing the last two terms with the standard expression of CF rate, it is easy to see that these present similar behavior with the minor difference that the relay codeword has been replaced with $(X,X_1)$. It is also worth mentioning that  by treating the CF relay as noisy, e.g., whenever its link is too noisy, or by using NNC which improves the constraint \eqref{CFcondition}, Corollary~\ref{thm:tworelay} improves over the results in~\cite{Kramer2005}. 

Based on  Corollary~\ref{thm:tworelay} we can state an achievable result for the composite relay channel that is a direct  consequence of Corollary~\ref{thm:tworelay} and some additional subtleties which are addressed in Appendix~\ref{Sec-SCSproof}. First, we emphasize on the fact that the coding strategy used in Corollary~\ref{thm:tworelay} is also well adapted to the composite relay channel. Basically, the relay may dispose of two set of codebooks, namely $X_1$ and $X_2$, and it sends either $X_{1\theta_r}=X_1$ (corresponding to DF scheme) when condition $\theta_r\in \mc{D}_{ \textrm{DF}}$ holds or $X_{1\theta_r}=X_2$ (corresponding to CF scheme) elsewhere. Therefore, since the error probability is made arbitrary small simultaneously for both relays, the source does not need to know the specific relay function implemented. With this coding, the relay can select the coding strategy according to its instantaneous channel measurement $\theta_r$. Secondly, we remark that for the CF relay there may be the 
additional  condition \eqref{CFcondition} for decoding. The destination is assumed to know $\theta$ and consequently is aware if condition \eqref{CFcondition} does hold or not. In the case where it fails, destination will treat the relay input as interference --without perform its decoding-- and then the  condition for unsuccessful decoding simple becomes $\mathbf{1}\left\{r>I_\uptheta(X;Y)\right\}$. We refer to this coding scheme as to ``selective coding strategy'' (SCS). In the next section, we show that it can further improve  the asymptotic error probability. \vspace{1mm}

\begin{proposition}[SCS with partial CSI at relay]\label{corollary1}
The average error probability of the composite relay channel with partial CSI $\theta_r$ at the relay can be upper bounded by 
\begin{align}
&\bar{\epsilon}(r) \leq\min_{p(x,x_1,q)}\inf_{\mc{D}_{\textrm{DF}}\subseteq\Theta_r}
\prob[E]_{\uptheta_r} \Big\{ \prob_{\uptheta|\uptheta_r}\big[r> I_{\textrm{DF}}(\uptheta)\,,\,\uptheta_r \in \mc{D}_{ \textrm{DF}}  \big | \uptheta_r\big] \nonumber\\
&{ + \displaystyle\min_{{p(x_{2}|q)p(\hat{y}_{2}|x_{2},y_{1},q)} }  \prob_{\uptheta|\uptheta_r}\big[r > I_{\textrm{CF}}(\uptheta)\,,\, \uptheta_r \notin \mc{D}_{ \textrm{DF}} \big| \uptheta_r\big] \Big\}}\,,
\label{SCSoutage-1}   
\end{align}
where $(X_{1},X_{2})$ denote the relay inputs corresponding to each strategy selected as follows
\begin{equation*}
X_{1\theta_r}=\left\{\begin{array}{lll}
X_{1}& \, & \textrm{if $\theta_r\in \mc{D}_{ \textrm{DF}}$}\\
X_{2}& \, & \textrm{if $\theta_r\notin \mc{D}_{ \textrm{DF}}$}
\end{array}
\right.
\end{equation*}
and the quantities $I_{\textrm{DF}}, I_{\textrm{CF}}$ are defined by
\begin{IEEEeqnarray*}{rCl}
I_{\textrm{DF}}(\uptheta)&\triangleq& \min\big\{I_{\uptheta_r}(X;Y_{1}|X_{1}Q)\,,\,I_{\uptheta}(XX_{1};Y|Q)\big\}\ ,\\
I_{\textrm{CF}}(\uptheta)&\triangleq &\max\big\{\min\big[ I_{\uptheta}(X;\hat{Y}_{2}Y|X_{2}Q),I_{\uptheta}(XX_{2};Y)\nonumber \\
&&{-I_{\uptheta_r}(Y_{1};\hat{Y}_{2}|YXX_{2}Q)\big] ,I_{\uptheta}(X;Y)\big\} }\ .
\end{IEEEeqnarray*}
\end{proposition}

Consider an index draw $\theta_r$ such that $\mathbf{1}\big\{r>I_{\theta_r}(X;Y_{1}|X_1)\big\}$ $=1$, i.e., the relay is not able to decode the message. Then, DF scheme would lead to an outage event while CF scheme does not necessarily yield to such event and so the best guess of the relay would be to use CF scheme. The question that  arises here is what the proper guess would be if the relay can decode the message. As a matter of fact, if the relay decodes and uses DF scheme, an outage event may still occur if $\mathbf{1}\left\{r>I_{\uptheta}(XX_1;Y)\right\}=1$. However, since $X_{2}$ is independent of $X$ while $X$ is in general dependent on $X_1$, for Gaussian inputs,  we have that  $I_\uptheta(XX_1;Y)\geq I_\uptheta(XX_{2};Y)$. This implies that if an outage event occurs with DF scheme while the relay has the message then the event will happen anyway with CF scheme. Note that the preceding inequality is not true in general, e.g., consider the case of binary RVs where correlation 
does not necessarily increase mutual information. \vspace{1mm}

\begin{remark}[\emph{Optimizing the decision region}]
The optimal decision region when the inputs \eqref{SCSoutage-1} are jointly Gaussian is given by the set 
\begin{IEEEeqnarray}{rCl}
\mc{D}_{\textrm{DF}}^\star\triangleq \big\{ \theta_r\in \Theta_r\, \big |\,  I_{\uptheta_r}(X;Y_{1}|X_{1}Q)> r \big\}\,. 
\label{eq:optimal-DR-CRC}
\end{IEEEeqnarray}
\end{remark}
Although the knowledge of $\theta_r$ at the relay is enough to select the adequate coding strategy, full CSI $(\theta_r,\theta_d)$ further improves the description that the relay sends to the destination and yields the following extension of Proposition~\ref{corollary1}.\vspace{1mm}

\begin{proposition}[SCS with full CSI at relay]\label{corollary2}
The average error probability of the composite relay channel with full CSI $\theta=(\theta_r,\theta_d)$ at the relay can be upper bounded by
\begin{align}
\bar{\epsilon}(r) \leq &\min_{p(x,x_1,q)}\inf_{\mc{D}_{\textrm{DF}}\subseteq\Theta_r}
 \left\{ \prob_{\uptheta}\big[r>I_{\textrm{DF}}(\uptheta) \,, \,\uptheta_r \in \mc{D}_{ \textrm{DF}}\big]\right.\nonumber\\
 &\left.+ \displaystyle  \prob_{\uptheta}\big[r >I_{\textrm{CF}}(\uptheta)\, ,\, \uptheta_r \notin \mc{D}_{ \textrm{DF}}\big]\right\}\,,
 \label{SCSoutage}   
\end{align}
where $(X_{1},X_{2})$ denote the relay inputs corresponding to each strategy selected as follows
\begin{equation}
X_{1\theta_r}=\left\{\begin{array}{lll}
X_{1}& \, & \textrm{if $\theta_r\in \mc{D}_{ \textrm{DF}}$}\\
X_{2}& \, & \textrm{if $\theta_r\notin \mc{D}_{ \textrm{DF}}$}
\end{array}
\right.
\end{equation}
and the quantities $I_{\textrm{DF}}, I_{\textrm{CF}}$ are defined by
\begin{IEEEeqnarray}{rCl}
I_{\textrm{DF}}(\uptheta)&\triangleq& \min\big\{I_{\uptheta_r}(X;Y_{1}|X_{1}Q)\,,\,I_\uptheta(XX_{1};Y|Q)\big\}\ ,\\
I_{\textrm{CF}}(\uptheta)& \triangleq& \max_{{p(x_{2}|q)p(\hat{y}_{2}|x_{2},y_{1},q)} } \min\big\{ I_\uptheta(X;\hat{Y}_{2}Y|X_{2}Q)\ ,\nonumber\\
&&I_\uptheta(XX_{2};Y|Q){ -I_{\uptheta_r}(Y_{1};\hat{Y}_{2}|YXX_{2}Q)\big\}\ . }
\end{IEEEeqnarray}
\label{prop:2}
\end{proposition}
The proof follows along the same lines than that of Proposition~\ref{corollary1}. It is worth mentioning that since full CSI is here available at the relay, the relay input can be optimized over $\theta=(\theta_r,\theta_d)$ and then $I_{\textrm{CF}}$ can never be less than the capacity of the source-to-destination channel. Similarly, the optimal decision region assuming Gaussian inputs reads as \eqref{eq:optimal-DR-CRC}. 


\subsection{Composite Cooperative Unicast Networks}\label{SectionIV-2}
\begin{figure} [t]
\centering  
\includegraphics [width=.45 \textwidth] {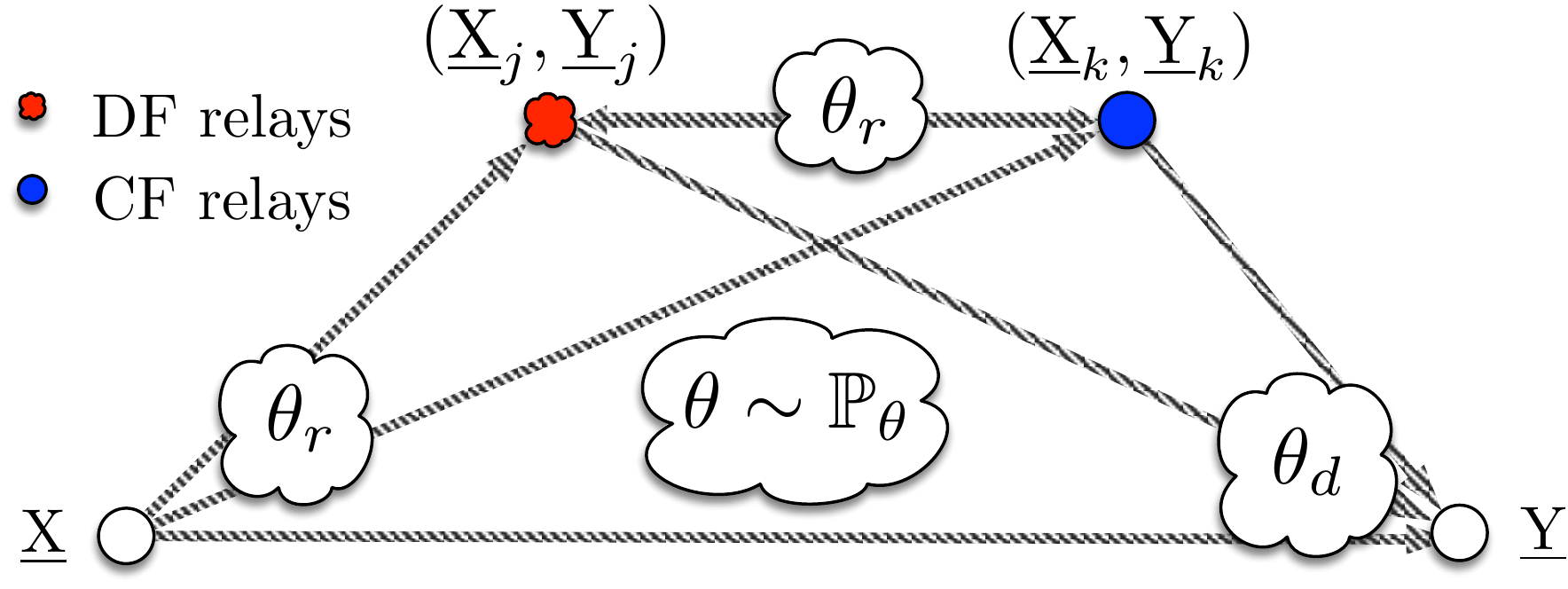}    
\caption{Composite Cooperative Unicast Network.}  
\label{fig:CUMRN}
\end{figure}

\begin{figure*} [!b]
\hrule
\begin{IEEEeqnarray}{lll}
\mc{W}_\Theta &\triangleq & 
\Big\{P_{Y^{n} Y_{1}^{n}\dots  Y_{N}^{n}|X^{n} X^{n}_{1}\dots   X^{n}_{N};\theta}(\underline{y},\underline{y}_1, \dots, \underline{y}_N|\underline{x}, \underline{x}_1,\dots, \underline{x}_N;\theta)\,\Big |\, \underline{x}\in \mathcal{X}^n, \,\underline{x_1} \in \mathcal{X}_1^n,\dots, \underline{x_N} \in \mathcal{X}_N^n, \nonumber\\
\IEEEeqnarraymulticol{3}{r}{
\underline{y}\in \mathcal{Y}^n, \underline{y}_1\in \mathcal{Y}_1^n, \dots, \underline{y}_N\in \mathcal{Y}_N^n ,\theta\in\Theta \Big\}^{\infty}_{n=1}\ .}
\label{eq:compuninet}
\end{IEEEeqnarray}
\end{figure*}

Consider the composite cooperative unicast network as described in Fig.~\ref{fig:CUMRN}, where the probability distribution characterizing the network is indexed with parameters $\theta\in\Theta$. This network can be defined as a set of memoryless probability distributions in \eqref{eq:compuninet}.
This family of networks corresponds to the definition of the compound cooperative unicast network for which the channels are chosen in an arbitrary manner but remain fix during the communication. Similar to the case of the composite relay channel, we introduce a probability measure $\prob_\uptheta$ over $\Theta$, then each index $\theta$ is present with probability $\prob_\uptheta$. The vectors of parameters is $\theta=(\theta_d,\theta_r)\in {\Theta}$ with $\theta_r$ denoting all parameters that affect the relays' outputs, and $\theta_d$ are the remaining parameters involved in the communication, as shown  in Fig.~\ref{fig:CUMRN}. More precisely, the marginal PDs read as:
\begin{IEEEeqnarray*}{rCl}
P_{Y_{1}^{n}\dots  Y^n_N|X^{n} X^{n}_{1}\dots  X^n_N;\uptheta} & = & P_{Y_{1}^{n}\dots  Y^n_N|X^{n} X^{n}_{1}\dots  X^n_N;\uptheta_r}\ ,\\
P_{Y^{n}|X^{n} X^{n}_{1}\dots  X^n_N;\uptheta} &= & P_{Y^{n}|X^{n} X^{n}_{1}\dots  X^n_N;\uptheta_d}\ .
\end{IEEEeqnarray*}
The specific draw of $\theta=(\theta_d,\theta_r)$ is assumed to be unknown at the source and fully known at the destination while the relays only know $\theta_r$. Again the notion of capacity-versus-outage shall be used to characterize the performance of this network.\vspace{1mm}

\begin{definition}[code and achievability] \label{def-code-CMR}
A code-$\mc{C}(n,M_n,r)$ for the composite cooperative unicast network with $(\mc{W}_\Theta, \prob_\uptheta)$  consists of:
\begin{itemize}
\item An encoder mapping  $\left \{\varphi:\mc{M}_{n} \longmapsto \mathcal{X}^n  \right\}$, 
\item A decoder mapping $\left\{ \phi:\mathcal{Y}^n\times\Theta \longmapsto \mc{M}_{n}\right\}$,
\item A set of relay functions $\left\{ f^{(k)}_{i} :\mathcal{Y}_k^{i-1}\times\Theta_r  \longmapsto \mathcal{X}_{k} \right \}_{i=1}^n$ for $k\in\mc{N}$.
\end{itemize}
An error probability $0\leq \epsilon<1$  is said to be $r$-achievable, or the rate $r$ is said to be $\epsilon$-achievable, if there exists a code-$\mc{C}(n,M_n,r)$ with rate satisfying
\begin{equation*}
\liminf\limits_{n\rightarrow \infty}\frac{1}{n} \log M_n \geq r 
\end{equation*} 
and average error probability 
\begin{equation*}
 \limsup\limits_{n\rightarrow \infty}\, \esp_{\uptheta}\left[\Pr\left\{\phi(Y^n_{\uptheta},{\uptheta}) \neq W\big | \uptheta\right\}\right]\leq \epsilon\ .
\label{errorprob_def-CMR}   
\end{equation*}    
The infimum of all $r$-achievable error probabilities $\bar{\epsilon}(r)$ is defined as 
\begin{equation*}
\bar{\epsilon}(r)\triangleq \inf\left\{0\leq \epsilon<1\,:\, \textrm{$\epsilon$ is $r$-achievable}\right\}\ .
\label{errorprob_def2-CMR}   
\end{equation*} 
\end{definition}
It should be worth mentioning here that the definition $\epsilon$-achievability based on \eqref{errorprob_def-CMR} becomes equivalent to that of averaged channels~\cite{Ahlswede1968,Han2003}, where
\begin{equation*}
C_{\epsilon}\triangleq \sup\left\{0\leq r \,:\, \textrm{$r$ is $\epsilon$-achievable}\right\}\ .
\label{epscapacity-def-CMR}   
\end{equation*}

Assume that the destination is equipped with outage and full error identification functions $I$ and $J$, given by expressions \eqref{eq:funI} and \eqref{eq:funJ}, respectively. The average error of a given code can be bounded by following the same arguments as before and thus,
\begin{IEEEeqnarray*}{rCl}
\Pr\{J(\uptheta)=0\} & \leq & \limsup\limits_{n\rightarrow \infty}\, \esp_{\uptheta}\left[\Pr\left\{\phi(Y^n,{\uptheta}) \neq W\big | \uptheta\right\}\right] \\
&\leq& \Pr\{I(\uptheta)=0\}\,.
\end{IEEEeqnarray*}
Because the preceding lower bound is valid for arbitrary codes, it provides a converse bound on $\bar{\epsilon}(r)$.  

\begin{remark}
The outage and full error identification functions $I$ and $J$ can be equally used to bound the $\epsilon$-capacity of averaged networks with parameter $\uptheta$. Consider a code with maximal achievable rate $r$ and outage probability given by $\Pr\{I(\uptheta)=0\}$. Since the outage probability provides an upper bound on the expected error of the code, the rate $r$ is $\epsilon$-achievable for all $0\leq \epsilon<1$ exceeding the outage probability. Hence, all codes with outage probability less than $\epsilon$ are $\epsilon$-achievable, i.e., 
\begin{IEEEeqnarray}{rCl}
& & C_\epsilon\geq \sup\Big\{r\geq 0 \,:\, \textrm{$\exists$ a code-$\mc{C}(n,M_n,r)$ } \nonumber\\
\IEEEeqnarraymulticol{3}{r}{\textrm{with $\Pr\{I(\uptheta)=0\}\leq\epsilon$ and }\,\,\liminf\limits_{n\rightarrow \infty}\frac{1}{n} \log M_n \geq r \Big\}  \ .}\,\,\,
\label{epscapacity-LB}   
\end{IEEEeqnarray}
Let us assume any code with rate $r$ and full error identification function $J$ such that $\Pr\{J(\uptheta)=0\}> \epsilon$, then we know that the code is not $\epsilon$-achievable. This condition does not imply any specific connection between the rate $r$ and the $\epsilon$-capacity because there may exist another code with rate $r$ which is $\epsilon$-achievable. To remove this possibility, we can consider only those rates with $\Pr\{J(\uptheta)=0\}> \epsilon$ such that that there is no code with this rate which is $\epsilon$-achievable. The set of these rates is non-empty in general because there is certain rate limit over which every code with that rate  will have the error probability asymptotically tending to one.
For instance, every rate beyond the capacity of discrete memoryless satisfies such condition and hence it is not $\epsilon$-achievable for all $\epsilon<1$. Formally, if the rate $r$ is such that for all code-$\mc{C}(n,M_n,r)$ with 
 $\liminf\limits_{n\rightarrow \infty}\frac{1}{n} \log M_n \geq r$ the error probability satisfies $\Pr\{J(\uptheta)=0\}> \epsilon$,  then $r$ provides an upper bound on the $\epsilon$-capacity of the composite cooperative unicast network, i.e.,
\begin{IEEEeqnarray}{rCl}
&&C_\epsilon \leq  \inf\Big\{r\geq 0 \,:\, \textrm{$\forall$ code-$\mc{C}(n,M_n,r)$\,\, }\nonumber\\
\IEEEeqnarraymulticol{3}{r}{\textrm{ if \,\,$\liminf\limits_{n\rightarrow \infty}\frac{1}{n} \log M_n \geq r $  then} \,\, \Pr\{J(\uptheta)=0\}>\epsilon \Big\}\ .}
\label{epscapacity-UB}   
\end{IEEEeqnarray}
For all codes with rate at least $r$, an identification function is given by~\cite{6404645} 
\begin{equation*}
J(\theta)\triangleq \mathbf{1}[r>C_{\textrm{CB}}(\theta)]\ ,
\end{equation*} 
 where $C_{\textrm{CB}}(\theta)$ denotes the cut-set bound of the cooperative unicast network with index $\theta$. The $\epsilon$-capacity of the averaged composite cooperative unicast network is bounded by
\begin{equation}
C_\epsilon\leq \inf\left\{r\geq 0 \,:\, \Pr\{r>C_{\textrm{CB}}(\uptheta)\}>\epsilon\right\} \ .
\label{epscapacity-UB-SR}   
\end{equation}
\end{remark}

In the rest of this section, we upper bound the average error probability based on the outage probability of  the ``Selective Coding Strategy" (SCS). Indeed, we first derive an upper bound using Theorem~\ref{thm:MNNC}. Let us select a set of nodes $\mc{V}\subseteq\mc{N}$ and a probability distribution $P_{QVX}$ that is independent of the specific draw $\theta$, which is not available at the source. Besides relays can adapt $\{P_{X_k\vert QV}\}$ to the  parameters involved in $\theta_r$ for which the identification function reads as:
\begin{equation*}
I(\theta)\triangleq \mathbf{1}[r \leq I_{\textrm{MNNC}}(\mc{V},\uptheta)]\ ,
\end{equation*}  
where $I_{\textrm{MNNC}}(\mc{V},\uptheta)$ is defined by
\begin{IEEEeqnarray}{rCl} 
I_{\textrm{MNNC}}(\mc{V},\uptheta)\triangleq &\min\left(\max_{\mc{T}\in\Upsilon(\mc{N})}\,\min_{\mc{V}^c \subseteq\mc{S}\subseteq\mc{T}} R_{\mc{T}}(\mc{S},\uptheta)\ ,\right.\nonumber\\
&\left.\,
\min_{k\in\mc{V}^c}\max_{\mc{T}_k\in\Upsilon_k(\mc{N})}\,\min_{\mc{S}\subseteq\mc{T}_k} R^{(k)}_{\mc{T}_k}(\mc{S},\uptheta_r)\right)\ , 
\label{MNNCrate}
\end{IEEEeqnarray}   
with $\mc{V}^c\triangleq \mc{T}-\mc{V}$ and
\begin{IEEEeqnarray*}{rCl}
R_{\mc{T}}(\mc{S},\uptheta) &\triangleq& I_\uptheta(XX_{\mc{S}};\hat{Y}_{\mc{S}^c}Y|X_{\mc{S}^c}Q)\nonumber\\
\IEEEeqnarraymulticol{3}{r}{-I_\uptheta(\hat{Y}_{\mc{S}};{Y}_{\mc{S}}|XX_{\mc{T}}\hat{Y}_{\mc{S}^c}YQ)\ , }	\\
R^{(k)}_{\mc{T}_k}(\mc{S},\uptheta_r)&\triangleq &
I_{\uptheta_r}(X;\hat{Y}_{\mc{T}_k}Y_k|VX_kX_{\mc{T}_k}Q)\nonumber\\
\IEEEeqnarraymulticol{3}{r}{+I_{\uptheta_r}(X_{\mc{S}};Y_k|VX_kX_{\mc{S}^c}Q)}\nonumber\\
\IEEEeqnarraymulticol{3}{r}{-I_{\uptheta_r}(\hat{Y}_{\mc{S}};Y_{\mc{S}}|VX_kX_{\mc{T}_k}\hat{Y}_{\mc{S}^c}Y_kQ)\ . }
\end{IEEEeqnarray*}
The sets $\Upsilon(\mc{N})$ and $\Upsilon_k(\mc{N})$ are given by
\begin{IEEEeqnarray*}{rCl}
 \Upsilon(\mc{N})&\triangleq& \big\{\mc{T}\subseteq\mc{N}\,: \,\,\forall\,\, \mc{S}\subseteq\mc{T}\,, \,Q_{\mc{T}}(\mc{S},{\uptheta})\geq 0\big\}\ , \\
\Upsilon_k(\mc{N})&\triangleq&\big\{\mc{T}\subseteq\mc{N}-\{k\}\,: \,\,\forall\,\,\mc{S}\subseteq\mc{T}\ , \nonumber\\
\IEEEeqnarraymulticol{3}{c}{\,Q^{(k)}_{\mc{T}}(\mc{S},{\uptheta_r})\geq 0\big\}\ ,}
\end{IEEEeqnarray*}
where $Q_{\mc{T}}(\mc{S},{\uptheta})$ and  $Q^{(k)}_{\mc{T}}(\mc{S},{\uptheta_r})$ are defined as:
\begin{IEEEeqnarray*}{rCl}
Q_{\mc{T}}(\mc{S},{\uptheta}) &\triangleq &I_{\uptheta}(X_{\mc{S}};\hat{Y}_{\mc{S}^c}Y|VXX_{\mc{S}^c}Q)\nonumber\\
\IEEEeqnarraymulticol{3}{r}{-I_{\uptheta}(\hat{Y}_{\mc{S}};{Y}_{\mc{S}}|VXX_{\mc{T}}\hat{Y}_{\mc{S}^c}YQ)\ ,}\\
Q^{(k)}_{\mc{T}}(\mc{S},{\uptheta_r})&\triangleq &I_{\uptheta_r}(X_{\mc{S}};Y_k|VX_kX_{\mc{S}^c}Q)\nonumber\\
\IEEEeqnarraymulticol{3}{r}{-I_{\uptheta_r}(\hat{Y}_{\mc{S}};Y_{\mc{S}}|VXX_kX_{\mc{T}}\hat{Y}_{\mc{S}^c}Y_kQ)\ .}
\end{IEEEeqnarray*}
The following upper bound on the expected error probability of the composite cooperative unicast network with partial CSI $\theta_r$ at the relays holds: 
\begin{equation}
\bar{\epsilon}(r) \leq \min_{p(x,v,q)}\, \inf_{\mc{V}\subseteq\mc{N}}
\, \prob[E]_{\uptheta_r} 
\Big\{
\min_{p(\cdot)\in\mc{Q}}  \, 
\prob_{\uptheta|\uptheta_r}\big[
I_{\textrm{MNNC}}(\mc{V},\uptheta)
 \big| \uptheta_r\big] 
\Big\}\,,
\label{CMNNCoutage}   
\end{equation}
where the set  of all admissible PDs $\mc{Q}$ is given by
\begin{equation*}
p(\cdot)=\prod_{j\in \mc{V}}p({x_j|q})p({\hat{y}_j|x_jy_jq})\prod_{j\in \mc{V}^c}p({x_j|vq})p({\hat{y}_j|vx_jy_jq})\ .
\end{equation*}

We can also further exploit the SCS presented in the previous subsection. Again the general idea is that, based on the channel parameters $\theta_r$, each relay is allowed to use either CF or DF scheme. In this setting, the relays must have  two set of codebooks: one set is intended to the case in which the relay uses CF scheme and the other set is for the case of DF scheme.  Without loss of generality, we assume that $\theta_r$ is known to all relay terminals. Each relay has a decision region, say $\mc{D}^{(k)}_{\textrm{DF}}$. So that relay $k$ can decide for $\theta_r\in\mc{D}^{(k)}_{\textrm{DF}}$ to use DF scheme and otherwise it would use CF scheme. Let $\mc{V}\subseteq\mc{N}$ and define $\mc{D}_{\mc{V}}$ as follows:
\begin{equation*}
\displaystyle\mc{D}_{\mc{V}}\triangleq \left(\bigcap_{k\in\mc{V}^c}{\mc{D}^{(k)}_{\textrm{DF}}}\right) \cap \left(\bigcap_{k\in\mc{V}}{\mc{D}^{(k)}_{\textrm{DF}}}^c\right)\ .
\end{equation*}
By consequence, if $\theta_r\in\mc{D}_{\mc{V}}$ then $\theta_r\notin\mc{D}^{(k)}_{\textrm{DF}}$ for all $k\in\mc{V}$, and $\theta_r\in\mc{D}^{(k)}_{\textrm{DF}}$ for all $k\in\mc{V}^c$. Relay $k$, for $k\in\mc{V}$, uses CF scheme while relay $k^\prime$, for $k^\prime\in\mc{V}^c$, employes DF scheme. The ensemble of decision regions of the relays given by the regions $\mc{D}_{\mc{V}}$, which are mutually disjoint, form a partition of the space of parameters $\Theta_r$. Notice that for some $\mc{V}$, the set $\mc{D}_{\mc{V}}$ may be empty. Indeed, a decision region $\left\{\mc{D}_{\mc{V}}\,:\,\mc{V}\subseteq\mc{N}\right\}$ is a set of individual decision region $\mc{D}_{\mc{V}}$ satisfying: 
$$
\bigcup_{\mc{V}\subseteq\mc{N}} \mc{D}_{\mc{V}}=\Theta_r 
$$
for $\mc{V}\neq\mc{V}^\prime$  with  $\mc{D}_{\mc{V}}\cap\mc{D}_{\mc{V}^\prime}=\emptyset$ .

\begin{figure} [t]
\centering  
\includegraphics [width=.4 \textwidth] {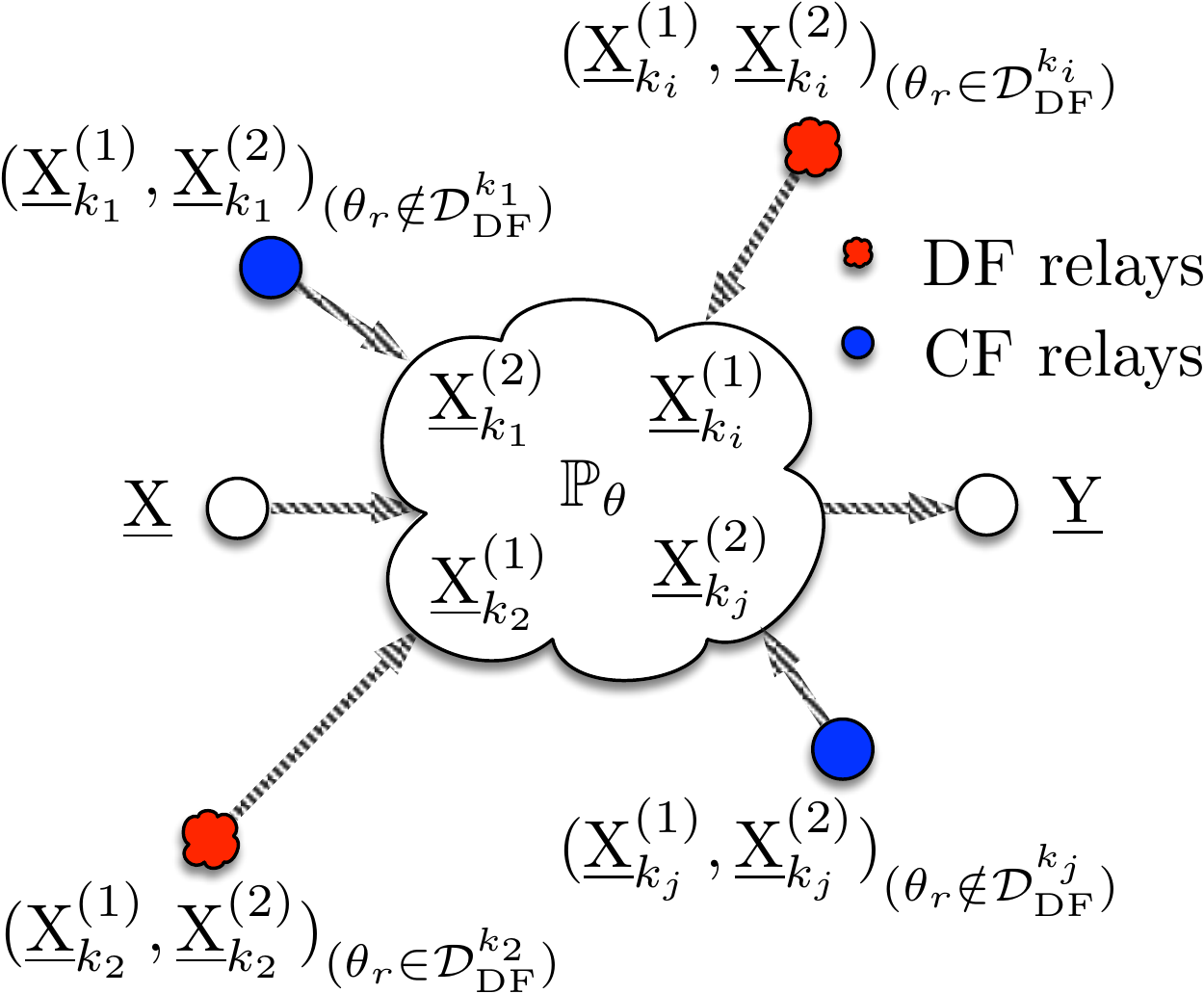}    
\caption{Selective coding strategy (SCS) over cooperative unicast networks.}
\label{fig:CMRN}
\end{figure}

The indexed partitions over $\Theta_r$ which are at most $2^N$ subsets\footnote{It is due to the fact that each partition subset is indexed by $\mc{V}$.} are refereed to as the decision region. In other words, the very same partitioning $\pi(\Theta_r)$ over $\Theta_r$ can be indexed differently with subsets of $\mc{N}$ and so it can lead to different sets of decision regions. We denote by $\Pi\left(\Theta_r,\mc{N}\right)$  the set of all possible indexed partitions on $\Theta_r$. Hence, if $\theta_r\in\mc{D}_{\mc{V}}$ we have a cooperative unicast network where the relays in $\mc{V}$ use CF scheme while the others relays use DF scheme, as shown in Fig.~\ref{fig:CMRN}. Each relay has two codebooks based on $\underline{x}^{(1)}_{(k)}$ and $\underline{x}^{(2)}_{(k)}$. The first codebook with codewords $\underline{x}^{(1)}_{(k)}$ (DF scheme) is transmitted when $\theta_r\in\mc{D}^{(k)}_{\textrm{DF}}$, so relay $k$ decodes the source message and transmits it to the destination with the index according to 
MNNC scheme. The source not knowing whether the relay $k$ is decoding or not uses superposition coding over $\underline{x}^{(1)}_{(k)}$.  Otherwise, if $\theta_r\notin\mc{D}^{(k)}_{\textrm{DF}}$ then $\underline{x}^{(2)}_{(k)}$ (CF scheme) is transmitted. Notice that unlike DF, $\underline{x}^{(2)}_{(k)}$ is independent of the source codewords and so its probability distribution can be chosen adaptively based on $\theta_r$. The optimization over $\mc{D}_{\mc{V}}$ would potentially improve the outage probability compared to the case in which every relay uses a fix coding strategy for all $\theta_r$. A careful evaluation of Theorem~\ref{thm:MNNC} yields the following proposition.\vspace{1mm}
\begin{proposition} [SCS with partial CSI]
\label{thm:SCS-MNNC}
The average error probability of the composite unicast network with partial CSI $\theta_r$ at the relays can be upper bounded by 
\begin{IEEEeqnarray}{rCl}
&&\bar{\epsilon}(r) \leq \min_{p(x,v,q)}\,
\inf_{\left\{\mc{D}_{\mc{V}}\,:\,\mc{V}\subseteq\mc{N}\right\}\in \Pi\left(\Theta_r,N\right)}\,\displaystyle\sum_{\mc{V}\subseteq\mc{N}}
\nonumber\\
&&
\prob[E]_{\uptheta_r} 
\Big\{ \min_{p(\cdot)\in \mc{Q}}\,\prob_{\uptheta|\uptheta_r}\big[r > I_{\textrm{MNNC}}(\mc{V},\uptheta), \uptheta_r \in \mc{D}_{\mc{V}} \big| \uptheta_r\big] 
\Big\}\,, \,\,\,\,\,\,\,\,\,
\label{SCSoutage-MR-1}   
\end{IEEEeqnarray}
where the set  of all admissible PDs $\mc{Q}$ is given by 
\begin{equation*}
p(\cdot)=\prod_{j\in \mc{V}}p({x_j|q})p({\hat{y}_j|x_jy_jq})\prod_{j\in \mc{V}^c}p({x_j|vq})p({\hat{y}_j|vx_jy_jq})
\end{equation*}
and $\Pi\left(\Theta_r,N\right)$ is the set of all indexed partitions over $\Theta_r$ into at most $2^N$ disjoint sets and $I_{\textrm{MNNC}}(\mc{V},\uptheta)$ is defined by expression \eqref{MNNCrate}. 
\end{proposition}
\begin{IEEEproof}
This proposition is shown in Appendix~\ref{proof:SCS-MNNC}.
\end{IEEEproof}

It is worth noting that \eqref{SCSoutage-MR-1} reaches at least the same performance as \eqref{CMNNCoutage}. In \eqref{CMNNCoutage}, the set $\mc{V}$ is chosen beforehand independent of $\uptheta_r$. Now if one choose $\mc{D}_\mc{V}=\Theta_r$ for this $\mc{V}$ and $\mc{D}_\mc{V^\prime}=\emptyset$ for all $\mc{V}^\prime\neq\mc{V}$ in \eqref{SCSoutage-MR-1}, then  \eqref{CMNNCoutage} is obtained as special case. The advantage of \eqref{SCSoutage-MR-1} is in choosing the set of CF relays, $\mc{V}$, adaptively based on channel state information of relays. Finally, if full CSI is available at all relays then this will simply give the following proposition.\vspace{1mm}

\begin{proposition} [SCS with full CSI]
\label{col:SCS-MNNC}
The average error probability of the composite cooperative network with full CSI $\theta_r$ at the relays can be upper bounded by 
\begin{IEEEeqnarray}{rCl}
\bar{\epsilon}(r) &\leq & \min_{p(x,v,q)}
\,\inf_{\left\{\mc{D}_{\mc{V}}\,:\,\mc{V}\subseteq\mc{N}\right\}\in \Pi\left(\Theta_r,N\right)} \nonumber\\
\IEEEeqnarraymulticol{3}{r}{\displaystyle\sum_{\mc{V}\subseteq\mc{N}} \prob_{\uptheta}\big[r > I_{\textrm{MNNC}}(\mc{V},\uptheta), \uptheta_r \in \mc{D}_{\mc{V}} \big]}\ ,
\label{SCSoutage-MR-fullCSI}   
\end{IEEEeqnarray}
where $\Pi\left(\Theta_r,N\right)$ is the set of all indexed partitions over $\Theta_r$ into at most $2^N$ disjoint sets and $I_{\textrm{MNNC}}(\mc{V},\uptheta)$ is defined by
\begin{IEEEeqnarray*}{rCl}
I_{\textrm{MNNC}}(\mc{V},\uptheta) &\triangleq & \max_{p(\cdot)\in\mc{Q}} \min\left(\max_{\mc{T}\in\Upsilon(\mc{N})}\min_{\mc{T}-\mc{V}\subseteq\mc{S}\subseteq\mc{T}} R_{\mc{T}}(\mc{S},\uptheta)\,,\right.\nonumber\\
&&\left.\,\min_{k\in\mc{V}^c}\max_{\mc{T}_k\in\Upsilon_k(\mc{N})}\min_{\mc{S}\subseteq\mc{T}_k} R^{(k)}_{\mc{T}_k}(\mc{S},\uptheta_r)\right) \ ,
\end{IEEEeqnarray*}
where the set  of all admissible PDs $\mc{Q}$ is given by 
\begin{equation*}
p(\cdot)=\prod_{j\in \mc{V}}p({x_j|q})p({\hat{y}_j|x_jy_jq})\prod_{j\in \mc{V}^c}p({x_j|vq})p({\hat{y}_j|vx_jy_jq})\ .
\end{equation*}
\end{proposition}


Finally, we present a similar result to that of Theorem~\ref{thm:NC-MNNC} where there is no cooperation among the relay nodes.  \vspace{1mm}

\begin{proposition} [SCS with partial CSI]
\label{thm:SCS-NC-MNNC}
The average error probability of the composite non-cooperative unicast network  with partial CSI $\theta_r$ at the relays can be upper bounded by 
\begin{IEEEeqnarray}{rCl}
&&\bar{\epsilon}(r) \leq \min_{p(x,x^{(1)}_{\mc{N}},q)}\,
\inf_{\left\{\mc{D}_{\mc{V}},\mc{V}\subseteq\mc{N}\right\}\in \Pi\left(\Theta_r,N\right)}\nonumber\\
\IEEEeqnarraymulticol{3}{r}{
\displaystyle\sum_{\mc{V}\subseteq\mc{N}}\prob[E]_{\uptheta_r} 
\Big\{
\min_{p(\cdot)}  
\prob_{\uptheta|\uptheta_r}\big[r > I_{\textrm{MNNC}}(\mc{V},\uptheta), \uptheta_r \in \mc{D}_{\mc{V}} \big| \uptheta_r\big] 
\Big\}}\ ,  \,\,\,\,\,
\label{SCSoutage-NC-MR-1}   
\end{IEEEeqnarray}
where the set of all admissible PDs $\mc{Q}$ is given by all PDs decomposing as:
\begin{equation}
p(\cdot)=\prod_{j\in \mc{V}}p({x^{(2)}_j|q})p({\hat{y}_j|x^{(2)}_jy_jq})
\end{equation}
and  $\Pi\left(\Theta_r,N\right)$ is the set of all indexed partitions over $\Theta_r$ into at most $2^N$ disjoint sets and $I_{\textrm{MNNC}}(\mc{V},\uptheta)$ is defined by
\begin{IEEEeqnarray}{rCl}
I_{\textrm{MNNC}}(\mc{V},\uptheta)&\triangleq&  \max_{\mc{T}\subseteq\mc{V}}\min\Big\{\min_{\mc{S}\subseteq\mc{T}} R_{\mc{T}}(\mc{S},\uptheta)\ ,\nonumber\\
\IEEEeqnarraymulticol{3}{r}{\min_{i\in \mc{V}^c}I_{\uptheta_r}(X;Y_{i}|X^{(1)}_{\mc{N}}Q)\Big\}}
\label{NC-MNNC-composite}
\end{IEEEeqnarray}
with 
\begin{align*}
R_{\mc{T}}(\mc{S},\uptheta)\triangleq & I_\uptheta(XX^{(1)}_{\mc{V}^c}X^{(2)}_{\mc{S}};\hat{Y}_{\mc{S}^c}Y|X^{(2)}_{\mc{S}^c}Q)\nonumber\\
&-I_\uptheta(Y_{\mc{S}\uptheta_r};\hat{Y}_{\mc{S}}|XX^{(2)}_{\mc{T}}X^{(1)}_{\mc{V}^c}\hat{Y}_{\mc{S}^c}YQ)\ ,
\end{align*}
where $(X^{(1)}_{k},X^{(2)}_{k})$ denote the corresponding relay inputs selected as follows 
\begin{equation*}
X_{k\theta_r}=\left\{\begin{array}{lll}
X^{(1)}_{k}& \, & \textrm{if $\displaystyle\theta_r\in\mc{D}^{k}_{\textrm{DF}} $}\\
X^{(2)}_{k}& \, & \textrm{if $\displaystyle\theta_r\notin\mc{D}^{k}_{\textrm{DF}} $}
\end{array}
\right.
\end{equation*}
\begin{equation*}
 \,\,\textrm{ with } \,\, \displaystyle\mc{D}^{k}_{\textrm{DF}}=\bigcup_{\mc{V}\subset\mc{N}\,,\,k\notin\mc{V}}\mc{D}_{\mc{V}}\,.
\end{equation*}

For $\uptheta_r\in\mc{D}_{\mc{V}}$, the following Markov chain holds:
\begin{equation*}
(X^{(1)}_{\mc{V}},X^{(2)}_{\mc{V}^c}) \minuso (X,X^{(1)}_{\mc{V}^c},X^{(2)}_{\mc{V}}) \minuso (Y,Y_{\mc{N}})\,.
\end{equation*}
\end{proposition}
\begin{IEEEproof}
The proof is presented in Appendix~\ref{proof:SCS-NC-MNNC}.
\end{IEEEproof}
It can be checked that the use of superposition coding does not change the rate $R_{\mc{T}}(\mc{S},\uptheta)$, but unlike the case of the composite relay channel, the condition of correct decoding at DF relays is changed from $I_{\uptheta_r}(X;Y_{i}|X^{(1)}_{\mc{V}^c}Q)$ to $I_\uptheta(X;Y_{i}|X^{(1)}_{\mc{N}}Q)$.

\section{Gaussian Fading Relay Channel} \label{SectionV}

We now consider an application example of SCS to the fading Gaussian relay channel defined by the following destination and relay output, respectively, 
\begin{IEEEeqnarray*}{rCl}
Y& = & g_1 X+g_3 X_1+ \mathpzc{V}_1\ ,\\
Y_1 & = & g_2 X + \mathpzc{V}_2\ , 
\end{IEEEeqnarray*}
where $ \mathpzc{V}_1$ and $ \mathpzc{V}_2$ are independent  complex Gaussian noises with zero-mean and unit variance; the channel gains $(g_1,g_2,g_3)$ are independent  complex Gaussian with zero-mean and unit variance; and the inputs must not exceed the average powers $P$ and $P_1$, respectively. It is assumed that the source is not aware of the channel measurements $\uptheta \triangleq (g_1,g_2,g_3)$, the relay only knows $\uptheta_r\triangleq g_2$ and the destination is fully aware of all fading coefficients $\uptheta$, and thus  $\uptheta_d\triangleq (g_1,g_3)$. This model is special case of the composite relay channel described in Section~\ref{SectionIV-1}. We aim to evaluate the asymptotic error probability based on the bounds derived in Propositions~\ref{corollary1} and~\ref{corollary2}, and compare them to the upper bounds corresponding to DF and CF schemes~\eqref{usual-upperbounds}, and the cut-set based lower bound. In this case, the expression for the DF rate reads as:  
\begin{align}
I_{\textrm{DF}}(\uptheta) &\triangleq  \min\Bigg\{\mc{C}\left({\beta |g_2|^2  P}\right), \nonumber\\
&\mc{C}\left(|g_1|^2 P+|g_3|^2 P_1+2\sqrt{\overline\beta PP_1}\mathds{R}e \{g_1g_3^\star\}\right)\Bigg\}
\label{eq:18B}
\end{align}
where $0\leq \beta \leq 1$ and $ \mc{C} (x)\triangleq \log_2(1+x)$. The CF rate is given by
\begin{IEEEeqnarray}{rCl}
I_{\textrm{CF}} (\uptheta)& \triangleq &  \max\left\{I_{\textrm{CF}}^\prime (\uptheta), \mc{C} \left(  \frac{|g_1|^2   P}{|g_3|^2   P_1+1}\right)\right\},\\
I_{\textrm{CF}}^\prime (\uptheta) &\triangleq &  \min\left\{ \mc{C} \left( |g_1|^2 P+\frac{|g_2|^2   P}{\widehat{N}_2+1}\right),\right.\nonumber\\
&&\left. \mc{C}\left(|g_1|^2 P +|g_3|^2 P_1 \right)-\mc{C}\left(\frac{1}{\widehat{N}_2}\right)\right\}\,,
\end{IEEEeqnarray}
where the description $\hat{Y}_1$ is generated by adding independent  complex Gaussian noise with zero-mean and variance $\widehat{N}_2$. Finally, the asymptotically error probability based on DF and CF schemes can be easily derived by using expressions \eqref{DFoutage-2} and \eqref{CFoutage}, respectively. If full CSI is available at the relay, then $\widehat{N}_2$ can be optimally chosen as:
\begin{equation}
\widehat{N}_2^{\textrm{opt}} \triangleq \frac{P\left(|g_1|^2 + |g_2|^2\right)+1}{{|g_3|^2  P_1}}\,.
 \label{eq:12}
 \end{equation}
Whereas, if only $g_2$ is available at the relay, a constant $\widehat{N}_2$ must be selected to minimize the outage probability. We have shown that based on $| g_2 |$, SCS allows the relay to select the proper coding strategy. In this case, if the source-to-relay channel is not good enough for decoding  the message, the relay uses CF scheme while otherwise DF scheme would be the best choice. It turns out that the optimum decision region $\mc{D}_{\textrm{DF}}$ is given by the set $\mc{D}_{\textrm{DF}}^\star\triangleq \left\{g_2 \,:\,r \leq \mc{C}\left({\beta |g_2|^2  P} \right)\right\}$. 
\begin{figure} [t]
\centering  
\includegraphics [width=.55 \textwidth] {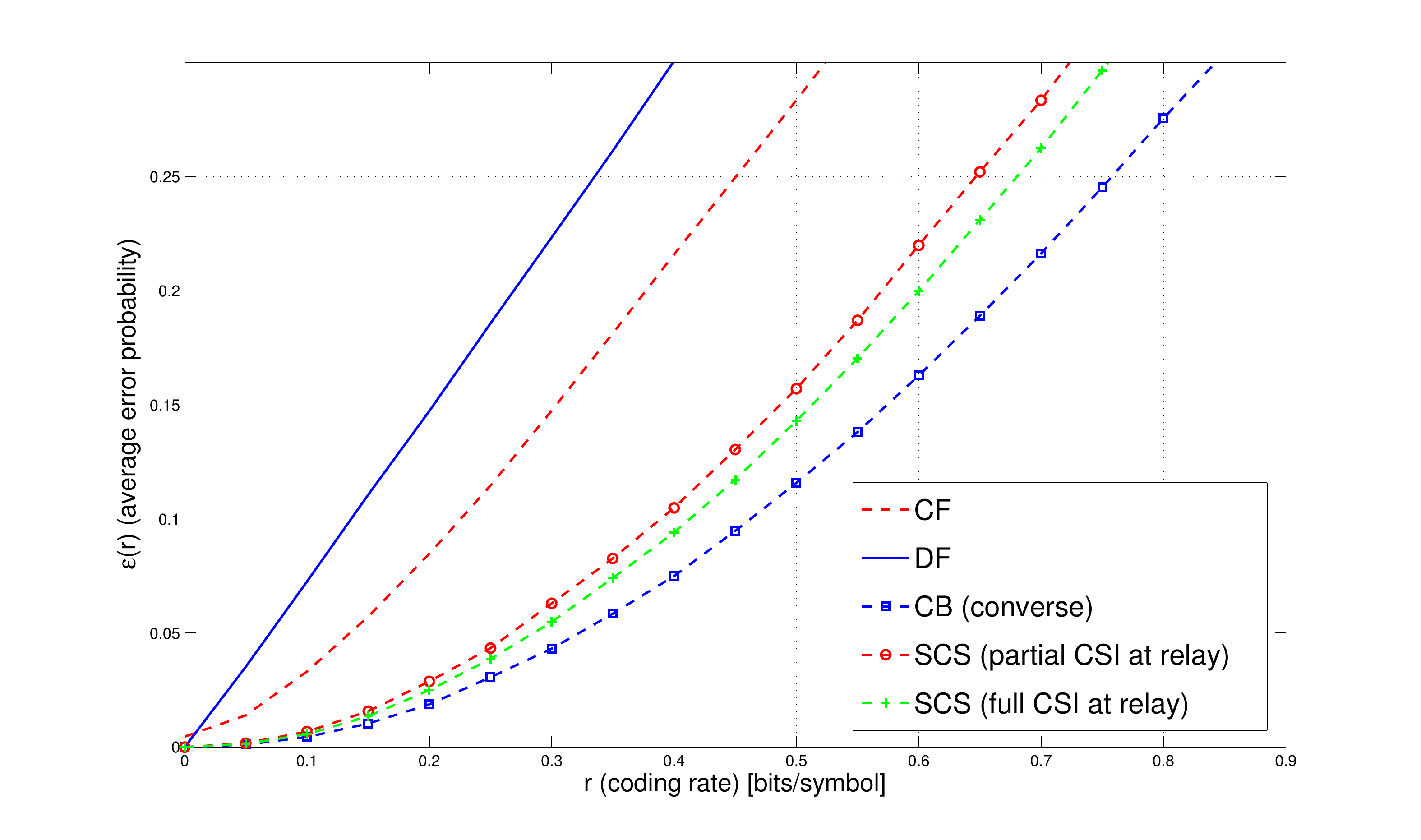}    
\caption{Asymptotic error probability $\bar{\epsilon}(r)$ vs. the coding rate $r$.}
\label{fig:2}
\end{figure}

Fig.~\ref{fig:2} presents numerical plots of the asymptotic error probability with $P_1=P=1$. For the case of partially CSI at the relay, we observe that SCS clearly outperforms the naive DF and CF schemes. Moreover, notice that full CSI $(g_1,g_2,g_3)$ at the relay improves the error probability only through the choice of the best possible compression noise $\widehat{N}_2$. Besides this guarantee that CF scheme can never perform less good than direct transmission. Finally, it can be seen that the upper bound (achievable error probability) resulting from SCS is close to the cut-set lower bound, and so to the best average error probability.

\begin{figure} [t]
\includegraphics [width=.53 \textwidth] {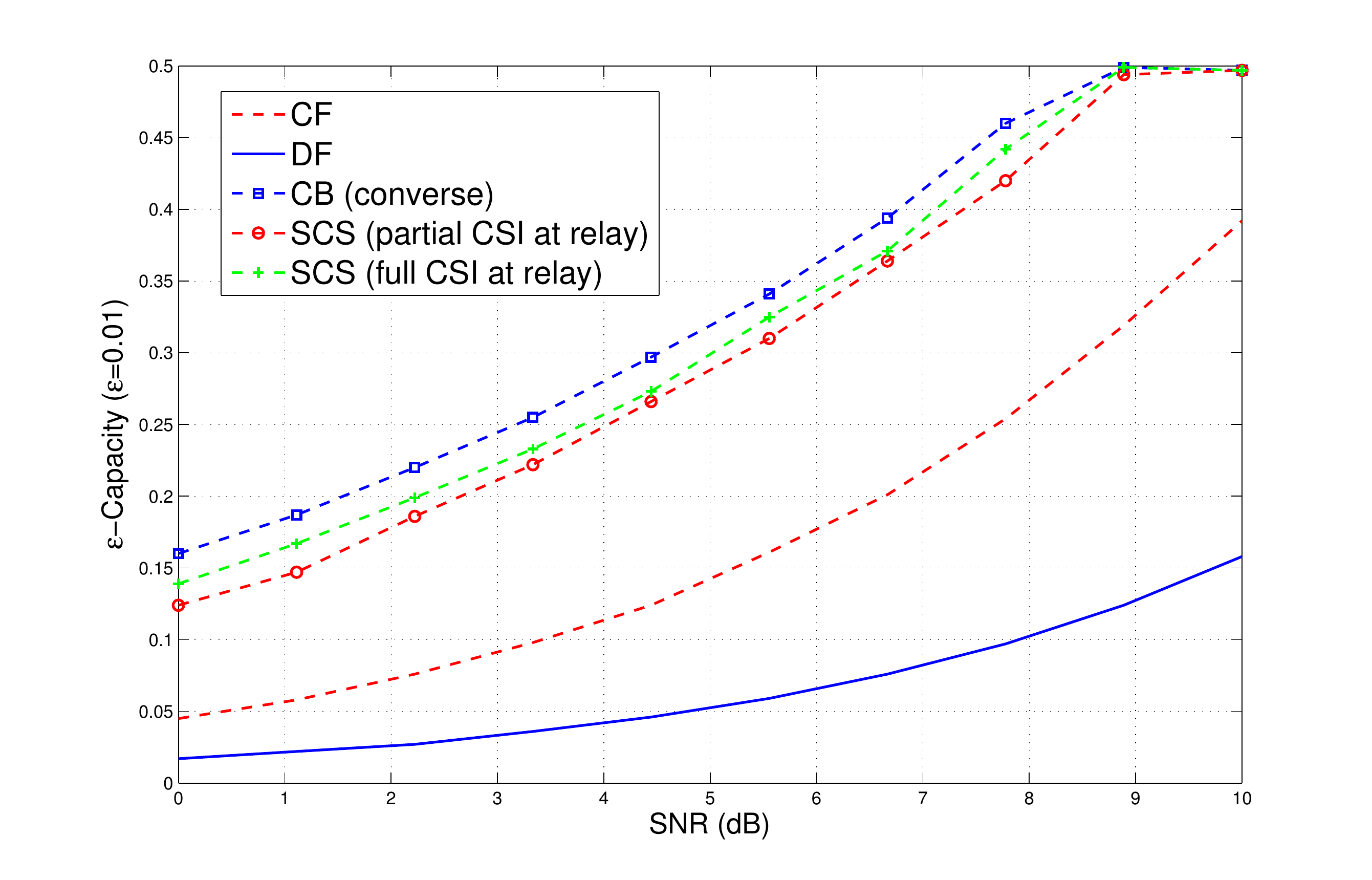}    
\caption{Bounds on the $\epsilon$-capacity vs. SNR.}
\label{fig:epscap}
\end{figure}

Fig.~\ref{fig:epscap} presents bounds on $\epsilon$-capacity of the corresponding averaged channel for $\epsilon=0.01$ based on the signal-to-noise ratio (SNR). We set $P_1=1$ while $P$  is varying with SNR. Indeed, the $\epsilon$-capacity represents the maximum achievable rate subject to satisfy  an expected error probability  less or equal than $\epsilon$.  The bounds illustrated in \eqref{epscapacity-LB} and \eqref{epscapacity-UB} have been used here.  Observe that again  the $\epsilon$-capacity is clearly enlarged by using our SCS and is not far from the upper bound.

\section{Summary and Concluding Remarks}

In this paper, we investigated the problem of communicating a single message to a single destination in presence of multiple relay nodes.  We considered a general framework of composite cooperative networks where the channel parameters are randomly drawn from a probability distribution and  each draw is assumed to be unknown at the source and fully known at the destination, but  only partly known at the relay nodes. Within this framework, we introduced ``Mixed Noisy Network Coding" (MNNC) where nodes are allowed to decode and forward messages while all nodes  transmit noisy descriptions of their observations. We further extended MNNC to multi-hopping networks that we referred to as ``Layered MNNC" (LMNNC) where DF relays are organized into disjoint groups,  each of them representing one hop in the network.

Perhaps the main feature of MNNC scheme relies on the meaningful concept of ``Selective Coding Strategy" (SCS)~\cite{6217861} that enables relays to decide dynamically, e.g., based on the channel measurements, whether in addition to communicate descriptions that are potentially exploited at the destination and the DF relays, would be possible to decode and forward messages. This guarantees full cooperation among all nodes, and in particular, without the requirement of any hierarchy between them. It is demonstrated for additive white Gaussian noise (AWGN) networks that MNNC improves over all previously established constant gaps to the cut-set bound, and for the slow-fading AWGN relay channel that SCS clearly outperforms the asymptotic average error probability of previous coding schemes. 

An important direction of future work is to investigate more sophisticated wireless models within our composite framework, e.g., the effects of user mobility,  network geometry and shadowing can be incorporated into the model~\cite{6284264}. It remains to investigate network scaling exponents~\cite{1638558} in the limit of large network size (number of nodes) where MNNC may substantially improve the scaling of the capacity with the number of nodes. In this setting, the multi-hopping scenario is in general challenging and needs further study since it is not clear how DF relays should choose their layering according to partial channel observation. Another important direction is to investigate how MNNC could be exploited in half-duplex networks.  With half-duplex constraints, the nodes are permitted --within each slot of time-- to either transmits or receives information. Therefore, it would be of interest to study cooperation protocols based on MNNC with slots of variable time durations, where nodes can 
dynamically select not only whether decode and forward messages in addition to send noisy descriptions of their observations, but also the time allowed to receive and transmit data~\cite{1542409,5075880}. We believe this problem may be quite rich because it yields several other scenarios of practical importance.


\appendices

\section{Proof of Theorem~\ref{thm:MNNC}}\label{proof:MNNC}

\begin{figure} [t]
\centering  
\includegraphics [width=.25 \textwidth] {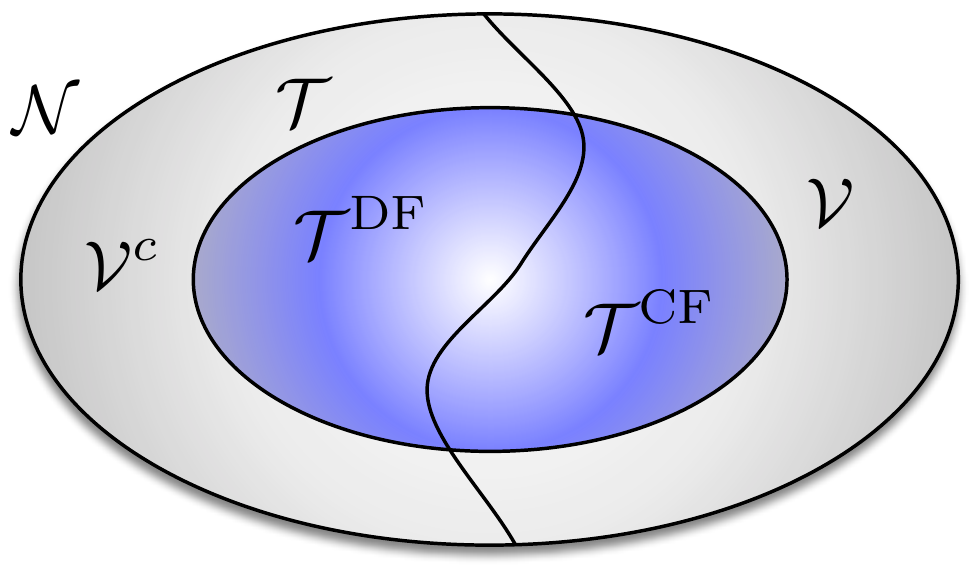}    
\caption{Groups of relay nodes.}
\label{fig:DR}
\end{figure}

First, we divide the relay nodes into two disjoint groups, namely, $\mc{V}$ and $\mc{V}^c=\mc{N}-\mc{V}$, as shown in Fig.~\ref{fig:DR}. The relays in $\mc{V}$ will use Compress-and-Forward (CF) scheme while the others relays will use Decode-and-Forward (DF) scheme, which are simply refereed to DF and CF relays. The DF relays transmit the compressed version of their observations, superimposed over the source message of the previous block. In this sense, the compressed version of the observation of each relay is transmitted to the other nodes. The $k$-th DF relay with $k\in\mc{V}^c$ decodes the source message of block $i$ by exploiting the descriptions sent by the others relays. Because the relays transmit the descriptions (or compression index) related to the block $i$ in block $i+1$, the $k$-th relay has to wait until the end of block $i+1$ to decode it and therefore DF relays has to wait until the block $i+2$ to forward the source message of the $i$-th block. Moreover, the $k$-th relay  exploits only the 
compression index of relays in $\mc{T}_k\subseteq\mc{N}-\{k\}$. Similarly, the destination decodes only the compression index of relays in $\mc{T}\subseteq\mc{N}$. It is shown that, by selecting a subset of relays, we contribute to increase the total rate provided that certain conditions are satisfied. 

For simplicity,  we adopt the following notation:
\begin{equation*}
\begin{array}{lll}
 \mc{T}^{\text{DF}}_k \triangleq \mc{T}_k\cap\mc{V}^c\ , & \mc{T}^{\text{DF}} \triangleq  \mc{T}\cap\mc{V}^c\ ,\\
\mc{T}^{\text{CF}}_k \triangleq \mc{T}_k\cap\mc{V}\ , & \mc{T}^{\text{CF}} \triangleq \mc{T}\cap\mc{V}\ .
\end{array}
\end{equation*}
Moreover, for any arbitrary message $w_i$, the notation $w_{\mc{S}}$ is used to denote the set of indices $\{w_i\,:\,i\in\mc{S}\}$. We now provide the code generation, encoding and decoding procedures.
Table \ref{tab:2} presents  details of the transmission schedule for a case with one DF relay and one CF relay. $\underline{x}_1$ and $\underline{x}_2$ represent the code of DF and CF relays with $\hat{y}_{1}$ and $\hat{y}_{2}$ as the compressed version of their observation. $\underline{v}$ is used for coherent transmission between the relays and the source. $\underline{x}$ and $\underline{y}$ are source codeword and destination received message. As discussed before, the DF relay delays the decoding of source message at $b=1$ to the end of block $b=2$ so that it can jointly decode the compression index $l_{11}$ and $w_1$ using $\underline{y}_{1}(1)$ and $\underline{y}_{1}(2)$. The destination on the other hand decodes backwardly. It first finds $l_{1(B+2)}$ and $l_{2(B+2)}$ and then jointly decodes $(w_{B},l_{1(B+1)},l_{2(B+1)})$. It continues this process until it finds all the messages. 

\begin{table*}[th] 
	\caption{Transmission schedule for ``Mixed Noisy Network Coding'' (MNNC).}
	\label{tab:2}
	 \centering
\begin{tabular}{| l | l | l | l | l | l | l | l | } 
\hline
$b=1$ & $b=2$ & $b=3$ &\dots  & $b=B+2$ & $b=B+3$ &\dots  & $b=B+L$ \\ 
\hline
$\underline{v}(1)$ & $\underline{v}(1)$ & $\underline{v}(w_1)$ &\dots    & $\underline{v}(w_{B})$ & $\underline{v}(1)$ &\dots  & $\underline{v}(1)$ 
\\
\hline
$\underline{x}_1(1,1)$ & $\underline{x}_1(1,l_{11})$ & $\underline{x}_1(w_1,l_{12})$ &\dots    & $\underline{x}_1(w_{B},l_{1(B+1)})$ & $\underline{x}_1(1,l_{1(B+2)})$ &\dots  & $\underline{x}_1(1,l_{1(B+2)})$ 
\\
\hline
$\underline{x}(1,w_1)$ & $\underline{x}(1,w_2)$ & $\underline{x}(w_1,w_3)$ & \dots    & $\underline{x}(w_{B},1)$ & $\underline{x}(1,1)$ &\dots  & $\underline{x}(1,1)$ 
\\
\hline
$\underline{x}_2(1)$ & $\underline{x}_2(l_{21})$ & $\underline{x}_2(l_{22})$ &\dots    & $\underline{x}_2(l_{2(B+1)})$ & $\underline{x}_2(l_{2(B+2)})$ &\dots  & $\underline{x}_2(l_{2(B+2)})$ 
\\
\hline
\hline
$\underline{\hat{y}}_{1}(1,1,l_{11})$  & $\underline{\hat{y}}_{1}(1,l_{11},l_{12})$ & $\underline{\hat{y}}_{1}(w_1,l_{12},l_{13})$ & \dots   & $\underline{\hat{y}}_{1}(w_B,l_{1(B+1)},l_{1(B+2)})$ & $\star$ & $\star$ & $\star$
\\
\hline
$\underline{\hat{y}}_{2}(1,l_{21})$  & $\underline{\hat{y}}_{2}(l_{21},l_{22})$ & $\underline{\hat{y}}_{2}(l_{22},l_{23})$ & \dots   &  $\underline{\hat{y}}_{2}(l_{2(B+1)},l_{2(B+2)})$ & $\star$ & $\star$ &$\star$
\\
\hline
\hline
$\underline{y}_{1}(1)$  & $\underline{y}_{1}(2)$ & $\underline{y}_{1}(3)$ &\dots   & $\underline{y}_{1}(B+2)$ & $\underline{y}_{1}(B+3)$ &\dots  & $\underline{y}_{1}(B+L)$
\\
\hline 
$\underline{y}_{2}(1)$  & $\underline{y}_{2}(2)$ & $\underline{y}_{2}(3)$ &\dots    & $\underline{y}_{2}(B+2)$ & $\underline{y}_{2}(B+3)$ &\dots  & $\underline{y}_{2}(B+L)$
\\
\hline 
$\underline{y}(1)$ & $\underline{y}(2)$ & $\underline{y}(3)$ &\dots    &  $\underline{y}(B+2)$ & $\underline{y}(B+3)$ &\dots  & $\underline{y}(B+L)$
\\
\hline
\end{tabular} 
\end{table*}

\subsection*{Code Generation:}
\begin{enumerate} [(i)]
\item Randomly and independently generate $2^{nR}$ sequences $\underline{v}$ drawn i.i.d. from
\begin{equation*}
P_{V}^{n}(\underline{v})=\prod\limits_{j=1}^n P_{V}(v_{j})\ .
\end{equation*}
Index them as $\underline{v}(w_0)$ with index $w_0\in \left[1,2^{nR}\right]$. 
\item For each $k\in\mc{V}^c$ and each $\underline{v}(w_0)$, randomly and independently generate  $2^{n\hat{R}_k}$ sequences $\underline{x}_k$ drawn i.i.d. from 
\begin{equation*}
P_{X_k\vert V}^n(\underline{x}_k|\underline{v}(w_0))=\prod\limits_{j=1}^n P_{X_k\vert V}(x_{kj}\vert v_{j}(w_0))\ . 
\end{equation*}
Index them as $\underline{x}_k(w_0,l_{0k})$, where $l_{0k}\in \big[1,2^{n\hat{R}_k}\big]$ for $\hat{R}_k \triangleq I(Y_k;\hat{Y}_k|X_k,V)+\epsilon$ with $k\in \mc{V}^c$. This is the codebook for DF relays.
\item For each $k\in\mc{V}$, randomly and independently generate  $2^{n\hat{R}_k}$ sequences $\underline{x}_k$ drawn i.i.d. from 
\begin{equation*}
P_{X_k}^n(\underline{x}_k)=\prod\limits_{j=1}^n P_{X_k}(x_{kj})\,.
\end{equation*}
Index them as $\underline{x}_k(l_{0k})$, where $l_{0k}\in \big[1,2^{n\hat{R}_k}\big]$ for $\hat{R}_k \triangleq I(Y_k;\hat{Y}_k|X_k)+\epsilon$ with $k\in \mc{V}$. This is the codebook for CF relays.
\item For each $\underline{v}(w_0)$, randomly and conditionally independently generate $2^{nR}$ sequences $\underline{x}$ drawn i.i.d. from 
\begin{equation*}
P_{X|V}^n(\underline{x}\vert \underline{v}(w_0))=\prod\limits_{j=1}^n P_{X|V}(x_{j}|v_{j}(w_0))\ .
\end{equation*}
Index them as $\underline{x}(w_0,w)$, where $w\in \big[1,2^{nR}\big]$. This is the source codebook.
\item For each $k\in\mc{V}^c$ and each $\underline{v}(w_0),\underline{x}_k(w_0,l_{0k})$, randomly and conditionally independently generate $2^{n\hat{R}_k}$ sequences  $\underline{\hat{y}}_k$ each with probability 
\begin{align*}
P_{\hat{Y}_k|X_kV}^n &(\underline{\hat{y}}_k\vert \underline{x}_k(w_0,l_{0k}),\underline{v}(w_0))\nonumber\\
&= \prod\limits_{j=1}^n P_{\hat{Y}_k\vert X_kV}(\hat{y}_{kj}\vert x_{kj}(w_0,l_{0k}),{v}_j(w_0))\ .
\end{align*}
Index them as  $\underline{\hat{y}}_k(w_0,l_{0k},l_{k})$, where $l_k\in \big[1,2^{n\hat{R}_k}\big]$ with $k\in \mc{V}^c$. This is the compressed version of DF relays output.
\item For each $k\in\mc{V}$ and each $\underline{x}_k(l_{0k})$, randomly and conditionally independently generate $2^{n\hat{R}_k}$ sequences  $\underline{\hat{y}}_k$ each with probability 
\begin{equation*}
P_{\hat{Y}_k|X_k}^n (\underline{\hat{y}}_k\vert \underline{x}_k(l_{0k}))= \prod\limits_{j=1}^n P_{\hat{Y}_k\vert X_k}(\hat{y}_{kj}\vert x_{kj}(l_{0k}))\ .
\end{equation*}
Index them as  $\underline{\hat{y}}_k(l_{0k},l_k)$, where $l_k\in \big[1,2^{n\hat{R}_k}\big]$ with $k\in \mc{V}$. This is the compressed version of CF relays output.
\item Provide the corresponding codebooks to the relays, the encoder and the decoder ends. 
\end{enumerate}
\subsection*{Encoding:} 
\begin{enumerate}[(i)]
\item
In every block $i=[1: B]$, the source sends $w_{i}$ using $\underline{x}\big(w_{(i-2)},w_i\big)$, where we have defined $w_{0}=w_{-1}=1$. Moreover, for blocks $i=[B+1:B+L]$, the source sends the dummy message $w_{i}=1$ known to all users.
\item
For every block $i=[1:B+L]$, and each $k\in\mc{V}^c$, the relay $k$ knows $w_{(i-2)}$ by assumption and $w_0=w_{-1}=1$, so it picks up $\underline{v}\big(w_{(i-2)}\big)$. For each $i=[1: B+2]$, the relay $k$ after receiving $\underline{y}_k(i)$, searches for at least one index $l_{ki}$ with $l_{k0}=1$ such that 
\begin{IEEEeqnarray}{rCl}
&&\big(\underline{v}(w_{(i-2)}),\underline{x}_{k}(w_{(i-2)},l_{k(i-1)}),\underline{y}_{k}(i),\underline{\hat{y}}_{k}(w_{(i-2)},\nonumber\\
&&l_{k(i-1)},l_{ki})\big)\in \mc{A}^n_\epsilon[VX_kY_k\hat{Y}_k]\,.
\end{IEEEeqnarray}
The probability of finding such $l_{ki}$ goes to one as $n$ goes to infinity  due to our adequate choice of the rate  $\hat{R}_k$ for compression. 
\item
For $i=[1:B+2]$ and $k\in \mc{V}^c $, relay $k$ knows from the previous block ${l}_{k(i-1)}$ and $w_{(i-2)}$ and it sends $\underline{x}_k(w_{(i-2)},l_{k(i-1)})$. Moreover, $k$-th relay  repeats $l_{k(B+2)}$ for $i=[B+3: B+L]$,  i.e. for $L-2$ blocks.
\item 
For each $i=[1: B+2]$, each $k\in\mc{V}$, the relay $k$ after receiving $\underline{y}_k(i)$, searches for at least one index $l_{ki}$ with $l_{k0}=1$ such that 
\begin{equation}
\big(\underline{x}_{k}(l_{k(i-1)}),\underline{y}_{k}(i),\underline{\hat{y}}_{k}(l_{k(i-1)},l_{ki})\big)\in \mc{A}^n_\epsilon[X_kY_k\hat{Y}_k]\,.
\end{equation}
The probability of finding such $l_{ki}$ goes to one as $n$ goes to infinity due to our adequate choice of the rate  $\hat{R}_k$ for compression. 
\item
For $i=[1:B+2]$ and $k\in\mc{V}$, relay $k$ knows from the previous block ${l}_{k(i-1)}$ and it sends $\underline{x}_k(l_{k(i-1)})$. Moreover, $k$-th relay  repeats $l_{k(B+2)}$ for $i=[B+3: B+L]$, i.e., for $L-2$ blocks.
\end{enumerate}
\subsection*{Decoding:} 
\begin{enumerate}[(i)]
\item After the transmission of the block  $i+1=[1:B+1]$ and for each $k\in\mc{V}^c$, the relay $k$ decodes the message $w_i$ and the compression index $l_{\mc{T}_ki}$, i.e., the compression indices  for the block $i$ of all relays in $\mc{T}_k$, with the assumption that all messages and compression indices up to block $i-1$ have been correctly decoded. We emphasize that there are two kind of relays in $\mc{T}_k$, those who employ DF scheme and those who are using CF scheme. Relay $k$ knows the message $(w_{(i-2)},w_{(i-1)})$, and so $\underline{v}\big(w_{(i-2)}\big)$ and $\underline{v}\big(w_{(i-1)}\big)$. Let us define the sequences:
\begin{IEEEeqnarray}{rCl}
\mc{E}_k\big(\hat{w}_b,\hat{l}_{\mc{T}_kb}\big) &\triangleq& \nonumber\\
\IEEEeqnarraymulticol{3}{r}{\Big(\underline{x}({w}_{(b-2)},\hat{w}_{b}),\underline{v}({w}_{(b-2)}),\underline{x}_{k}({w}_{(b-2)},l_{k(b-1)}),\underline{y}_{k}(b),}\nonumber\\
\IEEEeqnarraymulticol{3}{r}{\big(\underline{x}_{i}(w_{(b-2)},{l}_{i(b-1)}),\underline{\hat{y}}_{i}(w_{(b-2)},{l}_{i(b-1)},\hat{l}_{ib})\big)_{i\in\mc{T}^{\text{DF}}_k},}\nonumber\\
\IEEEeqnarraymulticol{3}{r}{\big(\underline{x}_{i}({l}_{i(b-1)}),\underline{\hat{y}}_{i}({l}_{i(b-1)},\hat{l}_{ib})\big)_{i\in\mc{T}^{\text{CF}}_k}\Big)}\nonumber\\
\mc{E}_k\big(\hat{l}_{\mc{T}_kb}\big) &\triangleq & \Big(\underline{v}({w}_{(b-1)}),\underline{x}_{k}({w}_{(b-1)},l_{kb}),\underline{y}_{k}(b+1),\nonumber\\
\IEEEeqnarraymulticol{3}{r}{\big(\underline{x}_{i}(w_{(b-1)},\hat{l}_{ib})\big)_{i\in\mc{T}^{\text{DF}}_k},{\big(\underline{x}_{i}(\hat{l}_{ib})\big)_{i\in\mc{T}^{\text{CF}}_k}\Big)\ .}}\,\,\,\,\,\,\,\,
\label{eq:FWDdecoding}
\end{IEEEeqnarray}
By looking at two consecutive blocks $(b,b+1)$, the $k$-th relay searches for the unique indices $(\hat{w}_{b},\hat{l}_{\mc{T}_kb})$ such that:
\begin{IEEEeqnarray}{rCl}
\mc{E}_k(\hat{w}_b,\hat{l}_{\mc{T}_kb}) & \in& \mc{A}^n_\epsilon[VXX_kX_{\mc{T}_k}\hat{Y}_{\mc{T}_k}Y_k] \,\, \text{ and } \nonumber\\
 \mc{E}_k(\hat{l}_{\mc{T}_kb}) & \in &\mc{A}^n_\epsilon[VX_{\mc{T}_k}X_kY_k]\ .
\label{DFrelaydecoding}
\end{IEEEeqnarray}
Given the sets $\mc{S}\subseteq\mc{T}_k$, and $\mc{S}^c\triangleq \mc{T}_k-\mc{S}$ and assuming that the correct messages were $({w}_b,{l}_{\mc{T}_kb})$, we define the following events:
\begin{IEEEeqnarray}{rCl}
\mb{E}_0&: &\Big\{\mc{E}_k({w}_b,{l}_{\mc{T}_kb})  \notin \mc{A}^n_\epsilon[VXX_kX_{\mc{T}_k}\hat{Y}_{\mc{T}_k}Y_k] \,\,\text{ or } \,\,\nonumber\\
&&\mc{E}_k({l}_{\mc{T}_kb})  \notin \mc{A}^n_\epsilon[VX_{\mc{T}_k}X_kY_k]  \Big\}  \,, \\
\mb{E}_{\mc{S}}& : & \Big\{\mc{E}_k({w}_b,\hat{l}_{\mc{T}_kb})  \in \mc{A}^n_\epsilon[VXX_kX_{\mc{T}_k}\hat{Y}_{\mc{T}_k}Y_k] \,\, \text{ and } \nonumber\\
&&\,\,\mc{E}_k(\hat{l}_{\mc{T}_kb})  \in \mc{A}^n_\epsilon[VX_{\mc{T}_k}X_kY_k]\text{ for some }\nonumber \\
&& \hat{l}_{kb}\neq l_{kb}\,, \,k\in\mc{S}\,\,\text{ and }\,\, \hat{l}_{kb}=l_{kb}\,,\,  k\in\mc{S}^c\Big\} \ , \\
\mb{E}_{w,\mc{S}}& : &  \Big\{\mc{E}_k(\hat{w}_b,\hat{l}_{\mc{T}_kb})  \in \mc{A}^n_\epsilon[VXX_kX_{\mc{T}_k}\hat{Y}_{\mc{T}_k}Y_k] \,\, \text{ and } \nonumber\\
&&\mc{E}_k(\hat{l}_{\mc{T}_kb})  \in \mc{A}^n_\epsilon[VX_{\mc{T}_k}X_kY_k] \text{ for some }\,\,\hat{w}_b\neq w_{b}\,,\nonumber \\
&& \,\hat{l}_{kb}\neq l_{kb}\,, \, k\in\mc{S}\,\, \text{ and } \,\,\hat{l}_{kb}=l_{kb}\,, \, k\in\mc{S}^c \Big\}\ .
\end{IEEEeqnarray}
Hence, the error probability can be bounded as follows:
\begin{IEEEeqnarray*}{rCl}
\Pr\Big((\hat{w}_{b},\hat{l}_{\mc{T}_kb}) &\neq & ({w}_{b},{l}_{\mc{T}_kb}) \Big)\leq \Pr(\mb{E}_0)\nonumber\\
&+&\sum_{\mc{S}\subseteq\mc{T}}\left[\Pr(\mb{E}_{\mc{S}})+\Pr(\mb{E}_{w,\mc{S}})\right]\ ,
\end{IEEEeqnarray*}
where $\Pr(\mb{E}_0)$ goes to zero as $n\to\infty$, provided by the code generation and the encoding process. As the next step, we bound the probability:
\begin{IEEEeqnarray*}{rCl}
\Pr(\mb{E}_{\mc{S}}) & \leq & \displaystyle\sum_{\hat{l}_{kb}  \neq l_{kb}\,,\,k\in\mc{S}}\Pr\Big[ \mc{E}_k({w}_b,\hat{l}_{\mc{T}_kb})  \in \nonumber\\
&&  \mc{A}^n_\epsilon[VXX_kX_{\mc{T}_k}\hat{Y}_{\mc{T}_k}Y_k] \,\, \text{ and } \nonumber\\ 
&&{  \mc{E}_k(\hat{l}_{\mc{T}_kb})  \in \mc{A}^n_\epsilon[VX_{\mc{T}_k}X_kY_k] }
\Big]\\
&\leq& \prod_{j\in\mc{S}}\left(2^{n\hat{R}_j}-1\right) 2^{n(\Delta_1 +\Delta_2)}\ ,
\end{IEEEeqnarray*}
where   
\begin{IEEEeqnarray*}{rCl}
\Delta_1 & \triangleq &  H(VXY_k X_kX_{\mc{T}_k}\hat{Y}_{\mc{T}_k}) -\sum_{j\in\mc{S}\cap\mc{T}^{\text{CF}}_k} H(\hat{Y}_j|X_j)
\end{IEEEeqnarray*}

\begin{IEEEeqnarray*}{rCl}
&-& \sum_{j\in\mc{S}\cap\mc{T}^{\text{DF}}_k} H(\hat{Y}_j|X_jV)-  H(VXX_kY_kX_{\mc{T}_k}\hat{Y}_{\mc{S}^c}) \ , \\
\Delta_2 & \triangleq & H(VX_kY_kX_{\mc{T}_k}) -\sum_{j\in\mc{S}\cap\mc{T}^{\text{CF}}_k}H(X_j)\nonumber\\
&-& \sum_{j\in\mc{S}\cap\mc{T}^{\text{DF}}_k} H(X_j|V) -H(VX_kY_kX_{\mc{S}^c}) +\epsilon_1 \ .
\end{IEEEeqnarray*}
To guarantee that the probability $\Pr(\mb{E}_{\mc{S}})$ is arbitrarily small, the following inequality needs to hold:  
\begin{IEEEeqnarray*}{rCl}
&\sum_{j\in\mc{S}\cap\mc{T}^{\text{CF}}_k}&I(\hat{Y}_j;Y_j|X_j)+\sum_{j\in\mc{S}\cap\mc{T}^{\text{DF}}_k}I(\hat{Y}_j;Y_j|X_jV)
+\epsilon_1 \nonumber\\
& &<  \sum_{j\in\mc{S}\cap\mc{T}^{\text{CF}}_k}H(\hat{Y}_j|X_j)+\sum_{j\in\mc{S}\cap\mc{T}^{\text{CF}}_k}H(X_j) \nonumber\\
&&+\sum_{j\in\mc{S}\cap\mc{T}^{\text{DF}}_k}H(\hat{Y}_j|X_jV)+ \sum_{j\in\mc{S}\cap\mc{T}^{\text{DF}}_k} H(X_j|V)\nonumber\\
&& + H(VXX_kY_kX_{\mc{T}_k}\hat{Y}_{\mc{S}^c})+H(VX_kY_kX_{\mc{S}^c}) \nonumber\\
&&-H(VXY_k X_kX_{\mc{T}_k}\hat{Y}_{\mc{T}_k})-H(VX_kY_kX_{\mc{T}_k})
\end{IEEEeqnarray*}
from which we can have that
\begin{IEEEeqnarray}{rCl}
\epsilon_1 & < & \sum_{j\in\mc{S}\cap\mc{T}^{\text{CF}}_k}H(\hat{Y}_j|Y_jX_j)+\sum_{j\in\mc{S}\cap\mc{T}^{\text{DF}}_k}H(\hat{Y}_j|Y_jX_jV)\nonumber\\
\IEEEeqnarraymulticol{3}{l}{+ I(X_{\mc{S}};VX_kX_{\mc{S}^c}Y_k) -H(\hat{Y}_{\mc{S}}|VXX_kX_{\mc{T}_k}\hat{Y}_{\mc{S}^c}Y_k)\ .}\,\,\,\,\,\,\,\,\,\,\label{eq:preceding2}
\end{IEEEeqnarray}
Indeed, \eqref{eq:preceding2} can be further simplified by using the fact that $\hat{Y}_j$ is independent of all the other random variables, given $(X_j,Y_j)$ for $j\in\mc{T}^{\text{CF}}_k$ and $(V,X_j,Y_j)$ for $j\in\mc{T}^{\text{DF}}_k$:
\begin{IEEEeqnarray*}{rCl}
&&\sum_{j\in\mc{S}\cap\mc{T}^{\text{CF}}_k}H(\hat{Y}_j|Y_jX_j)+\sum_{j\in\mc{S}\cap\mc{T}^{\text{DF}}_k} H(\hat{Y}_j|Y_jX_jV)\nonumber\\
\IEEEeqnarraymulticol{3}{r}{+I(X_{\mc{S}};VX_kX_{\mc{S}^c}Y_k) -H(\hat{Y}_{\mc{S}}|VXX_kX_{\mc{T}_k}\hat{Y}_{\mc{S}^c}Y_k)}\nonumber\\
&=&\sum_{j\in\mc{S}} H(\hat{Y}_j|Y_jX_jV)+I(X_{\mc{S}};VX_kX_{\mc{S}^c}Y_k)\nonumber\\
\IEEEeqnarraymulticol{3}{r}{-H(\hat{Y}_{\mc{S}}|VXX_kX_{\mc{T}_k}\hat{Y}_{\mc{S}^c}Y_k)}\\
&=&\sum_{j\in\mc{S}}H(\hat{Y}_j|Y_{\mc{S}}X_{\mc{S}}V)+I(X_{\mc{S}};VX_kX_{\mc{S}^c}Y_k)\nonumber\\
\IEEEeqnarraymulticol{3}{r}{-H(\hat{Y}_{\mc{S}}|VXX_kX_{\mc{T}_k}\hat{Y}_{\mc{S}^c}Y_k)}\\
&=& \sum^{\left|\mc{S}\right|}_{j=1}H(\hat{Y}_{o(j)}|Y_{\mc{S}}X_{\mc{S}}V)
+I(X_{\mc{S}};VX_kX_{\mc{S}^c}Y_k)\nonumber\\
\IEEEeqnarraymulticol{3}{r}{-H(\hat{Y}_{\mc{S}}|VXX_kX_{\mc{T}_k}\hat{Y}_{\mc{S}^c}Y_k)} \\
&=& \sum^{\left|\mc{S}\right|}_{j=1}H(\hat{Y}_{o(j)}|\hat{Y}_{o(1)}\dots  \hat{Y}_{o(j-1)}Y_{\mc{S}}X_{\mc{S}}V)\nonumber\\
&&+I(X_{\mc{S}};VX_kX_{\mc{S}^c}Y_k)-H(\hat{Y}_{\mc{S}}|VXX_kX_{\mc{T}_k}\hat{Y}_{\mc{S}^c}Y_k)\nonumber\\
&=& H(\hat{Y}_{\mc{S}}|Y_{\mc{S}}X_{\mc{S}}V)+I(X_{\mc{S}};VX_kX_{\mc{S}^c}Y_k)\nonumber\\
\IEEEeqnarraymulticol{3}{r}{-H(\hat{Y}_{\mc{S}}|VXX_kX_{\mc{T}_k}\hat{Y}_{\mc{S}^c}Y_k)\ ,}
\end{IEEEeqnarray*}
where $o:[1:\left|\mc{S}\right|] \longmapsto\mc{S}$ is an arbitrary ordering over $\mc{S}$. This manipulation provides us the next condition that must be satisfied:
\begin{IEEEeqnarray}{rCl}
\epsilon_1<&& 	I(X_{\mc{S}};Y_k|VX_kX_{\mc{S}^c})\nonumber\\
&&-I(\hat{Y}_{\mc{S}};Y_{\mc{S}}|VXX_kX_{\mc{T}_k}\hat{Y}_{\mc{S}^c}Y_k)\ .
 \label{Compression-relay-MNNC}
\end{IEEEeqnarray}
However, given the fact that $\mc{T}_k\in\Upsilon_k(\mc{N})$, inequality \eqref{Compression-relay-MNNC} holds for every subset $\mc{S}\subseteq\mc{T}_k$. 
Finally, the probability $\Pr(\mb{E}_{w,\mc{S}})$ can be bounded by following the very same steps as before and thus $\Pr(\mb{E}_{w,\mc{S}})$ goes to zero as $n\to\infty$ provided that
\begin{IEEEeqnarray}{rCl}
R&+&\sum_{j\in\mc{S}\cap\mc{T}^{\text{CF}}_k}I(\hat{Y}_j;Y_j|X_j)+\sum_{j\in\mc{S}\cap\mc{T}^{\text{DF}}_k}I(\hat{Y}_j;Y_j|X_jV)+\epsilon_2 \nonumber\\
&<& \sum_{j\in\mc{S}\cap\mc{T}^{\text{CF}}_k}H(\hat{Y}_j|X_j)+\sum_{j\in\mc{S}\cap\mc{T}^{\text{CF}}_k}H(X_j)\nonumber\\
&+&\sum_{j\in\mc{S}\cap\mc{T}^{\text{DF}}_k}H(\hat{Y}_j|X_jV) +\sum_{j\in\mc{S}\cap\mc{T}^{\text{DF}}_k}H(X_j|V)\nonumber\\
&+& H(X|V)+H(VX_kY_kX_{\mc{T}_k}\hat{Y}_{\mc{S}^c})+H(VX_kY_kX_{\mc{S}^c})\nonumber\\
&&H(VXY_k X_kX_{\mc{T}_k}\hat{Y}_{\mc{T}_k}) -H(VX_kY_kX_{\mc{T}_k})\ .
\end{IEEEeqnarray}
From which we obtain the last condition: 
\begin{IEEEeqnarray}{rCl}
R+\epsilon_2& <& I(X;\hat{Y}_{\mc{S}^c}Y_k|VX_kX_{\mc{T}_k})+I(X_{\mc{S}};Y_k|VX_kX_{\mc{S}^c})  \nonumber\\
\IEEEeqnarraymulticol{3}{c}{-I(\hat{Y}_{\mc{S}};Y_{\mc{S}}|VXX_kX_{\mc{T}_k}\hat{Y}_{\mc{S}^c}Y_k)}\nonumber\\
&=&I(X;\hat{Y}_{\mc{T}_k}Y_k|VX_kX_{\mc{T}_k})+I(X_{\mc{S}};Y_k|VX_kX_{\mc{S}^c}) \nonumber\\
\IEEEeqnarraymulticol{3}{c}{-I(\hat{Y}_{\mc{S}};Y_{\mc{S}}|VX_kX_{\mc{T}_k}\hat{Y}_{\mc{S}^c}Y_k)\ .}
\label{DFdecoding-relay-MNNC}
\end{IEEEeqnarray}
\item Decoding at the destination is done backwardly. After the last block, the decoder jointly searches for the unique indices $\big(\hat{l}_{k(B+2)}\big)_{k\in\mc{T}}$ such that for all $b=[B+3:B+L]$ the following condition holds:
\begin{IEEEeqnarray*}{rCl}
&&\Big(\big(\underline{x}_{k}(\hat{l}_{k(B+2)})\big)_{k\in\mc{T}^{\text{CF}}},\big(\underline{x}_{k}(1,\hat{l}_{k(B+2)})\big)_{k\in\mc{T}^{\text{DF}}},\underline{x}(1,1),\nonumber\\
\IEEEeqnarraymulticol{3}{r}{\underline{v}(1),\underline{y}(b)\Big)\in \mc{A}^n_\epsilon[VXX_{\mc{T}}Y]\ .}
\end{IEEEeqnarray*}
The probability of error goes to zero as $n$ goes to infinity provided that
\begin{IEEEeqnarray*}{rCl}
\sum_{k\in\mc{S}\cap\mc{T}^{\text{CF}}} &I(\hat{Y}_k;Y_k|X_k)+\sum_{k\in\mc{S}\cap\mc{T}^{\text{DF}}} I(\hat{Y}_k;Y_k|X_kV)+\epsilon_2\nonumber\\
&\leq (L-2)I(X_{\mc{S}};VXX_{\mc{S}^c}Y)\ ,
\label{last-compression-MNNC}
\end{IEEEeqnarray*}
for all subsets $\mc{S}\subseteq\mc{T}$.
\item 
After finding the correct index $l_{k(B+2)}$, for each $k\in\mc{T}$, and using the fact that $w_{(B+1)}=1$, the destination decodes jointly the message and all compression indices $(w_b,l_{\mc{T}(b+1)})$, for each block $b=[1:B]$, where we define $l_{\mc{T}b}\triangleq \left(l_{kb}\right)_{k\in\mc{T}}$. Decoding is performed  backwardly with the assumption that $(w_{b+2},l_{\mc{T}(b+2)})$ have been correctly decoded. Let us define the following sequence: 
\begin{IEEEeqnarray*}{rCl}
\mc{E}(\hat{w}_b,\hat{l}_{\mc{T}(b+1)}) &\triangleq & \Big(\underline{x}(\hat{w}_{b},w_{(b+2)}),\underline{v}(\hat{w}_{b}),\underline{y}(b+2),\nonumber\\
\IEEEeqnarraymulticol{3}{c}{\big(\underline{x}_{k}(\hat{l}_{k(b+1)}),\underline{\hat{y}}_{k}(\hat{l}_{k(b+1)},l_{k(b+2)})\big)_{k\in\mc{T}^{\text{CF}}},\big(\underline{x}_{k}(\hat{w}_{b},\hat{l}_{k(b+1)}),}\nonumber \\
\IEEEeqnarraymulticol{3}{c}{{\underline{\hat{y}}_{k}(\hat{w}_{b},\hat{l}_{k(b+1)},l_{k(b+2)})\big)_{k\in\mc{T}^{\text{DF}}}\Big)\ .}}
\end{IEEEeqnarray*}
The destination finds the unique pair of indices $(\hat{w}_b,\hat{l}_{\mc{T}(b+1)})$ such that 
\begin{equation*}
\mc{E}(\hat{w}_b,\hat{l}_{\mc{T}(b+1)})\in \mc{A}^n_\epsilon[VXX_{\mc{T}}\hat{Y}_{\mc{T}}Y]\,.
\label{Destdecoding}
\end{equation*}
For  every $\mc{S}\subseteq\mc{T}$  and $\mc{S}^c\triangleq \mc{T}-\mc{S}$, we consider the error events associated with this step which are given by
\begin{IEEEeqnarray*}{rCl}
\mb{E}_0&:  &\Big\{ \mc{E}({w}_b,{l}_{\mc{T}(b+1)})
 \notin  \mc{A}^n_\epsilon[VXX_{\mc{T}}\hat{Y}_{\mc{T}}Y] \Big\} \ ,\\
\mb{E}_{\mc{S}}&: & \Big\{\mc{E}({w}_b,\hat{l}_{\mc{T}(b+1)}) \in \mc{A}^n_\epsilon[VXX_{\mc{T}}\hat{Y}_{\mc{T}}Y] \text{ for some } \nonumber\\
&& \hat{l}_{kb}\neq l_{kb}\,,\,  k\in\mc{S} \text{ and } \,\, \hat{l}_{kb}=l_{kb}\,,\,  k\in\mc{S}^c\Big\}\ , \\
\mb{E}_{w,\mc{S}}&: &\Big\{\mc{E}(\hat{w}_b,\hat{l}_{\mc{T}(b+1)}) \in \mc{A}^n_\epsilon[VXX_{\mc{T}}\hat{Y}_{\mc{T}}Y] \text{ for some }\nonumber\\
&& \hat{w}_b\neq w_{b}\,,\, \hat{l}_{kb}\neq l_{kb}\,, \, k\in\mc{S}, \,\,  \hat{l}_{kb}=l_{kb}\,, \, k\in\mc{S}^c\Big\} \ , 
\end{IEEEeqnarray*}
where $\mb{E}_{\mc{S}}$ denotes the event  that there exist joint typical sequences for relays in $\mc{S}$ with correct message index but with wrong compression indices while $\mb{E}_{w,\mc{S}}$ denotes the event that there exist joint typical sequences for relays in $\mc{S}$ with both wrong, message and compression indices. Hence, the error probability of this step is bounded by
\begin{IEEEeqnarray*}{rCl}
\Pr\left[(\hat{w}_{b},\hat{l}_{\mc{T}b})\neq ({w}_{b},{l}_{\mc{T}b}) \right]&\leq &\Pr(\mb{E}_0)\nonumber\\
\IEEEeqnarraymulticol{3}{r}{+\sum_{\mc{S}\subseteq\mc{T}}\left[\Pr(\mb{E}_{\mc{S}})+\Pr(\mb{E}_{w,\mc{S}})\right]\ .}
\end{IEEEeqnarray*}
The first probability, on the right-hand side goes to zero as $n\to\infty$. On the other hand, $\Pr(\mb{E}_{\mc{S}})$ goes to zero as $n\to\infty$ provided that
\begin{IEEEeqnarray*}{rCl}
&&\sum_{j\in\mc{S}\cap\mc{T}^{\text{CF}}}I(\hat{Y}_j;Y_j|X_j) +\sum_{j\in\mc{S}\cap\mc{T}^{\text{DF}}}I(\hat{Y}_j;Y_j|X_jV)+\epsilon_3 \nonumber\\
&&<\sum_{j\in\mc{S}\cap\mc{T}^{\text{CF}}}H(\hat{Y}_jX_j)+ \sum_{j\in\mc{S}\cap\mc{T}^{\text{DF}}} H(\hat{Y}_jX_j|V)\nonumber\\
&&+H(VXX_{\mc{S}^c}\hat{Y}_{\mc{S}^c}Y)-H(VXX_{\mc{T}}\hat{Y}_{\mc{T}}Y)\,,
\end{IEEEeqnarray*}
which also reads as:
\begin{IEEEeqnarray}{lCl}
\epsilon_3 & <& \sum_{j\in\mc{S}\cap\mc{T}^{\text{CF}}}H(\hat{Y}_j|Y_jX_j)+ \sum_{j\in\mc{S}\cap\mc{T}^{\text{DF}}}H(\hat{Y}_j|Y_jX_jV)\nonumber\\
\IEEEeqnarraymulticol{3}{r}{+H(XX_{\mc{S}^c}\hat{Y}_{\mc{S}^c}Y|V)-H(XX_{\mc{S}^c}\hat{Y}_{\mc{T}}Y|X_{\mc{S}}V)\ .}\,\,\,\,\,\,
\label{eq:inequalityA1}
\end{IEEEeqnarray}
Indeed, inequality \eqref{eq:inequalityA1} can be further simplified by using the same method as before:
\begin{IEEEeqnarray}{lCl}
&&\sum_{j\in\mc{S}\cap\mc{T}^{\text{CF}}}H(\hat{Y}_j|Y_jX_j)+\sum_{j\in\mc{S}\cap\mc{T}^{\text{DF}}}H(\hat{Y}_j|Y_jX_jV) \nonumber\\
\IEEEeqnarraymulticol{3}{r}{+H(XX_{\mc{S}^c}\hat{Y}_{\mc{S}^c}Y|V)-H(XX_{\mc{S}^c}\hat{Y}_{\mc{T}}Y|X_{\mc{S}}V)}\nonumber\\
\IEEEeqnarraymulticol{3}{r}{= H(\hat{Y}_{\mc{S}}|Y_{\mc{S}}X_{\mc{S}}V) +H(XX_{\mc{S}^c}\hat{Y}_{\mc{S}^c}Y|V)}\nonumber\\
\IEEEeqnarraymulticol{3}{r}{-H(XX_{\mc{S}^c}\hat{Y}_{\mc{T}}Y|X_{\mc{S}}V)\ .}
\end{IEEEeqnarray}
This manipulation yields the following expression: 
\begin{IEEEeqnarray}{rCl}
\epsilon_3 &<& H(\hat{Y}_{\mc{S}}|{Y}_{\mc{S}}X_{\mc{S}}V) +H(XX_{\mc{S}^c}\hat{Y}_{\mc{S}^c}Y|V)\nonumber\\
\IEEEeqnarraymulticol{3}{r}{-H(XX_{\mc{S}^c}\hat{Y}_{\mc{T}}Y|X_{\mc{S}}V)} \nonumber\\
&=&  H(\hat{Y}_{\mc{S}}|{Y}_{\mc{S}}X_{\mc{S}}V)-H(\hat{Y}_{\mc{S}}|XX_{\mc{S}^c}\hat{Y}_{\mc{S}^c}Y X_{\mc{S}}V)\nonumber\\
\IEEEeqnarraymulticol{3}{r}{+H(XX_{\mc{S}^c}\hat{Y}_{\mc{S}^c}Y|V)-H(XX_{\mc{S}^c}\hat{Y}_{\mc{S}^c}Y|X_{\mc{S}}V)}\nonumber\\
&=& I(X_{\mc{S}};XX_{\mc{S}^c}\hat{Y}_{\mc{S}^c}Y|V)\nonumber\\
\IEEEeqnarraymulticol{3}{r}{-I(\hat{Y}_{\mc{S}};{Y}_{\mc{S}}|VXX_{\mc{T}}\hat{Y}_{\mc{S}^c}Y)}\nonumber\\ 
&=& I(X_{\mc{S}};\hat{Y}_{\mc{S}^c}Y|XX_{\mc{S}^c}V)\nonumber\\
\IEEEeqnarraymulticol{3}{r}{-I(\hat{Y}_{\mc{S}};{Y}_{\mc{S}}|VXX_{\mc{T}}\hat{Y}_{\mc{S}^c}Y)\ .}\,\,\,\,
\label{Compression-MNNC}
\end{IEEEeqnarray}
Given the fact that $\mc{T}\in\Upsilon(\mc{N})$,  inequality \eqref{Compression-MNNC} holds for each $\mc{S}\subseteq\mc{T}$.  Finally, $\Pr(\mb{E}_{w,\mc{S}})$ tends to zero as $n$ goes to infinity provided that:
\begin{IEEEeqnarray}{rCl}
R&+&\sum_{j\in\mc{S}\cap\mc{T}^{\text{CF}}}I(\hat{Y}_j;Y_j|X_j)+\sum_{j\in\mc{S}\cap\mc{T}^{\text{DF}}}I(\hat{Y}_j;Y_j|X_jV)+\epsilon_4 \nonumber\\
&<&\sum_{j\in\mc{S}\cap\mc{T}^{\text{CF}}}H(\hat{Y}_jX_j)+ \sum_{j\in\mc{T}^{\text{DF}}}H(\hat{Y}_jX_j|V) + H(VX)\nonumber\\
&+&H(X_{\mc{T}^{\text{CF}}\cap\mc{S}^c}\hat{Y}_{\mc{T}^{\text{CF}}\cap\mc{S}^c}Y)-H(VXX_{\mc{T}}\hat{Y}_{\mc{T}}Y)\ .\,\,\,
\label{eq:inequalityA2}
\end{IEEEeqnarray}

It is worth mentioning here that the right-hand side of \eqref{eq:inequalityA2} is independent of $\mc{S}\cap\mc{T}^{\text{DF}}$ and thus, if we take the set $\mc{S}$ such that $\mc{S}\cap\mc{T}^{\text{DF}}=\mc{T}^{\text{DF}}$, then \eqref{eq:inequalityA2} implies similar inequalities for all other $\mc{S}$ with $\mc{S}\cap\mc{T}^{\text{DF}}\subset\mc{T}^{\text{DF}}$. In fact, we continue the proof based on this choice that leads to

\begin{IEEEeqnarray}{rCl}
R&+&\sum_{j\in\mc{S}\cap\mc{T}^{\text{CF}}}I(\hat{Y}_j;Y_j|X_j)+\sum_{j\in\mc{T}^{\text{DF}}}I(\hat{Y}_j;Y_j|X_jV)+\epsilon_4 \nonumber\\
&<&\sum_{j\in\mc{S}\cap\mc{T}^{\text{CF}}}H(\hat{Y}_jX_j)+ \sum_{j\in\mc{T}^{\text{DF}}}H(\hat{Y}_jX_j|V)+H(VX)\nonumber\\
&+&H(X_{\mc{T}^{\text{CF}}\cap\mc{S}^c}\hat{Y}_{\mc{T}^{\text{CF}}\cap\mc{S}^c}Y)-H(VXX_{\mc{T}}\hat{Y}_{\mc{T}}Y)\,,\,\,\,
\end{IEEEeqnarray}
and thus
\begin{IEEEeqnarray}{rCl}
&&R+\epsilon_4  <  \sum_{j\in\mc{S}\cap\mc{T}^{\text{CF}}}H(\hat{Y}_j|Y_jX_j)\nonumber\\
&+& \sum_{j\in\mc{T}^{\text{DF}}}H(\hat{Y}_j|Y_jX_jV)
+H(X|V)\nonumber\\
&+&H(X_{\mc{T}^{\text{CF}}\cap\mc{S}^c}\hat{Y}_{\mc{T}^{\text{CF}}\cap\mc{S}^c}Y)\nonumber\\
&-&H(XX_{\mc{S}^c\cap\mc{T}^{\text{CF}}}\hat{Y}_{\mc{T}}Y|X_{\mc{S}\cup\mc{T}^{\text{DF}}}V) \nonumber\\
&=&H(\hat{Y}_{\mc{S}\cup\mc{T}^{\text{DF}}}|Y_{\mc{S}\cup\mc{T}^{\text{DF}}}X_{\mc{S}\cup\mc{T}^{\text{DF}}}V)\nonumber\\
&+&H(X|V)+H(X_{\mc{T}^{\text{CF}}\cap\mc{S}^c}\hat{Y}_{\mc{T}^{\text{CF}}\cap\mc{S}^c}Y)\nonumber\\
&-& H(XX_{\mc{S}^c\cap\mc{T}^{\text{CF}}}\hat{Y}_{\mc{T}}Y|X_{\mc{S}\cup\mc{T}^{\text{DF}}}V)\nonumber\\
&=&H(\hat{Y}_{\mc{S}\cup\mc{T}^{\text{DF}}}|Y_{\mc{S}\cup\mc{T}^{\text{DF}}}X_{\mc{S}\cup\mc{T}^{\text{DF}}}V)\nonumber\\
&-&H(\hat{Y}_{\mc{S}\cup\mc{T}^{\text{DF}}}|X_{\mc{S}\cup\mc{T}^{\text{DF}}}VXX_{\mc{S}^c\cap\mc{T}^{\text{CF}}}\hat{Y}_{\mc{T}^{\text{CF}}\cap\mc{S}^c}Y)\nonumber\\
&+&H(X_{\mc{T}^{\text{CF}}\cap\mc{S}^c}\hat{Y}_{\mc{T}^{\text{CF}}\cap\mc{S}^c}Y)\nonumber\\
&-&H(X_{\mc{T}^{\text{CF}}\cap\mc{S}^c}\hat{Y}_{\mc{T}^{\text{CF}}\cap\mc{S}^c}Y|XX_{\mc{S}\cup\mc{T}^{\text{DF}}}V)\nonumber\\
&=&I(VXX_{\mc{S}\cup\mc{T}^{\text{DF}}};\hat{Y}_{\mc{T}^{\text{CF}}\cap\mc{S}^c}Y|X_{\mc{T}^{\text{CF}}\cap\mc{S}^c})\nonumber\\
&-&I(\hat{Y}_{\mc{S}\cup\mc{T}^{\text{DF}}};{Y}_{\mc{S}\cup\mc{T}^{\text{DF}}}|X_{\mc{S}\cup\mc{T}^{\text{DF}}}VXX_{\mc{S}^c\cap\mc{T}^{\text{CF}}}\hat{Y}_{\mc{T}^{\text{CF}}\cap\mc{S}^c}Y)\nonumber\\
&=&I(XX_{\mc{S}\cup\mc{T}^{\text{DF}}};\hat{Y}_{\mc{T}^{\text{CF}}\cap\mc{S}^c}Y|X_{\mc{T}^{\text{CF}}\cap\mc{S}^c})\nonumber\\
&-&I(\hat{Y}_{\mc{S}\cup\mc{T}^{\text{DF}}};{Y}_{\mc{S}\cup\mc{T}^{\text{DF}}}|XX_{\mc{T}}\hat{Y}_{\mc{T}^{\text{CF}}\cap\mc{S}^c}Y) \label{DFdecoding-step-1} \\
&=&I(XX_{\mc{S}};\hat{Y}_{\mc{S}^c}Y|X_{\mc{S}^c})-I(\hat{Y}_{\mc{S}};{Y}_{\mc{S}}|XX_{\mc{T}}\hat{Y}_{\mc{S}^c}Y)\,\,\,\,\,\,
\label{DFdecoding-MNNC}
\end{IEEEeqnarray}
where step \eqref{DFdecoding-step-1} comes from the fact $\mc{S}$ has been selected satisfying $\mc{T}^{\text{DF}}\subseteq\mc{S}$. 

By choosing finite $L$ but large enough, inequalities \eqref{Compression-relay-MNNC}, \eqref{Compression-MNNC}, \eqref{DFdecoding-MNNC} and \eqref{DFdecoding-relay-MNNC} prove Theorem \ref{thm:MNNC}, where the final rate is achieved by letting $(B,n)$ tend to infinity. At the end, a time sharing random variable $Q$ can be added over all expressions, concluding the proof. 
\end{enumerate}
\section{Proof of Theorem \ref{thm:NC-MNNC}}\label{proof:NC-MNNC}

Let $p$ be an arbitrary probability distribution satisfying the conditions in expression \eqref{NC-MNNC}, and let two sets $\mc{V}$ and $\mc{T}$ maximizing the right-hand side of \eqref{NC-MNNC}. Let  $\mc{M}_n$ be a set of messages of size $2^{nR}$ with an index $W$ to be transmitted. Transmission is done in $B+L$ blocks, each of them of length $n$, and decoding at the destination is done backwardly. At the last $L-1$ blocks, the last compression index is first decoded and then all compression indices and transmitted messages are jointly decoded. By the vector notation $\underline{x}_{\mc{S}}$ we denote the collection $\left(\underline{x}_{i}\right)_{i\in\mc{S}}$.

\subsection*{Code generation:}
\begin{enumerate} [(i)]
\item Randomly and independently generate $2^{nR}$ sequences $\underline{x}_{\mc{V}^c}$ drawn i.i.d. from
\begin{equation*}
P_{X_{\mc{V}^c}}^{n}(\underline{x}_{\mc{V}^c})=\prod\limits_{j=1}^n P_{X_{\mc{V}^c}}\big(x_{{\mc{V}^c}j}\big)\ .
\end{equation*}
Index them as $\underline{x}_{\mc{V}^c}(w_0)$ with index $w_0\in \big[1,2^{nR}\big]$. This step will provide $\left|\mc{V}^c\right|$ different codebooks $\left(\underline{x}_{k}(w_0), w_0\in[1,2^{nR}]\right)$ for each $k\in\mc{V}^c$, every having $2^{nR}$ codewords. However, the codewords in each codebook corresponding to an index are jointly generated based on $P_{X_{\mc{V}^c}}^{n}$ and  in general are not independent.
\item For each $\underline{x}_{\mc{V}^c}(w_0)$, randomly and conditionally independently generate $2^{nR}$ sequences $\underline{x}$ drawn i.i.d. from  
\begin{equation*}
P_{X|X_{\mc{V}^c}}^n(\underline{x}\vert \underline{x}_{\mc{V}^c}(w_0))=\prod\limits_{j=1}^n P_{X|X_{\mc{V}^c}}\big(x_{j}|x_{{\mc{V}^c}j}(w_0)\big)\ .
\end{equation*}
Index them as $\underline{x}(w_0,w)$ with  $w\in \big[1,2^{nR}\big]$.
\item For each $k\in\mc{T}$, randomly and independently generate  $2^{n\hat{R}_k}$ sequences $\underline{x}_k$ drawn i.i.d. from 
\begin{equation*}
P_{X_k}^n(\underline{x}_k)=\prod\limits_{j=1}^n P_{X_k}\big(x_{kj}\big)\ . 
\end{equation*}
Index them as $\underline{x}_k(l_{0k})$ with $l_{0k}\in \big[1,2^{n\hat{R}_k}\big]$ for $\hat{R}_k\triangleq I(Y_k;\hat{Y}_k|X_k)+\epsilon$.
\item For each $k\in\mc{T}$ and each $\underline{x}_k(l_{0k})$, randomly and conditionally independently generate $2^{n\hat{R}_k}$ sequences  $\underline{\hat{y}}_k$ each with probability 
\begin{equation*}
P_{\hat{Y}_k|X_k}^n (\underline{\hat{y}}_k\vert \underline{x}_k(l_{0k}))= \prod\limits_{j=1}^n P_{\hat{Y}_k\vert X_k}\big(\hat{y}_{kj}\vert x_{kj}(l_{0k})\big)\ .
\end{equation*}
Index them as $\underline{\hat{y}}_k(l_{0k},l_k)$ with $l_k\in \big[1,2^{n\hat{R}_k}\big]$.

\item Provide the corresponding codebooks to the relays, the encoder and the decoder ends. \\
\end{enumerate}
\subsection*{Encoding:} 
\begin{enumerate}[(i)]
\item
In every block $i=[1: B]$, the source sends $w_{i}$ based on $\underline{x}\big(w_{(i-1)},w_i\big)$. Moreover, for blocks $i=[B+1:B+L]$, the source sends the dummy message $w_{i}=1$ known to all nodes.
\item
For every block $i=[1:B+L]$, and each $k\in\mc{V}^c$, relay $k$ knows $w_{(i-1)}$ by assumption and $w_0=1$, so it sends $\underline{x}_k\big(w_{(i-1)}\big)$. 
\item 
For each $i=[1: B+1]$, each $k\in\mc{T}$, relay $k$ after receiving $\underline{y}_k(i)$, searches for at least one index $l_{ki}$ with $l_{k0}=1$ such that 
\begin{equation*}
\big(\underline{x}_{k}(l_{k(i-1)}),\underline{y}_{k}(i),\underline{\hat{y}}_{k}(l_{k(i-1)},l_{ki})\big)\in \mc{A}^n_\epsilon[X_kY_k\hat{Y}_k]\ .
\end{equation*}
The probability of finding such $l_{ki}$ goes to one as $n$ goes to infinity provided by  our choice of the rate $\hat{R}_k$. 
\item
For $i=[1:B+1]$ and $k\in\mc{T}$, relay $k$ knows the index ${l}_{k(i-1)}$ from the previous block and it sends $\underline{x}_k(l_{k(i-1)})$. Moreover, relay $k$ repeats $l_{k(B+1)}$ for all $i=[B+2: B+L]$.
\end{enumerate}


\subsection*{Decoding:} 
\begin{enumerate}[(i)]
\item After transmission of the block  $i=[1:B]$ is accomplished and for each $k\in\mc{V}^c$, relay $k$ decodes the message of block $i$ with the assumption that all messages up to block $i-1$ have been correctly decoded. Since the relay $k$ knows the message $w_{(i-1)}$ and so $\underline{x}_k\big(w_{(i-1)}\big)$, it also knows because of the code generation all others codewords $\underline{x}_{k'}\big(w_{(i-1)}\big)$ for $k'\in\mc{V}^c$. Relay $k$ searches for the unique index $\hat{w}_{i}\in\mc{M}_n$ such that:
\begin{IEEEeqnarray*}{lCl}
&& \left(\underline{x}\big(w_{(i-1)},\hat{w}_{i}\big),\underline{x}_{\mc{V}^c}\big(w_{(i-1)}\big),\underline{y}_{k}(i)\right)\in \mc{A}^n_\epsilon[XX_{\mc{V}^c}Y_1]\ .
\end{IEEEeqnarray*}
By following similar arguments to those in~\cite{Cover1979}, the probability of error goes to zero as $n$ goes to infinity provided that:
\begin{equation}
R  < I(X;Y_k\vert X_{\mc{V}^c})\ . 
\label{DFdecoding-NC-MNNC}
\end{equation}
\item Decoding at destination is done backwardly. First, the destination decodes all last compression indices sent by the relays in $\mc{T}$ then it waits until the last block  to jointly search for unique indices $\big(\hat{l}_{k(B+1)}\big)_{k\in\mc{T}}$ such that for all $b=[B+2:B+L]$ the following condition holds:
\begin{IEEEeqnarray*}{lCl}
&&\left(\big(\underline{x}_{k}(\hat{l}_{k(B+1)})\big)_{k\in\mc{T}},\underline{x}(1,1),\underline{x}_{\mc{V}^c}(1),\underline{y}(b)\right)\nonumber\\
\IEEEeqnarraymulticol{3}{r}{ \in \mc{A}^n_\epsilon[XX_{\mc{T}}X_{\mc{V}^c}Y]\ .}
\end{IEEEeqnarray*}
Let us define the following events that can cause an error in the previous decoding step:
\begin{IEEEeqnarray*}{lCl}
 \mb{E}_0 &: &\Big\{ \left(\left(\underline{x}_{k}({l}_{k(B+1)})\right)_{k\in\mc{T}},\underline{x}(1,1),\underline{x}_{\mc{V}^c}(1),\underline{y}(b)\right)\nonumber\\
\IEEEeqnarraymulticol{3}{c}{ \notin \mc{A}^n_\epsilon[XX_{\mc{T}}X_{\mc{V}^c}Y]\Big\}\ , }\nonumber\\
\mb{E}_{\mc{S}} &: & \Big\{\Big(
\big(\underline{x}_{k}(\hat{l}_{k(B+1)})\big)_{k\in\mc{S}},\big(\underline{x}_{k}(l_{k(B+1)})\big)_{k\in\mc{S}^c},\\
&&\underline{x}(1,1),\underline{x}_{\mc{V}^c}(1),\underline{y}(b)\Big)\in \mc{A}^n_\epsilon[XX_{\mc{T}\cup\mc{V}^c}Y]\text{ for some }\nonumber \\
\IEEEeqnarraymulticol{3}{c}{\hat{l}_{k(B+1)}\neq l_{k(B+1)}\text{ and }\forall \, b=[B+2:B+L] \Big\}\ .} 
\end{IEEEeqnarray*}
The last event is the event that there exist joint typical sequences that have correct indices for the relays in $\mc{S}^c=\mc{T}-\mc{S}$ and wrong indices for the relays in $\mc{S}$. The probability of error is bounded as follows
\begin{equation*}
 \Pr\big(\hat{l}_{B+1}\neq l_{B+1}\big)\leq \Pr(\mb{E}_0)+\sum_{\mc{S}\subseteq\mc{T}}\Pr({\mb{E}_{\mc{S}}})\ .
\end{equation*}
The first probability on the right-hand side goes to zero as $n\to\infty$,  and the second probability can be bounded as follows: 
\begin{IEEEeqnarray*}{lCl}
\Pr({\mb{E}_{\mc{S}}}) &\leq & \sum\limits_{\hat{l}_{k(B+1)}\neq l_{k(B+1)}\,,\,k\in\mc{S}}\Pr\Big[\bigcap\limits_{b=[B+2:B+L]} \nonumber\\
\IEEEeqnarraymulticol{3}{l}{\Big\{\big(\underline{x}_{k}(\hat{l}_{k(B+1)})\big)_{k\in\mc{S}},\left(\underline{x}_{k}({l}_{k(B+1)})\right)_{k\in\mc{S}^c},} \nonumber\\ 
\IEEEeqnarraymulticol{3}{l}{\underline{x}(1,1),\underline{x}_{\mc{V}^c}(1),\underline{y}(b)\Big\}\in \mc{A}^n_\epsilon[XX_{\mc{T}}X_{\mc{V}^c}Y]\Big]} \nonumber\\
\IEEEeqnarraymulticol{3}{l}{\leq \prod_{k\in\mc{S}}\left(2^{n\hat{R}_k}-1\right)\left[2^{-n(I(X_{\mc{S}};XX_{\mc{S}^c\cup\mc{V}^c}Y)-\epsilon_1)}\right]^{L-1}\ . }
\end{IEEEeqnarray*}
This probability goes to zero as $n$ goes to infinity provided that for all $\mc{S}\subseteq\mc{T}$:
\begin{equation*} 
\sum_{k\in\mc{S}} I(\hat{Y}_k;Y_k|X_k)+\epsilon_2\leq (L-1)I(X_{\mc{S}};XX_{\mc{S}^c\cup\mc{V}^c}Y)\ .
\label{Compression-1-NC-MNNC}
\end{equation*}
\item 
After finding correct index $l_{k(B+1)}$ for all $k\in\mc{T}$ and since $w_{(B+1)}=1$, the destination decodes jointly the message and all the compression indices $(w_b,l_{\mc{T}b})$ for each $b=[1:B]$  where $l_{\mc{T}b}=\left(l_{kb}\right)_{k\in\mc{T}}$. Decoding is performed  backwardly with the assumption that $(w_{b+1},l_{\mc{T}(b+1)})$ have been correctly decoded. Define the following event:
\begin{IEEEeqnarray*}{lCl}
\mc{E}(\hat{w}_b,\hat{l}_{\mc{T}b})& \triangleq & \Big\{ \Big(\underline{x}(\hat{w}_{b},w_{(b+1)}),\underline{x}_{\mc{V}^c}(\hat{w}_{b}),\underline{y}(b+1),\nonumber\\
\IEEEeqnarraymulticol{3}{r}{\big(\underline{x}_{k}(\hat{l}_{kb}),\underline{\hat{y}}_{k}(\hat{l}_{kb},l_{k(b+1)})\big)_{k\in\mc{T}}\Big)\Big\} \ .}
\end{IEEEeqnarray*}
The destination finds the unique pair  $(\hat{w}_b,\hat{l}_{\mc{T}b})$ such that 
\begin{equation*}
\mc{E}(\hat{w}_b,\hat{l}_{\mc{T}b}) \in \mc{A}^n_\epsilon[XX_{\mc{T}\cup\mc{V}^c}\hat{Y}_{\mc{T}}Y]\ .
\end{equation*}
Consider the following error events associated with this step ($\mc{S}\subseteq\mc{T},\mc{S}^c=\mc{T}-\mc{S}$):
\begin{IEEEeqnarray*}{rCl}
\mb{E}_0&: & \Big\{\mc{E}({w}_b,{l}_{\mc{T}b})
\notin \mc{A}^n_\epsilon[XX_{\mc{T}\cup\mc{V}^c}\hat{Y}_{\mc{T}}Y]\Big\}\ ,\nonumber \\
\mb{E}_{\mc{S}} &: & 
\Big\{\mc{E}({w}_b,\hat{l}_{\mc{T}b})
\in \mc{A}^n_\epsilon[XX_{\mc{T}\cup\mc{V}^c}\hat{Y}_{\mc{T}}Y] \text{ for some }\nonumber\\
\IEEEeqnarraymulticol{3}{r}{\hat{l}_{kb}\neq l_{kb}, k\in\mc{S} \text{ and } \hat{l}_{kb}= l_{kb}\,,\, k\in\mc{S}^c\Big\}\ ,} \\
\mb{E}_{w,\mc{S}} &:& 
\Big\{ \mc{E}(\hat{w}_b,\hat{l}_{\mc{T}b}) \in \mc{A}^n_\epsilon[XX_{\mc{T}\cup\mc{V}^c}\hat{Y}_{\mc{T}}Y]\text{ for some }\nonumber\\
\IEEEeqnarraymulticol{3}{r}{\hat{w}\neq w_{b}\,,\, \hat{l}_{kb}\neq l_{kb},  \,\,{k\in\mc{S}\text{ and }\hat{l}_{kb}= l_{kb}\,, \,k\in\mc{S}^c\Big\}\ .}}
\end{IEEEeqnarray*}
The event $\mc{E}_{\mc{S}}$ represents the event that there exist jointly typical sequences with correct message index but wrong compression indices for the relays in $\mc{S}$. On the other hand $\mb{E}_{w,\mc{S}}$ is the event that there is jointly typical codes with wrong message index and wrong compression indices for the relays in $\mc{S}$. The error probability of this step is bounded by
\begin{IEEEeqnarray*}{rCl}
\Pr\Big((\hat{w}_{b},\hat{l}_{\mc{T}b})\neq ({w}_{b},{l}_{\mc{T}b}) \Big)&\leq& \Pr(\mb{E}_0)\nonumber\\
\IEEEeqnarraymulticol{3}{r}{+ \sum_{\mc{S}\subseteq\mc{T}}\left[\Pr(\mb{E}_{\mc{S}})+\Pr(\mb{E}_{w,\mc{S}})\right]\ .}
\end{IEEEeqnarray*}
The first term on the right-hand side and $\Pr(\mb{E}_{\mc{S}})$ go to zero as $n\to\infty$ provided that
\begin{IEEEeqnarray*}{rCl}
\sum_{k\in\mc{S}}I(\hat{Y}_k;Y_k|X_k)+\epsilon_3  &<& \sum_{k\in\mc{S}}H(\hat{Y}_k|X_k) \nonumber\\
\IEEEeqnarraymulticol{3}{r}{+H(XX_{\mc{V}^c\cup\mc{S}^c}\hat{Y}_{\mc{S}^c}Y)-H(XX_{\mc{V}^c\cup\mc{S}^c}\hat{Y}_{\mc{T}}Y|X_{\mc{S}})\ ,}
\end{IEEEeqnarray*}
which can be written as:
\begin{IEEEeqnarray*}{rCl}
\epsilon_3 &<& \sum_{k\in\mc{S}}H(\hat{Y}_k|Y_kX_k)+H(XX_{\mc{V}^c\cup\mc{S}^c}\hat{Y}_{\mc{S}^c}Y)\nonumber\\
\IEEEeqnarraymulticol{3}{r}{-H(XX_{\mc{V}^c\cup\mc{S}^c}\hat{Y}_{\mc{T}}Y|X_{\mc{S}})\ .}
\end{IEEEeqnarray*}
The preceding inequality can be simplified by using the fact that $\hat{Y}_k$ is independent of the other random variables given $(X_k,Y_k)$. The standard manipulation introduced before gives us the following:
\begin{IEEEeqnarray}{rCl}
\epsilon_3 & < &  I(XX_{\mc{V}^c\cup\mc{S}^c}\hat{Y}_{\mc{S}^c}Y;X_{\mc{S}})-I(\hat{Y}_{\mc{S}};Y_{\mc{S}}|XX_{\mc{V}^c\cup\mc{T}}\hat{Y}_{\mc{S}^c}Y)\nonumber\\
&=& I(\hat{Y}_{\mc{S}^c}Y;X_{\mc{S}}|XX_{\mc{V}^c\cup\mc{S}^c})\nonumber\\
\IEEEeqnarraymulticol{3}{c}{ -I(\hat{Y}_{\mc{S}};Y_{\mc{S}}|XX_{\mc{V}^c\cup\mc{T}}\hat{Y}_{\mc{S}^c}Y)\,.} \label{Compression-2-NC-MNNC}
\end{IEEEeqnarray}
Given the fact that $\mc{T}\in\Upsilon(\mc{V})$, the last inequality holds for each $\mc{S}\subseteq\mc{T}$.  As the next step, we bound the probability $\Pr(\mb{E}_{w,\mc{S}})$ as follows:
\begin{IEEEeqnarray}{rCl}
\Pr(\mb{E}_{w,\mc{S}})& \leq & \displaystyle\sum_{\hat{l}_{kb}\neq l_{kb}\,,\,\hat{w}\neq w_b}\Pr\Big[\Big(\underline{x}(\hat{w}_{b},w_{(b+1)}),\nonumber\\
\IEEEeqnarraymulticol{3}{c}{\underline{x}_{\mc{V}^c}(\hat{w}_{b}),\underline{y}(b+1),\big(\underline{x}_{k}(\hat{l}_{kb}),\underline{\hat{y}}_{k}(\hat{l}_{kb},l_{k(b+1)})\big)_{k\in\mc{S}},}\nonumber\\
\IEEEeqnarraymulticol{3}{c}{ \big(\underline{x}_{k}({l}_{kb}),\underline{\hat{y}}_{k}({l}_{kb},l_{k(b+1)})\big)_{k\in\mc{S}^c}\Big)}\nonumber\\
\IEEEeqnarraymulticol{3}{r}{\in \mc{A}^n_\epsilon[XX_{\mc{T}\cup\mc{V}^c}\hat{Y}_{\mc{T}}Y] \Big]}\nonumber\\
&\leq & \left(2^{nR}-1\right)\prod_{k\in\mc{S}}\left(2^{n\hat{R}_k}-1\right) 2^{n\Delta_3}\ ,\,\,\,\,\,\,\,\,\,
\end{IEEEeqnarray} 
where 
\begin{IEEEeqnarray}{rCl}
\Delta_3 & \triangleq& H(XX_{\mc{V}^c\cup\mc{S}^c}Y\hat{Y}_{\mc{T}}|X_{\mc{S}})-H(XX_{\mc{V}^c}) \nonumber\\
&-& H(Y\hat{Y}_{\mc{S}^c}X_{\mc{S}^c})-\sum_{k\in\mc{S}}H(\hat{Y}_k|X_k)+\epsilon_4) \ .\,\,\,\,\,\,\,\,\,
\end{IEEEeqnarray} 
From the following inequality, the last probability also tends to zero as $n\to\infty$,
\begin{IEEEeqnarray}{rCl}
R+\sum_{k\in\mc{S}}I(\hat{Y}_k;{Y}_k|X_k))+\epsilon_5 & \leq	&H(XX_{\mc{V}^c\cup\mc{S}^c}Y\hat{Y}_{\mc{T}}|X_{\mc{S}})\nonumber\\
+H(XX_{\mc{V}^c})+ H(Y\hat{Y}_{\mc{S}^c}X_{\mc{S}^c})&+&\sum_{k\in\mc{S}}H(\hat{Y}_k|X_k)\ . \,\,\,\,\,\,\,\,\,
\label{CFdecoding-NC-MNNC}
\end{IEEEeqnarray}
Inequality \eqref{CFdecoding-NC-MNNC} is then simplified and reads as
\begin{IEEEeqnarray}{rCl}
R+\epsilon_5 & \leq & -H(XX_{\mc{V}^c\cup\mc{S}^c}Y\hat{Y}_{\mc{T}}|X_{\mc{S}})+H(XX_{\mc{V}^c})\nonumber\\
\IEEEeqnarraymulticol{3}{r}{+H(Y\hat{Y}_{\mc{S}^c}X_{\mc{S}^c})+\sum_{k\in\mc{S}}H(\hat{Y}_k|{Y}_kX_k) }\nonumber\\
& = & -H(XX_{\mc{V}^c\cup\mc{S}^c}Y\hat{Y}_{\mc{T}}|X_{\mc{S}})+H(XX_{\mc{V}^c}) \nonumber\\ 
\IEEEeqnarraymulticol{3}{r}{+H(Y\hat{Y}_{\mc{S}^c}X_{\mc{S}^c})+H(\hat{Y}_{\mc{S}}|{Y}_{\mc{S}}X_{\mc{S}})}\nonumber \\
& = & I(XX_{\mc{V}^c}X_{\mc{S}};Y\hat{Y}_{\mc{S}^c}|X_{\mc{S}^c})\nonumber\\
\IEEEeqnarraymulticol{3}{r}{-H(\hat{Y}_{\mc{S}}|XX_{\mc{V}^c\cup\mc{T}}Y\hat{Y}_{\mc{S}^c})+H(\hat{Y}_{\mc{S}}|{Y}_{\mc{S}}X_{\mc{S}})} \nonumber\\
& = & I(XX_{\mc{V}^c}X_{\mc{S}};Y\hat{Y}_{\mc{S}^c}|X_{\mc{S}^c})\nonumber\\
\IEEEeqnarraymulticol{3}{r}{-I(\hat{Y}_{\mc{S}};{Y}_{\mc{S}}|XX_{\mc{V}^c\cup\mc{T}}Y\hat{Y}_{\mc{S}^c})\ .}\,\,\,
\label{CFdecoding-2-NC-MNNC}
\end{IEEEeqnarray}
By choosing finite $L$ but large enough, inequalities \eqref{DFdecoding-NC-MNNC} and \eqref{CFdecoding-2-NC-MNNC} prove Theorem \ref{thm:NC-MNNC}, where the rate is achieved by letting $(B,n)$ tend to infinity. At the end, a time sharing random variable $Q$ can be added. 
\end{enumerate}

\section{Proof of Theorem \ref{thm:LMNNC}}\label{proof:LMNNC}
For the rest of this section we shall assume that  relays are ordered as $(\mc{L}_1,\dots,\mc{L}_T)$. The transmission is done in $B+T+L-1$ blocks where at  the end of block $B$ the source finished of transmitting messages. At  the end of this block, the relays at the layer $\mc{L}_t$ had only repeated messages until the block $B-T-2+t$, and therefore it continues transmission until the end of block $B+T+2-t$. Therefore, we need $B+T+1$ blocks for all relays to transmit all messages. From $B+T+2$ up to block $B+T+L-1$, namely for $L-2$ blocks, CF relays repeat their compression index of their last block $B+T+2$. In the Table~\ref{tab:3}, we illustrate this procedure by considering a two hop network with a single CF relay. As it can be seen, the codewords denoted by $\underline{x}_2$ of the second layer among DF relays are superimposed on those denoted by $\underline{x}_1$ of the first layer among DF relays. Notice that the number of DF relays can exceed the number of hops. The relays in the second 
layer starts to decode at the end of block $B+1$ and those in the first layer at the end of block $B+2$.

\begin{table*}[th] 
	\caption{Transmission schedule for ``Layered Mixed Noisy Network Coding'' (LMNNC).}
	\label{tab:3}
	 \centering
\begin{tabular}{| l | l | l | l  | l |  l | } 
\hline
$b=1$ & $b=2$ & $b=3$ & $b=B+2$ & $b=B+3$   & $b=B+L+1$ \\ 
\hline
$\underline{v}_1(1)$ & $\underline{v}_1(1)$ & $\underline{v}_1(1)$ &  $\underline{v}_1(w_{B-1})$ & $\underline{v}_1(w_B)$   & $\underline{v}_1(1)$ 
\\
\hline
$\underline{x}_1(1)$ & $\underline{x}_1(1)$ & $\underline{x}_1(1)$ &  $\underline{x}_1(w_{B-1})$ & $\underline{x}_1(w_B)$   & $\underline{x}_1(1)$ 
\\
\hline
$\underline{v}_2(1,1)$ & $\underline{v}_2(1,1)$ & $\underline{v}_2(1,w_1)$ &  $\underline{v}_2(w_{B-1},w_B)$ & $\underline{v}_2(w_B,1)$    & $\underline{v}_2(1,1)$ 
\\
\hline
$\underline{x}_2(1,1)$ & $\underline{x}_2(1,1)$ & $\underline{x}_2(1,w_{1})$ &  $\underline{x}_2(w_{B-1},w_B)$ & $\underline{x}_2(w_B,1)$    & $\underline{x}_2(1,1)$ 
\\
\hline
$\underline{x}(1,1,w_1)$ & $\underline{x}(1,1,w_2)$ & $\underline{x}(1,w_1,w_2)$ &   $\underline{x}(w_{B-1},w_B,1)$ & $\underline{x}(w_B,1,1)$    & $\underline{x}(1,1,1)$ 
\\
\hline
$\underline{x}_3(1)$ & $\underline{x}_3(l_{31})$ & $\underline{x}_3(l_{32})$ &  $\underline{x}_3(l_{3(B+1)})$ & $\underline{x}_3(l_{3(B+2)})$    & $\underline{x}_3(l_{3(B+3)})$ 
\\
\hline
\hline
$\underline{\hat{y}}_{3}(1,l_{31})$  & $\underline{\hat{y}}_{3}(l_{31},l_{32})$ & $\underline{\hat{y}}_{3}(l_{32},l_{33})$ &    $\underline{\hat{y}}_{3}(l_{3(B+1)},l_{3(B+2)})$ & $\underline{\hat{y}}_{3}(l_{3(B+2)},l_{3(B+3)})$  &  $\star$
\\
\hline
\hline
$\underline{y}_{1}(1)$  & $\underline{y}_{1}(2)$ & $\underline{y}_{1}(3)$ &  $\underline{y}_{1}(B+2)$ & $\underline{y}_{1}(B+3)$   & $\underline{y}_{1}(B+L+1)$
\\
\hline 
$\underline{y}_{2}(1)$  & $\underline{y}_{2}(2)$ & $\underline{y}_{2}(3)$ &  $\underline{y}_{2}(B+2)$ & $\underline{y}_{2}(B+3)$   & $\underline{y}_{2}(B+L+1)$
\\
\hline 
$\underline{y}_{3}(1)$  & $\underline{y}_{3}(2)$ & $\underline{y}_{3}(3)$ &  $\underline{y}_{3}(B+2)$ & $\underline{y}_{3}(B+3)$    & $\underline{y}_{3}(B+L+1)$
\\
\hline 
$\underline{y}(1)$ & $\underline{y}(2)$ & $\underline{y}(3)$ &   $\underline{y}(B+2)$ & $\underline{y}(B+3)$   & $\underline{y}(B+L)$
\\
\hline
\end{tabular} 
\end{table*}

At the of block $B+3$, the first relay sends $w_1$ while the second relay transmits both messages $(w_1,w_2)$. When the number of hops is larger than two, Table~\ref{tab:4} shows the order of message transmission across the different  hops. Briefly, the DF relays with higher layers start to decode sooner and transmit more messages in each transmission round.



\subsection*{Code generation:}
\begin{enumerate} [(i)]
\item Randomly and independently generate $2^{nR}$ sequences $\underline{v}_1$ drawn i.i.d. from
\begin{equation}
P_{V_1}^{n}(\underline{v}_1)=\prod\limits_{j=1}^n P_{V_1}(v_{1j})\ .
\end{equation}
Index them as $\underline{v}(w_1)$ with index $w_1\in \left[1,2^{nR}\right]$. 
\item For all hops $t=[2:T]$, randomly and independently generate $2^{nR}$ sequences $\underline{v}_t$ based on $\underline{v}_1(w_1),\dots \underline{v}_{t-1}(w_1,\dots, w_{t-1})$ i.i.d. from
\begin{IEEEeqnarray*}{rCl}
&&P_{V_t|V_{\leq t-1}}^{n}\big(\underline{v}_t|\underline{v}_1(w_1),\dots \underline{v}_{t-1}(w_1,\dots,w_{t-1})\big)\nonumber\\
\IEEEeqnarraymulticol{3}{c}{=\prod\limits_{j=1}^n P_{V_t| V_{\leq t-1}}\big(v_{tj}|{v}_{1j}(w_1),\dots}\nonumber\\
\IEEEeqnarraymulticol{3}{r}{\dots {v}_{(t-1)j}(w_1,\dots,w_{t-1})\big)\ .}
\end{IEEEeqnarray*}
Index them as $\underline{v}_t(w_1,\dots,w_t)$ with index $w_t\in \left[1,2^{nR}\right]$. 
\item For each $k\in\mc{L}_t$ and each tuple of codewords $\big(\underline{v}_1(w_1),\dots,\underline{v}_t(w_1,\dots,w_t)\big)$, randomly and independently generate $\underline{x}_k$ drawn i.i.d. from 
\begin{IEEEeqnarray*}{rCl}
&& P_{X_k\vert V_{\leq t}}^n\big(\underline{x}_k|\underline{v}_1(w_1),\dots,\underline{v}_t(w_1,\dots,w_t)\big)\nonumber\\
\IEEEeqnarraymulticol{3}{r}{=\prod\limits_{j=1}^n P_{X_k\vert  V_{\leq t}}\big(x_{kj}\vert\underline{v}_{1j}(w_1),\dots,\underline{v}_{tj}(w_1,\dots,w_t)\big)\ . }
\end{IEEEeqnarray*}
This is the codebook for DF relays in $\mc{L}_t$.
\item For each $k\in\mc{V}$, randomly and independently generate  $2^{n\hat{R}_k}$ sequences $\underline{x}_k$ drawn i.i.d. from 
\begin{equation*}
P_{X_k}^n(\underline{x}_k)=\prod\limits_{j=1}^n P_{X_k}(x_{kj})\ .
\end{equation*}
Index them as $\underline{x}_k(l_{0k})$, where $l_{0k}\in \big[1,2^{n\hat{R}_k}\big]$ for $\hat{R}_k \triangleq I(Y_k;\hat{Y}_k|X_k)+\epsilon$ with $k\in \mc{V}$. This is the codebook for CF relays.
\item For  each tuple $\big(\underline{v}_1(w_1),\dots,\underline{v}_T(w_1,\dots,w_T)\big)$, randomly and conditionally independently generate $2^{nR}$ sequences $\underline{x}$ drawn i.i.d. from 
\begin{IEEEeqnarray*}{rCl}
&& P_{X|V_{\leq T}}^n\big(\underline{x}\vert \underline{v}_1(w_1),\dots,\underline{v}_T(w_1,\dots,w_T) \big)\nonumber\\
\IEEEeqnarraymulticol{3}{r}{=\prod\limits_{j=1}^n P_{X|V_{\leq T}}\big(x_{j}|\underline{v}_{1j}(w_1),\dots,\underline{v}_{Tj}(w_1,\dots,w_T)\big)\ .}
\end{IEEEeqnarray*}
Index them as $\underline{x}(w_1,\dots,w_T,w)$, where $w\in \big[1,2^{nR}\big]$. This is the source codebook.
\item For each $k\in\mc{V}$ and each $\underline{x}_k(l_{0k})$, randomly and conditionally independently generate $2^{n\hat{R}_k}$ sequences  $\underline{\hat{y}}_k$ each with probability 
\begin{equation*}
P_{\hat{Y}_k|X_k}^n (\underline{\hat{y}}_k\vert \underline{x}_k(l_{0k}))= \prod\limits_{j=1}^n P_{\hat{Y}_k\vert X_k}(\hat{y}_{kj}\vert x_{kj}(l_{0k}))\ .
\end{equation*}
Index them as  $\underline{\hat{y}}_k(l_{0k},l_k)$, where $l_k\in \big[1,2^{n\hat{R}_k}\big]$ with $k\in \mc{V}$. This is the cookbook for the descriptions sent by CF relays.
\item Provide the corresponding codebooks to all relays, the encoder and the decoder ends. 
\end{enumerate}
\begin{table}[ht] 
	\caption{Coding for DF relays at each layer.}
	\label{tab:4}
	 \centering
\begin{tabular}{| l | l | l |} 
\hline
Layers & Block $b=T+2$ & Block $b$  
\\
\hline
$\mc{L}_1$ & $\underline{x}_1(w_1)$ & $\underline{x}_1(w_{b-T-1})$ 
\\
\hline
$\mc{L}_2$ & $\underline{x}_2(w_1,w_2)$ & $\underline{x}_2(w_{b-T-1},w_{b-T})$
\\
\hline
$\vdots$ & $\vdots$ & $\vdots$ 
\\
\hline
$\mc{L}_t$ & $\underline{x}_t(w_1,\dots,w_t)$ & $\underline{x}_t(w_{b-T-1},\dots,w_{b-T-2+t})$  
\\
\hline
$\vdots$ & $\vdots$  & $\vdots$ 
\\
\hline
$\mc{L}_T$ & $\underline{x}_T(w_1,\dots,w_T)$ & $\underline{x}_T(w_{b-T-1},\dots,w_{b-2})$  
\\
\hline
\hline
Source & $\underline{x}(w_1,\dots,w_T,w_{T+2})$ & $\underline{x}(w_{b-T-1},\dots,w_{b-2},w_{b})$ 
\\
\hline
\end{tabular} 
\end{table}

\subsection*{Encoding:} 
\begin{enumerate}[(i)]
\item
In every block $i\in[1: B]$, the source sends $w_{i}$ using $\underline{x}\big(w_{(i-T-1)},\dots,w_{(i-2)},w_i\big)$, where we have defined $w_{i}=1$ for $i\leq 0$. Moreover, for blocks $i=[B+1:B+T+L-1]$, the source sends the dummy message $w_{i}=1$ known to all users.
\item
For $i\in[1:B+T+2-t]$ and each $k\in\mc{L}_t$, the $k$-th relay knows $\big(w_{i-T-1},\dots,w_{i-T-2+t}\big)$ in block $i$, and it sends $\underline{x}_k(w_{i-T-1},\dots,w_{(i-T-2+t)})$.  Notice that at the of block $B+T+1$, all DF relays are transmitting $\underline{x}_k(w_{B},1,\dots,1)$.
\item 
For $i=[1: B+T+1]$, and each $k\in\mc{V}$, the $k$-th relay --after receiving $\underline{y}_k(i)$-- searches for at least one index $l_{ki}$ with $l_{k0}=1$ such that 
\begin{equation*}
\big(\underline{x}_{k}(l_{k(i-1)}),\underline{y}_{k}(i),\underline{\hat{y}}_{k}(l_{k(i-1)},l_{ki})\big)\in \mc{A}^n_\epsilon[X_kY_k\hat{Y}_k]\ .
\end{equation*}
The probability of finding such $l_{ki}$ goes to one as $n$ goes to infinity due to our adequate choice of the rate  $\hat{R}_k$ for compression. 
\item
For $i=[1:B+T+1]$ and $k\in\mc{V}$, relay $k$ knows from the previous block ${l}_{k(i-1)}$ and it sends $\underline{x}_k(l_{k(i-1)})$. Moreover, $k$-th relay repeats $l_{k(B+T+1)}$ for $i=[B+T+2: B+T+L-1]$, i.e., for $L-2$ blocks.
\end{enumerate}

\subsection*{Decoding:} 
\begin{enumerate}[(i)]

\item The DF relays in $\mc{L}_T$ are first to start decoding messages. After the transmission of the block  $i+1=[2:B+1]$ and for each $k\in\mc{L}_T$, the  $k$-th relay decodes the message $w_{i}$ and the compression index $l_{\mc{T}_ki}$ of block $i$ for all relays in $\mc{T}_k$, with the assumption that all messages and compression indices up to block $i-1$ have been  correctly decoded. We emphasize that in this case there are only CF relays in $\mc{T}_k$. The $k$-th relay knows the tuple of message $\big(w_{(i-T-2)},\dots,w_{(i-1)}\big)$, and so it knows all codewords indexed by them. Let us define:
\begin{IEEEeqnarray*}{rCl}
\mc{E}_k\big(\hat{w}_b,\hat{l}_{\mc{T}_kb},b\big) &\triangleq&\Big\{ \Big(\underline{x}({w}_{(b-T-1)},\dots,{w}_{(b-2)},\hat{w}_{b}), \nonumber\\
\IEEEeqnarraymulticol{3}{r}{\big(\underline{x}_{i}({l}_{i(b-1)}),\underline{\hat{y}}_{i}({l}_{i(b-1)},\hat{l}_{ib})\big)_{i\in\mc{T}_k},\underline{y}_{k}(b)}\nonumber\\
\IEEEeqnarraymulticol{3}{r}{\big(\underline{v}_t({w}_{(b-T-1)},\dots,{w}_{(b-T-2+t)}),}\nonumber\\
\IEEEeqnarraymulticol{3}{r}{\underline{x}_{\mc{L}_t}({w}_{(b-T-1)},\dots,{w}_{(b-T-2+t)})\big)_{t\in[1:T]}\Big)\Big\}\ ,}\nonumber\\
\mc{E}_k\big(\hat{l}_{\mc{T}_kb},b+1\big) &\triangleq & \Big\{\Big(\big(\underline{v}_t({w}_{(b-T)},\dots,{w}_{(b-T-1+t)}),\nonumber\\
\IEEEeqnarraymulticol{3}{r}{\underline{x}_{\mc{L}_t}({w}_{(b-T)},\dots,{w}_{(b-T-1+t)})\big)_{t\in[1:T]}, }\nonumber\\
\IEEEeqnarraymulticol{3}{r}{\underline{y}_{k}(b+1),\big(\underline{x}_{i}(\hat{l}_{ib})\big)_{i\in\mc{T}_k}\Big)\Big\}\ .}
\end{IEEEeqnarray*}
Therefore, the $k$-th relay, by  looking at two consecutive blocks $(b,b+1)$, searches for the unique indices $(\hat{w}_{b},\hat{l}_{\mc{T}_kb})$ such that:
\begin{IEEEeqnarray*}{rCl}
\mc{E}_k(\hat{w}_b,\hat{l}_{\mc{T}_kb},b) & \in& \mc{A}^n_\epsilon[V_{\leq T}XX_{\mc{V}^c}X_{\mc{T}_k}\hat{Y}_{\mc{T}_k}Y_k] \,\, \text{ and }  \,\,\nonumber\\
\mc{E}_k(\hat{l}_{\mc{T}_kb},b+1)  &\in &\mc{A}^n_\epsilon[V_{\leq T}X_{\mc{V}^c}X_{\mc{T}_k}Y_k]\ .
\label{DFrelaydecoding-NCLMNNC}
\end{IEEEeqnarray*}
We define the following error events: 
\begin{IEEEeqnarray*}{rCl}
\mb{E}_0&: &\Big\{\mc{E}_k({w}_b,{l}_{\mc{T}_kb},b)  \notin \mc{A}^n_\epsilon[V_{\leq T}XX_{\mc{V}^c}X_{\mc{T}_k}\hat{Y}_{\mc{T}_k}Y_k] \nonumber\\
\IEEEeqnarraymulticol{3}{r}{\,\, \text{ or } \,\,\mc{E}_k({l}_{\mc{T}_kb},b+1)  \notin \mc{A}^n_\epsilon[V_{\leq T}X_{\mc{V}^c}X_{\mc{T}_k}Y_k] \Big\} \ ,} \,\,\,\,\,\, \nonumber\\
\mb{E}_{\mc{S}}& : & \Big\{\mc{E}_k({w}_b,\hat{l}_{\mc{T}_kb},b)  \in \mc{A}^n_\epsilon[V_{\leq T}XX_{\mc{V}^c}X_{\mc{T}_k}\hat{Y}_{\mc{T}_k}Y_k] \nonumber\\
\IEEEeqnarraymulticol{3}{r}{ \text{ and } \,\,\mc{E}_k(\hat{l}_{\mc{T}_kb},b+1)  \in  \mc{A}^n_\epsilon[V_{\leq T}X_{\mc{V}^c}X_{\mc{T}_k}Y_k]\text{ for some } }\nonumber \\
\IEEEeqnarraymulticol{3}{c}{ \hat{l}_{kb}\neq l_{kb}\,, \,k\in\mc{S}\text{ and }\,\,\hat{l}_{kb}=l_{kb}\,,\,  k\in\mc{S}^c\Big\}  \ ,}\nonumber \\
\mb{E}_{w,\mc{S}}& : & \Big\{ \mc{E}_k(\hat{w}_b,\hat{l}_{\mc{T}_kb},b)  \in \mc{A}^n_\epsilon[V_{\leq T}XX_{\mc{V}^c}X_{\mc{T}_k}\hat{Y}_{\mc{T}_k}Y_k] \nonumber\\
\IEEEeqnarraymulticol{3}{c}{ \text{ and } \,\,\mc{E}_k(\hat{l}_{\mc{T}_kb},b+1)  \in  \mc{A}^n_\epsilon[V_{\leq T}X_{\mc{V}^c}X_{\mc{T}_k}Y_k] }\nonumber \\
\IEEEeqnarraymulticol{3}{c}{ \text{ for some }\,\, \hat{w}_b\neq w_{b}\,, \,\hat{l}_{kb}\neq l_{kb}\,, \, k\in\mc{S}}\nonumber\\
\IEEEeqnarraymulticol{3}{r}{ \,\, \text{ and } \,\,\hat{l}_{kb}=l_{kb}\,, \, k\in\mc{S}^c\Big\} \ .\,\,\,\,\,\,\,\,}
\end{IEEEeqnarray*}
The probability of error for these relays can be bounded as follows:
\begin{IEEEeqnarray*}{rCl}
\Pr\Big((\hat{w}_{b},\hat{l}_{\mc{T}_kb})\neq ({w}_{b},{l}_{\mc{T}_kb}) \Big) &\leq& \Pr(\mb{E}_0)\nonumber\\
\IEEEeqnarraymulticol{3}{r}{+\sum_{\mc{S}\subseteq\mc{T}}\left[\Pr(\mb{E}_{\mc{S}})+\Pr(\mb{E}_{w,\mc{S}})\right]
\ .}
\end{IEEEeqnarray*}
The probability $\Pr(\mb{E}_0)$ goes to zero as $n\to\infty$ given the code generation and the encoding process. We bound the other probabilities using the same technique as we already did in Appendix~\ref{proof:MNNC}. First of all, the probability $\Pr(\mb{E}_{\mc{S}}) $ is bounded as:
\begin{IEEEeqnarray*}{rCl}
\Pr(\mb{E}_{\mc{S}}) &\leq&  \prod_{k\in\mc{S}}\left(2^{n\hat{R}_k}-1\right) 2^{n(\Delta_1 +\Delta_2)}\ ,
\end{IEEEeqnarray*}
where
\begin{IEEEeqnarray*}{rCl}
\Delta_1 & \triangleq &  H(V_{\leq T}XX_{\mc{V}^c}X_{\mc{T}_k}\hat{Y}_{\mc{T}_k}Y_k) -\sum_{j\in\mc{S}} H(\hat{Y}_j|X_j) \nonumber\\
\IEEEeqnarraymulticol{3}{r}{-  H(V_{\leq T}XX_{\mc{V}^c}X_{\mc{T}_k}\hat{Y}_{\mc{S}^c}Y_k) \ , }\nonumber\\
\Delta_2 & \triangleq & H(V_{\leq T}X_{\mc{V}^c}X_{\mc{T}_k}Y_k) -\sum_{j\in\mc{S}}H(X_j)\nonumber\\
\IEEEeqnarraymulticol{3}{r}{ -H(V_{\leq T}X_{\mc{V}^c}X_{{\mc{S}^c}}Y_k)  +\epsilon_1 \ .}
\end{IEEEeqnarray*}
The probability $\Pr(\mb{E}_{\mc{S}}) $ goes to zero as $n$ goes to infinity if the exponent of the right hand side is also negative which yields:
\begin{IEEEeqnarray}{rCl}
\sum_{j\in\mc{S}}I(\hat{Y}_j;Y_j|X_j)+\epsilon_1&<& \sum_{j\in\mc{S}}H(\hat{Y}_j|X_j)+\sum_{j\in\mc{S}}H(X_j) \nonumber\\
&+& H(V_{\leq T}XX_{\mc{V}^c}X_{\mc{T}_k}\hat{Y}_{\mc{S}^c}Y_k)\nonumber\\
&+&H(V_{\leq T}X_{\mc{V}^c}X_{{\mc{S}^c}}Y_k) \nonumber\\
&-& H(V_{\leq T}XX_{\mc{V}^c}X_{\mc{T}_k}\hat{Y}_{\mc{T}_k}Y_k)\nonumber\\
&-&H(V_{\leq T}X_{\mc{V}^c}X_{\mc{T}_k}Y_k) \ .\nonumber
\end{IEEEeqnarray}
After simplification of both sides, in order to guarantee that the probability $\Pr(\mb{E}_{\mc{S}})$ is arbitrarily small, we need to add the next constraint: 
\begin{IEEEeqnarray*}{rCl}
\epsilon_1& <& I(X_{\mc{S}};Y_k|V_{\leq T}X_{\mc{V}^c\cup\mc{S}^c})\nonumber\\
&-&I(\hat{Y}_{\mc{S}};Y_{\mc{S}}|V_{\leq T}XX_{\mc{V}^c}X_{\mc{T}_k}\hat{Y}_{\mc{S}^c}Y_k)\ .
 \label{Compression-relay-LMNNC}
\end{IEEEeqnarray*}
However, given the fact that $\mc{T}_k\in\Upsilon_k(\mc{N})$, inequality \eqref{Compression-relay-LMNNC} holds for every subset $\mc{S}\subseteq\mc{T}_k$. 

As the next step, the probability $\Pr(\mb{E}_{w,\mc{S}})$ can be bounded by following the very same steps as before and thus $\Pr(\mb{E}_{w,\mc{S}})$ goes to zero as $n\to\infty$ provided that
\begin{IEEEeqnarray*}{rCl}
R+\sum_{j\in\mc{S}}I(\hat{Y}_j;Y_j|X_j)+\epsilon_2 &< & \sum_{j\in\mc{S}}H(\hat{Y}_j|X_j)\nonumber\\
\IEEEeqnarraymulticol{3}{l}{+\sum_{j\in\mc{S}}H(X_j)+ H(X|V_{\leq T}X_{\mc{V}^c})}\nonumber\\
\IEEEeqnarraymulticol{3}{l}{+H(V_{\leq T}X_{\mc{V}^c}X_{\mc{T}_k}\hat{Y}_{\mc{S}^c}Y_k)+H(V_{\leq T}X_{\mc{V}^c}X_{{\mc{S}^c}}Y_k)} \nonumber\\
\IEEEeqnarraymulticol{3}{l}{-H(V_{\leq T}XX_{\mc{V}^c}X_{\mc{T}_k}\hat{Y}_{\mc{T}_k}Y_k)-H(V_{\leq T}X_{\mc{V}^c}X_{\mc{T}_k}Y_k)\ .} \,\,\,\,\,\,\,\,
\end{IEEEeqnarray*}
From this, we obtain the following condition: 
\begin{IEEEeqnarray}{rCl}
&&R+\epsilon_2 < I(X;\hat{Y}_{\mc{S}^c}Y_k|V_{\leq T}X_{\mc{V}^c}X_{\mc{T}_k})\nonumber\\
\IEEEeqnarraymulticol{3}{l}{+I(X_{\mc{S}};Y_k|V_{\leq T}X_{\mc{V}^c\cup\mc{S}^c}) } \nonumber\\
\IEEEeqnarraymulticol{3}{l}{-I(\hat{Y}_{\mc{S}};Y_{\mc{S}}|V_{\leq T}XX_{\mc{V}^c}X_{\mc{T}_k}\hat{Y}_{\mc{S}^c}Y_k)}\nonumber\\
&=&I(X;\hat{Y}_{\mc{T}_k}Y_k|V_{\leq T}X_{\mc{V}^c}X_{\mc{T}_k})+I(X_{\mc{S}};Y_k|V_{\leq T}X_{\mc{V}^c\cup\mc{S}^c})  \nonumber\\
&&{-I(\hat{Y}_{\mc{S}};Y_{\mc{S}}|V_{\leq T}X_{\mc{V}^c}X_{\mc{T}_k}\hat{Y}_{\mc{S}^c}Y_k)}\ .
\label{DFdecoding-relay-LMNNC}
\end{IEEEeqnarray}
\item Decoding at DF relays in $\mc{L}_t$ is done as follows. As we already mentioned, these relays start to decode the fresh source message after all relays have decoded the same messages corresponding to the higher layers, namely $\{\mc{L}_{t+1},\dots,\mc{L}_T\}$. Therefore, those relays in higher layers act as relay for the relays in $\mc{L}_t$. Whereas the  lower layers cannot help the relays in $\mc{L}_t$, having decoded only the previous source messages. Indeed, they act as side information in the decoding process. 

After the transmission of the block  $i+1+T-t\in\{2+T-t,\dots,B+1+T-t\}$ and for each $k\in\mc{L}_t$, the $k$-th relay  decodes the message $w_i$ and the compression index $l_{\mc{T}_ki}$, i.e., the compression indices  for block $i$ of all relays in $\mc{T}_k$, with the assumption that all messages and compression indices up to block $i-1$ have been correctly decoded. To this end, this relay uses the blocks $[b:b+t]$. Let us define the sequences:
\begin{IEEEeqnarray*}{rcl}
&&\mc{E}_k \big(\hat{w}_b,\hat{l}_{\mc{T}_kb},b\big)\triangleq \Big\{\Big(\underline{x}({w}_{(b-T-1)},\dots,{w}_{(b-2)},\hat{w}_{b}), \nonumber\\
\IEEEeqnarraymulticol{3}{c}{\big(\underline{x}_{i}({l}_{i(b-1)}),\underline{\hat{y}}_{i}({l}_{i(b-1)},\hat{l}_{ib})\big)_{i\in\mc{T}_k},\underline{y}_{k}(b)}\nonumber\\
\IEEEeqnarraymulticol{3}{c}{\big(\underline{v}_t({w}_{(b-T-1)},\dots,{w}_{(b-T-2+t)}),}\nonumber\\
\IEEEeqnarraymulticol{3}{c}{\underline{x}_{\mc{L}_t}({w}_{(b-T-1)},\dots,{w}_{(b-T-2+t)})\big)_{t\in[1:T]}\Big) \Big\}\ ,}\nonumber\\
&&\mc{E}_k\big(\hat{l}_{\mc{T}_kb},b+1\big)\triangleq \Big\{\Big(\big(\underline{v}_t({w}_{(b-T)},\dots,{w}_{(b-T-1+t)}),\nonumber\\
\IEEEeqnarraymulticol{3}{c}{\underline{x}_{\mc{L}_t}({w}_{(b-T)},\dots,{w}_{(b-T-1+t)})\big)_{t\in[1:T]},}\nonumber\\
\IEEEeqnarraymulticol{3}{c}{\underline{y}_{k}(b+1),\big(\underline{x}_{i}(\hat{l}_{ib})\big)_{i\in\mc{T}_k}\Big)\Big\}\ ,}\nonumber\\
&&\mc{E}_k \left(\hat{w}_b,b+2\right) \triangleq \Big\{ \Big(\big(\underline{v}_t({w}_{(b-T+1)},\dots,{w}_{(b-T+t)}),\nonumber\\
\IEEEeqnarraymulticol{3}{c}{\underline{x}_{\mc{L}_t}({w}_{(b-T+1)},\dots,{w}_{(b-T+t)})\big)_{t\in[1:T-1]},}\nonumber\\
\IEEEeqnarraymulticol{3}{c}{\big(\underline{v}_T({w}_{(b-T+1)},\dots,\hat{w}_{b}),}\nonumber\\
\IEEEeqnarraymulticol{3}{r}{\underline{x}_{\mc{L}_{T}}({w}_{(b-T+1)},\dots,\hat{w}_{b})\big),\underline{y}_{k}(b+2)\Big)\Big\}\ ,}\nonumber\\
&& \mc{E}_k \left(\hat{w}_b,b+j\right)  \triangleq \Big\{ \Big(\big(\underline{v}_t({w}_{(b-T+j-1)},\dots,{w}_{(b-T-2+j+t)}),\nonumber\\
\IEEEeqnarraymulticol{3}{c}{{\underline{x}_{\mc{L}_t}({w}_{(b-T+j-1)},\dots,{w}_{(b-T-2+j+t)})\big)_{t\in[1:T-j+1]},}}
\nonumber\\
\IEEEeqnarraymulticol{3}{c}{\big(\underline{v}_{T-j+2}({w}_{(b-T+j-1)},\dots,\hat{w}_{b}),}\nonumber\\
\IEEEeqnarraymulticol{3}{r}{\underline{x}_{\mc{L}_{T-j+2}}({w}_{(b-T+j-1)},\dots,\hat{w}_{b})\big),\underline{y}_{k}(b+j)\Big)\Big\}\ ,}\\
&&\mc{E}_k \left(\hat{w}_b,b+1+T-t\right) \triangleq \Big(\big(\underline{v}_t({w}_{(b-t)},\dots,{w}_{(b-1)})\nonumber\\
\IEEEeqnarraymulticol{3}{c}{,\underline{x}_{\mc{L}_t}({w}_{(b-t)},\dots,{w}_{(b-1)})\big)_{t\in[1:t]},}
\nonumber\\
\IEEEeqnarraymulticol{3}{c}{\big(\underline{v}_{t+1}({w}_{(b-t)},\dots,\hat{w}_{b}),}\nonumber\\
\IEEEeqnarraymulticol{3}{c}{\underline{x}_{\mc{L}_{t+1}}({w}_{(b-t)},\dots,\hat{w}_{b})\big),\underline{y}_{k}(b+1+T-t)\Big)\ .}
\end{IEEEeqnarray*}

In this case, by  looking at $t+1$ consecutive blocks $(b,\dots,b+t)$, the $k$-th relay finds the unique pair of indices $(\hat{w}_{b},\hat{l}_{\mc{T}_kb})$ such that:
\begin{IEEEeqnarray}{rCl}
\mc{E}_k(\hat{w}_b,\hat{l}_{\mc{T}_kb},b)  &\in& \mc{A}^n_\epsilon[V_{\leq T}XX_{\mc{V}^c}X_{\mc{T}_k}\hat{Y}_{\mc{T}_k}Y_k] \text{ and } \nonumber\\
 \mc{E}_k(\hat{l}_{\mc{T}_kb},b+1)  &\in& \mc{A}^n_\epsilon[V_{\leq T}X_{\mc{V}^c}X_{\mc{T}_k}Y_k]\ .\nonumber\\
 \mc{E}_k(\hat{w}_{b},b+j)  &\in& \mc{A}^n_\epsilon[V_{\leq T-j+2}X_{\mc{L}_{\leq T-j+2}} Y_k] \nonumber \\
 &&\text{ for }  j=[2:T-t+1]\ .
\label{DFrelaydecodingII-PC-LMNNC}
\end{IEEEeqnarray}

We define the following error events:
\begin{IEEEeqnarray}{rCl}
\mb{E}_0&: &\Big\{\mc{E}_k({w}_b,{l}_{\mc{T}_kb},b)  \notin \mc{A}^n_\epsilon[V_{\leq T}XX_{\mc{V}^c}X_{\mc{T}_k}\hat{Y}_{\mc{T}_k}Y_k] \nonumber\\
&&\text{ or } \,\,\mc{E}_k({l}_{\mc{T}_kb},b+1)  \notin \mc{A}^n_\epsilon[V_{\leq T}X_{\mc{V}^c}X_{\mc{T}_k}Y_k]   \nonumber \\
&&\text{ or } \mc{E}_k({w}_{b},b+j)  \notin \mc{A}^n_\epsilon[V_{\leq T-j+2}X_{\mc{L}_{\leq T-j+2}} Y_k]\nonumber\\
&&\text{ for }  j=[2:T-t+1]\Big\}\ .\nonumber\\
\mb{E}_{\mc{S}}& : & \Big\{\mc{E}_k({w}_b,\hat{l}_{\mc{T}_kb},b)  \in \mc{A}^n_\epsilon[V_{\leq T}XX_{\mc{V}^c}X_{\mc{T}_k}\hat{Y}_{\mc{T}_k}Y_k]  \nonumber\\
&&\text{ and } \,\,\mc{E}_k(\hat{l}_{\mc{T}_kb},b+1)  \in  \mc{A}^n_\epsilon[V_{\leq T}X_{\mc{V}^c}X_{\mc{T}_k}Y_k] \nonumber \\
&&\text{ for some }\,\, \hat{l}_{kb}\neq l_{kb}\,, \,k\in\mc{S}\,\, \nonumber\\
&&\text{ and }\,\,\hat{l}_{kb}=l_{kb}\,,\,  k\in\mc{S}^c  \Big\}\ , 
\end{IEEEeqnarray}
\begin{IEEEeqnarray}{rCl}
\mb{E}_{w,\mc{S}}& : &  \Big\{\mc{E}_k(\hat{w}_b,\hat{l}_{\mc{T}_kb},b)  \in \mc{A}^n_\epsilon[V_{\leq T}XX_{\mc{V}^c}X_{\mc{T}_k}\hat{Y}_{\mc{T}_k}Y_k] \nonumber\\
&& \text{ and } \,\,\mc{E}_k(\hat{l}_{\mc{T}_kb},b+1)  \in  \mc{A}^n_\epsilon[V_{\leq T}X_{\mc{V}^c}X_{\mc{T}_k}Y_k] \nonumber \\
&&\text{ and } \mc{E}_k(\hat{w}_{b},b+j)  \in \mc{A}^n_\epsilon[V_{\leq T-j+2}X_{\mc{L}_{\leq T-j+2}} Y_k] \nonumber\\
&&\text{ for }  j=[2:T-t+1]\nonumber \\
&&\text{ for some }\,\, \hat{w}_b\neq w_{b}\,, \,\hat{l}_{kb}\neq l_{kb}\,, \, k\in\mc{S}\,\,\ \nonumber\\
&&\text{ and } \,\,\hat{l}_{kb}=l_{kb}\,, \, k\in\mc{S}^c\Big\} \ .
\end{IEEEeqnarray}
 
Hence, the error probability can be bounded as follows:
\begin{IEEEeqnarray*}{rCl}
\Pr\Big((\hat{w}_{b},\hat{l}_{\mc{T}_kb})&\neq& ({w}_{b},{l}_{\mc{T}_kb}) \Big)\leq \Pr(\mb{E}_0)\nonumber\\
\IEEEeqnarraymulticol{3}{c}{+\sum_{\mc{S}\subseteq\mc{T}}\left[\Pr(\mb{E}_{\mc{S}})+\Pr(\mb{E}_{w,\mc{S}})\right]\ .}
\end{IEEEeqnarray*}
It is easy to check that the probability $\Pr(\mb{E}_0)$ goes to zero as $n\to\infty$. As the next step, the probability $\Pr(\mb{E}_{\mc{S}})$ goes to zero with the exact same condition as~\eqref{Compression-relay-LMNNC}, namely:
\begin{IEEEeqnarray*}{rCl}
\epsilon_1 & < & I(X_{\mc{S}};Y_k|V_{\leq T}X_{\mc{V}^c\cup\mc{S}^c})\nonumber\\
&-&I(\hat{Y}_{\mc{S}};Y_{\mc{S}}|V_{\leq T}XX_{\mc{V}^c}X_{\mc{T}_k}\hat{Y}_{\mc{S}^c}Y_k)\ .
\end{IEEEeqnarray*}
The probability $\Pr(\mb{E}_{w,\mc{S}})$ can be bounded by following the very same steps as before and thus $\Pr(\mb{E}_{w,\mc{S}})$ goes to zero as $n\to\infty$ provided that
\begin{IEEEeqnarray*}{rCl}
R&+&\sum_{j\in\mc{S}}I(\hat{Y}_j;Y_j|X_j)+\epsilon_2 < \sum_{j\in\mc{S}}H(\hat{Y}_j|X_j)\nonumber\\
&+&\sum_{j\in\mc{S}}H(X_j)+ H(X|V_{\leq T}X_{\mc{V}^c})\nonumber\\
&+&H(V_{\leq T}X_{\mc{V}^c}X_{\mc{T}_k}\hat{Y}_{\mc{S}^c}Y_k)+H(V_{\leq T}X_{\mc{V}^c}X_{{\mc{S}^c}}Y_k) \nonumber\\
&-&{H(V_{\leq T}XX_{\mc{V}^c}X_{\mc{T}_k}\hat{Y}_{\mc{T}_k}Y_k)-H(V_{\leq T}X_{\mc{V}^c}X_{\mc{T}_k}Y_k) }\nonumber\\
&+&\sum_{j=2}^{T-t+1}I(V_{T-j+2}X_{\mc{L}_{T-j+2}};Y_k|V_{\leq T-j+1}X_{\mc{L}_{\leq T-j+1}})\ ,
\end{IEEEeqnarray*}
which is simplified to its equivalent condition given by 
\begin{IEEEeqnarray}{rCl}
&&R+\epsilon_2 < I(X;\hat{Y}_{\mc{S}^c}Y_k|V_{\leq T}X_{\mc{V}^c}X_{\mc{T}_k})\nonumber\\
\IEEEeqnarraymulticol{3}{l}{+I(X_{\mc{S}};Y_k|V_{\leq T}X_{\mc{V}^c}X_{\mc{S}^c}) -I(V_{>t}X_{\mc{L}_{>t}};Y_k|V_{\leq t}X_{\mc{L}_{\leq t}})}\nonumber\\
\IEEEeqnarraymulticol{3}{c}{-I(\hat{Y}_{\mc{S}};Y_{\mc{S}}|V_{\leq T}XX_{\mc{V}^c}X_{\mc{T}_k}\hat{Y}_{\mc{S}^c}Y_k)} \nonumber\\
&&=I(X;\hat{Y}_{\mc{T}_k}Y_k|V_{\leq T}X_{\mc{V}^c}X_{\mc{T}_k})+I(X_{\mc{S}};Y_k|V_{\leq T}X_{\mc{V}^c}X_{\mc{S}^c}) \nonumber\\
\IEEEeqnarraymulticol{3}{c}{-I(V_{>t}X_{\mc{L}_{>t}};Y_k|V_{\leq t}X_{\mc{L}_{\leq t}})} \nonumber\\
\IEEEeqnarraymulticol{3}{c}{-I(\hat{Y}_{\mc{S}};Y_{\mc{S}}|V_{\leq T}X_{\mc{V}^c}X_{\mc{T}_k}\hat{Y}_{\mc{S}^c}Y_k)}\nonumber\\
&&=I(XV_{>t}X_{\mc{L}_{>t}};Y_k|V_{\leq t}X_{\mc{L}_{\leq t}}) \nonumber\\
\IEEEeqnarraymulticol{3}{c}{+I(X_{\mc{S}};Y_k|V_{\leq T}X_{\mc{V}^c}X_{\mc{S}^c})}\nonumber\\
\IEEEeqnarraymulticol{3}{c}{  -I(\hat{Y}_{\mc{T}_k};Y_{\mc{S}}|V_{\leq T}X_{\mc{V}^c}X_{\mc{T}_k}Y_k)}\ .
\label{DFdecoding-relay-LMNNC-II}
\end{IEEEeqnarray}
%

\item Decoding at the destination is done backwardly and it first starts to decode the last compression index, namely the decoder jointly searches for the indices $\big(\hat{l}_{k(B+T+1)}\big)_{k\in\mc{T}}$ such that for all $b=[B+T+2:B+T+L-1]$ the following condition holds:
\begin{IEEEeqnarray*}{rCl}
\Big(\big(\underline{x}_{k}(\hat{l}_{k(B+T+1)}&&\big)_{k\in\mc{T}},\big(\underline{v}_t(1,\dots,1),\nonumber\\
\IEEEeqnarraymulticol{3}{r}{ \underline{x}_{\mc{L}_t}(1,\dots,1)\big)_{t\in[1:T]},\underline{x}(1,\dots,1),\underline{y}(b)\Big)}\nonumber \\
\IEEEeqnarraymulticol{3}{r}{ \in \mc{A}^n_\epsilon[V_{\leq T}X_{\mc{V}^c}X_{\mc{T}}Y]\ .}
\end{IEEEeqnarray*}
The probability of error is calculated similarly to previous theorems and it goes to zero as $n$ goes to infinity provided by
\begin{IEEEeqnarray*}{rCl}
\sum_{k\in\mc{S}\cap\mc{T}} I(\hat{Y}_k;Y_k|X_k)&+&\epsilon_2 \leq  \nonumber\\
\IEEEeqnarraymulticol{3}{r}{(L-2)I(X_{\mc{S}};V_{\leq T}XX_{\mc{V}^c\cup\mc{S}^c}Y)\ ,}
\label{last-compression-LMNNC}
\end{IEEEeqnarray*}
for all subsets $\mc{S}\subseteq\mc{T}$.
 
\item 
After finding the correct index $l_{k(B+T+1)}$, for each $k\in\mc{T}$, the destination starts decoding from the block $B+T+1$ backward. It decodes jointly the message and all compression indices $(w_b,l_{\mc{T}(b+T)})$, for each block $b=[1:B]$. The message $w_b$ is decoded at the block $b+T+1$ and it is assumed that $(w_{b+1},\dots,w_{b+T+1},l_{\mc{T}(b+T+1)})$ have been correctly decoded. Let us define the next event: 
\begin{IEEEeqnarray*}{rCl}
\mc{E}(\hat{w}_b,\hat{l}_{\mc{T}(b+1)}) &\triangleq& \Big(\underline{x}(\hat{w}_{b},\dots,w_{b+T-1},w_{(b+T+1)}),\nonumber\\
\IEEEeqnarraymulticol{3}{c}{ \big(\underline{x}_{k}(\hat{l}_{k(b+1)}),\underline{\hat{y}}_{k}(\hat{l}_{k(b+1)},l_{k(b+2)})\big)_{k\in\mc{T}},}\nonumber \\
\IEEEeqnarraymulticol{3}{c}{ \Big(\underline{v}_{t}(\hat{w}_{b},\dots,w_{b+t-1}), \underline{x}_{\mc{L}_t}(\hat{w}_{b},\dots,w_{b+t-1})\Big)_{t\in[1:T]},}\nonumber\\
\IEEEeqnarraymulticol{3}{r}{\underline{y}(b+T+1)\Big)\ .}
\end{IEEEeqnarray*}
The destination finds the unique pair of indices $(\hat{w}_b,\hat{l}_{\mc{T}(b+1)})$ such that 
\begin{equation}
\mc{E}(\hat{w}_b,\hat{l}_{\mc{T}(b+1)})\in \mc{A}^n_\epsilon[V_{\leq T}XX_{\mc{V}^c\cup\mc{T}}\hat{Y}_{\mc{T}}Y]\,.
\label{DestdecodingLMNNC}
\end{equation}
The error events associated with this step are characterized as follows ($\mc{S}\subseteq \mc{T}$):
\begin{IEEEeqnarray}{rCl}
\mb{E}_0&:  & \Big\{\mc{E}({w}_b,{l}_{\mc{T}(b+1)})
 \notin \mc{A}^n_\epsilon[V_{\leq T}XX_{\mc{V}^c\cup\mc{T}}\hat{Y}_{\mc{T}}Y]\Big\}\  \ ,\nonumber\\
\mb{E}_{\mc{S}}&: & \Big\{\mc{E}({w}_b,\hat{l}_{\mc{T}(b+1)}) \in \mc{A}^n_\epsilon[V_{\leq T}XX_{\mc{V}^c\cup\mc{T}}\hat{Y}_{\mc{T}}Y] \nonumber\\
&&\text{ for some } \,\,\hat{l}_{kb}\neq l_{kb}\,,\,  k\in\mc{S}  \nonumber\\
&&{\text{ and } \,\, \hat{l}_{kb}=l_{kb}\,,\,  k\in\mc{S}^c\Big\}\  , }\nonumber\\
\mb{E}_{w,\mc{S}}&: &\Big\{\mc{E}(\hat{w}_b,\hat{l}_{\mc{T}(b+1)}) \in\mc{A}^n_\epsilon[V_{\leq T}XX_{\mc{V}^c\cup\mc{T}}\hat{Y}_{\mc{T}}Y]\nonumber\\
&&\text{ for some }\,\, \hat{w}_b\neq w_{b}\,,\, \hat{l}_{kb}\neq l_{kb}\,, \, k\in\mc{S} \nonumber\\
&&{ \text{ and } \,\, \hat{l}_{kb}=l_{kb}\,, \, k\in\mc{S}^c\Big\}\ \ . }\nonumber
\end{IEEEeqnarray}

The error probability can be bounded as follows:
\begin{IEEEeqnarray*}{rCl}
\Pr\left[(\hat{w}_{b},\hat{l}_{\mc{T}b})\neq ({w}_{b},{l}_{\mc{T}b}) \right] &\leq &  \Pr(\mb{E}_0)\nonumber\\
\IEEEeqnarraymulticol{3}{r}{+ \sum_{\mc{S}\subseteq\mc{T}}\left[\Pr(\mb{E}_{\mc{S}})+\Pr(\mb{E}_{w,\mc{S}})\right]\ .}
\end{IEEEeqnarray*}
These probabilities are bounded using the same technique in Appendix \ref{proof:MNNC} and therefore we omit the details here. The probability $\Pr(\mb{E}_0)$ and  $\Pr(\mb{E}_{\mc{S}})$ go to zero as $n\to\infty$ provided that: 
\begin{IEEEeqnarray}{rCl}
\sum_{k\in\mc{S}}I(\hat{Y}_k;Y_k|X_k) &+& \epsilon_3  <  \sum_{k\in\mc{S}}H(\hat{Y}_kX_k)\nonumber\\
\IEEEeqnarraymulticol{3}{l}{+ H(V_{\leq T}XX_{\mc{V}^c\cup\mc{S}^c}\hat{Y}_{\mc{S}^c}Y)- H(V_{\leq T}XX_{\mc{V}^c\cup\mc{T}}\hat{Y}_{\mc{T}}Y)\ ,}\nonumber\\
\end{IEEEeqnarray}
which is equivalent to the following inequality after standard manipulation:
This manipulation yields the following expression: 
\begin{IEEEeqnarray}{rCl}
\epsilon_3 &<& I(X_{\mc{S}};\hat{Y}_{\mc{S}^c}Y|V_{\leq T}XX_{\mc{V}^c\cup\mc{S}^c})\nonumber\\
\IEEEeqnarraymulticol{3}{r}{ -I(\hat{Y}_{\mc{S}};{Y}_{\mc{S}}|V_{\leq T}XX_{\mc{V}^c\cup\mc{T}}\hat{Y}_{\mc{S}^c}Y)\ .}
\label{Compression-LMNNC}
\end{IEEEeqnarray}
By the choice $\mc{T}\in\Upsilon(\mc{V})$, inequality \eqref{Compression-LMNNC} holds for each $\mc{S}\subseteq\mc{T}$.  Now, the probability $\Pr(\mb{E}_{w,\mc{S}})$ tends to zero as $n$ goes to infinity provided that:
\begin{IEEEeqnarray}{rCl}
R+\sum_{k\in\mc{S}}I(\hat{Y}_k;Y_k|X_k)&+&\epsilon_4 < \sum_{k\in\mc{S}}H(\hat{Y}_kX_k)\nonumber\\
\IEEEeqnarraymulticol{3}{c}{+ H(V_{\leq T}XX_{\mc{V}^c})+ H(X_{\mc{S}^c}\hat{Y}_{\mc{S}^c}Y)}\nonumber\\
\IEEEeqnarraymulticol{3}{r}{-H(V_{\leq T}XX_{\mc{V}^c\cup\mc{T}}\hat{Y}_{\mc{T}}Y)\big]\ .}
\label{eq:inequalityA2-LMNNC}
\end{IEEEeqnarray}
We apply once again the standard simplification used in Appendix \ref{proof:MNNC} from which we get:
\begin{IEEEeqnarray}{rCl}
R+\epsilon_4 & < & I(XX_{\mc{V}^c\cup\mc{S}};\hat{Y}_{\mc{S}^c}Y|X_{\mc{S}^c})\nonumber\\
&&-I(\hat{Y}_{\mc{S}};{Y}_{\mc{S}}|XX_{\mc{V}^c\cup\mc{T}}\hat{Y}_{\mc{S}^c}Y)\ .
\label{DFdecoding-LMNNC}
\end{IEEEeqnarray}
 Notice that we can see that DF relays $X_{\mc{V}^c}$ appear in the mutual information part of the rate~\eqref{DFdecoding-LMNNC}, which means that they contribute to the increase the final rate by their cooperation. The proof is finalized by choosing finite but large enough $L$, and then the inequalities~\eqref{Compression-relay-LMNNC}, \eqref{Compression-LMNNC}, \eqref{DFdecoding-LMNNC}, \eqref{DFdecoding-relay-LMNNC-II}, and~\eqref{DFdecoding-relay-LMNNC} prove Theorem~\ref{thm:LMNNC}, where the final rate is achieved by letting $(B,n)$ tend to infinity. At the end, a time sharing random variable $Q$ can be added over all expressions which concludes the proof. 
\end{enumerate}


\section{Proof of Proposition~\ref{corollary1}}\label{Sec-SCSproof}
Consider now the composite relay channel with given parameters $\theta=(\theta_d,\theta_r)$ and target rate $r$.  Transmission takes place over $B+L$ blocks of length $n$.

\subsection*{Code generation:}
\begin{enumerate} [(i)]
\item The relay disposes of two different codebooks. First, randomly and independently generate $2^{nr}$ sequences $\underline{x}_1$ drawn i.i.d. from
\begin{equation}
P_{X_1}^{n}(\underline{x}_1)=\prod\limits_{j=1}^n P_{X_1}(x_{1j})\ .
\end{equation}
Index them as $\underline{x}_1(w_0)$ with index $w_0\in \left[1,2^{nr}\right]$. This codebook must be given to the source so it cannot depend on the specific draw $\theta_r$.
Then, for each $\theta_r\in\Theta_r$ randomly and independently generate  $2^{n\hat{R}_{\theta_r}}$ sequences $\underline{x}_{2}$ drawn i.i.d. from 
\begin{equation}
P_{X_{2}|\uptheta_r}^n(\underline{x}_{2}|\theta_r)=\prod\limits_{j=1}^n P_{X_{2}|\uptheta_r}(x_{2j}|\theta_r)\ .
\end{equation}
Index them as  $\underline{x}_{2}(l_0)$, where $l_0\in \left[1,2^{n\hat{R}_{\theta_r}}\right]$ for $\hat{R}_{\theta_r}\triangleq I_{\theta_r}(Y_{2};\hat{Y}_{2}|X_{2})+\epsilon$. 
\item For each $\underline{x}_{2}(l_0)$, randomly and conditionally independently generate $2^{n\hat{R}_{\theta_r}}$ sequences  $\underline{\hat{y}}_{2}$ each with probability 
\begin{align}
P_{\hat{Y}_{2}|X_{2};\uptheta_r}^n& (\underline{\hat{y}}_{2}\vert \underline{x}_{2}(l_0);\theta_r)\nonumber\\
&= \prod\limits_{j=1}^n P_{\hat{Y}_{2}\vert X_{2};\uptheta_r}(\hat{y}_{2j}\vert x_{2j}(l_0);\theta_r)\ .
\end{align}
Index them as $\underline{\hat{y}}_2(l_0,l)$, where $l\in \big[1,2^{n\hat{R}_{\theta_r} }\big]$. 
\item For each $\underline{x}_1(w_0)$, randomly and conditionally independently generate $2^{nr}$ sequences $\underline{x}$ drawn i.i.d. from  
\begin{equation}
P_{X|X_1}^n(\underline{x}\vert \underline{x}_1(w_0))=\prod\limits_{j=1}^n P_{X|X_1}(x_{j}|x_{1j}(w_0))\ .
\end{equation}
Index them as $\underline{x}(w_0,w)$, where $w\in \left[1,2^{nr}\right]$. This is the source codeword independent of the specific draw of $\theta$. Again note that $\underline{x}_1(r_1)$, which is used to generate the source code, does not depend on $\theta$. 
\item Provide the codebooks to every node available except for the collection of codebooks  $\{\underline{x}_{2}(l_0)\}$ that cannot be known to the source. 
\end{enumerate}


\subsection*{Encoding:} 
\begin{enumerate}[(i)]
\item
In every block $i\in[ 1: B]$, the source sends $w_{i}\in[1,2^{nr}]$ based on $\underline{x}\big(w_{(i-1)},w_i\big)$. Moreover, for blocks $i\in[B+1:B+L]$, the source sends the dummy message $w_{i}=1$ known to all nodes.

\item The parameter $\theta_r$ is available to the relay. If $\theta_r\in\mc{D}_{\textrm{DF}}$ the relay sends the codeword $\underline{x}_1$ from the first codebook  and uses it for the rest of the communication. In other words, the relay function for this choice is DF scheme. In block $i$, the relay uses its decoder output $\hat{w}_{(i-1)}$ ($w_0=1$) and sends the codeword $\underline{x}_1\big(\hat{w}_{(i-1)}\big)$. Otherwise, if $\theta_r\notin\mc{D}_{\textrm{DF}}$, then the relay picks the codebook of codewords $\underline{x}_2$ corresponding to  $\theta_r$. The relay function in this case is CF scheme. After receiving the corresponding output, namely $\underline{y}_{1}(i)$, the relay searches for at least one index $l_{i}$, where $l_0=1$, such that 
\begin{equation*}
\big(\underline{x}_{2}(l_{(i-1)}),\underline{y}_{1}(i),\underline{\hat{y}}_{2}(l_{(i-1)},l_{i})\big)\in \mc{A}^n_\epsilon[X_{2}Y_{1}\hat{Y}_{2}|\theta_r]\ ,
\end{equation*}
where $\mc{A}^n_\epsilon[X_{2}Y_{1}\hat{Y}_{2}|\theta_r]$ is the joint typical set indexed with $\theta_r$.
 The probability of finding such $l_{i}$ goes to one as $n$ goes to infinity. We remark that the typical set used for such coding is known to the relay because it knows $\theta_r$. For $i=[1:B+1]$, the relay knows from the previous block ${l}_{(i-1)}$ and it sends $\underline{x}_{2}(l_{(i-1)})$. Moreover, the relay repeats $l_{B+1}$ for $i=[B+2: B+L]$. 
\end{enumerate}

\subsection*{Decoding:} 
We assume that destination is equipped with an outage identification function $I$ that is given as follows: 
\begin{enumerate}
 \item For every block $i=[1:B+L]$, the relay decodes $w_i$ exactly similar to Theorem~\ref{thm:NC-MNNC}. For a fix $r$ and given $\uptheta_r$, if the next condition is satisfied the probability of error event will asymptotically tend to zero, i.e., 
\begin{equation}
r\leq I_{\uptheta_r}(X;Y_{1}|X_1)\ . 
\label{DFdecoding-csr}
\end{equation}
On the contrary, an error occurs when
\begin{equation}
r>I_{\uptheta_r}(X;Y_{1}|X_1)\ . 
\label{DFerror-1}
\end{equation}
In this case, the relay will transmit a random message $\hat{w}$ and the destination that knows $\theta$ will set $I(\uptheta)=0$. 
\item The decoder knows the channel index $\theta$ and hence $\theta_r$. If $\theta_r\in\mc{D}_{\textrm{DF}}$ and inequality \eqref{DFdecoding-csr} is satisfied, the destination decodes the message using  the DF codebook. Thus, decoding of this would not be successful if
\begin{equation}
r>I_\uptheta(X,X_1;Y)\ ,
\label{DFerror-2}
\end{equation}
for which the destination sets $I(\theta)=0$ and declares an outage event. Therefore, if $\theta_r\in\mc{D}_{\textrm{DF}}$, an outage event is declared when  $r>I_{\textrm{DF}}(\uptheta)$ with 
\begin{equation}
I_{\textrm{DF}}(\uptheta) \triangleq \min\big\{I_{\uptheta_r}(X;Y_{1}|X_1),I_\uptheta(X,X_1;Y)\big\}\ .
\label{DFerrorF}
\end{equation}
Consider the step for the case $\theta_r\notin\mc{D}_{\textrm{DF}}$ where the relay input is chosen from the second set of codebooks with distribution $X_2$. As a matter of fact,  for all draws $\theta$ yielding this case we have
\begin{equation}
X_1 \minuso X \minuso (Y,Y_{1},X_{2})\ .
\label{Markov1} 
\end{equation}
Moreover, the decoder knows whether the next inequality is satisfied subject to the Markov chain \eqref{Markov1}
\begin{equation}
I_{\theta}(X_{2};Y|XX_1)\geq I_{\theta}(Y_{1};\hat{Y}_{2}\vert YXX_1X_{2})\ .
\end{equation}
The decoder applies the exact same decoding procedure as in Theorem~\ref{thm:NC-MNNC}. It can be seen that the decoding conditions at destination do not change even if $X_1$ is not really transmitted. The only change is in the Markov chains, it can be seen that the previous inequality corresponds to \eqref{CFcondition} for the composite relay channel. The destination declares an outage event and $I(\theta)=0$ for $\theta_r\notin\mc{D}_{\textrm{DF}}$ if $r>I_{\textrm{CF}}(\uptheta)$, where 
\begin{IEEEeqnarray}{rCl}
I_{\textrm{CF}}(\uptheta)&\triangleq&\max\Big\{\min\Big[ I_\uptheta(XX_1;Y\hat{Y}_{2}|X_{2}), I_\uptheta(XX_1X_{2};Y)\nonumber\\
\IEEEeqnarraymulticol{3}{l}{-I_\uptheta(\hat{Y}_{2};Y_{1}|YXX_1X_{2}) \Big],I_{\uptheta}(X;Y)\Big\}\ .}
\label{CFerrorF}
\end{IEEEeqnarray}
Using \eqref{DFerrorF} and \eqref{CFerrorF}, the outage event denoted by the indicator function $\mathbf{1}_E$ is as follows
\begin{align*}
 \mathbf{1}_E&\triangleq \mathbf{1}[\uptheta_r\in\mc{D}_{\textrm{DF}}  \,\,  \text{ and }  \,\,  r> I_{\textrm{DF}}(\uptheta)]\nonumber\\
 &+\mathbf{1}[\uptheta_r\notin\mc{D}_{\textrm{DF}} \,\,\text{ and } \,\, r>I_{\textrm{CF}}(\uptheta)]\ .
\end{align*}
Taking the expected value from both sides lead to the outage probability. Indeed, the expected value is taken in two steps. For each $\theta_r$, the expected value is calculated using $\prob_{\uptheta|\uptheta_r}$. The relay, for each $\theta_r$ chooses the joint distribution of $(X_{2},\hat{Y}_{2})$ to minimize:
\begin{IEEEeqnarray*}{rCl}
&&\prob[E]_{\uptheta|\uptheta_r}[\mathbf{1}_E]=\nonumber\\
&&
\left\{ 
\begin{array} {ll}
\prob_{\uptheta|\uptheta_r}\left[r>I_{\textrm{DF}}(\uptheta)|\uptheta_r\right]& {\uptheta_r\in\mc{D}_{\textrm{DF}}}\\
\displaystyle\min_{
{p(x_{2})p(\hat{y}_{2}|x_{2},y_{1})}
} \prob_{\uptheta|\uptheta_r}\left[r> I_{\textrm{CF}}(\uptheta)|\uptheta_r\right] 
&{\uptheta_r\notin\mc{D}_{\textrm{DF}}}\ .
\end{array}
\right.
\end{IEEEeqnarray*}
At the next step, the expectation is taken over $\theta_r$ and is minimized over $\mc{D}_{\textrm{DF}}$ and $p(x,x_1)$, yielding 
\begin{IEEEeqnarray*}{lCl}
\overline{\epsilon}(r) &=&  \min_{p(x,x_1)}\inf_{\mc{D}_{\textrm{DF}}\subseteq\Theta_r}  \prob[E]_{\uptheta_r}\Big\{ \prob_{\uptheta|\uptheta_r}\big[r> I_{\textrm{DF}},\uptheta_r\in\mc{D}_{\textrm{DF}}|\uptheta_r\big] \nonumber\\
\IEEEeqnarraymulticol{3}{l}{+\displaystyle\min_{ {p(x_{2})p(\hat{y}_{2}|x_{2},y_{1\theta_r})} } \prob_{\uptheta|\uptheta_r}\big[r> I_{\textrm{CF}},\uptheta_r\notin\mc{D}_{\textrm{DF}}|\uptheta_r\big]\Big\}\ .}\,\,\,
\end{IEEEeqnarray*}
Finally, a time sharing random variable $Q$ can be added to the region, however the optimization should be done outside the expectation.
\end{enumerate}


\section{Outline of the Proof of Proposition~\ref{thm:SCS-MNNC}} \label{proof:SCS-MNNC}
Consider the composite unicast network with parameters $\uptheta=(\uptheta_d,\uptheta_r)$. Transmission is done over $B+L$ blocks. Suppose that every relay knows $\mc{D}_{\mc{V}}$, i.e., the decision region of the other relays for each $\theta_r$.
\subsection*{Code generation:}
\begin{enumerate} 
\item Randomly and independently generate $2^{nR}$ sequences $\underline{v}$ drawn i.i.d. from
\begin{equation*}
P_{V}^{n}(\underline{v})=\prod\limits_{j=1}^n P_{V}(v_j)\ .
\end{equation*}
Index them as $\underline{v}(w_0)$ with index $w_0\in \left[1,2^{nR}\right]$. This codebook must be given to the source so it cannot depend on the specific draw $\theta_r$.
\item Since all relays know $\theta_r$, for each $\theta_r$ generate two sets of codebooks:
\begin{enumerate}
\item Each codebook for $\theta_r$ in the first set is generated as follows. For each $\underline{v}(w_0)$, randomly and conditionally independently generate $2^{n\hat{R}_{k\theta_r}}$ sequences $\underline{x}_{k}$ drawn i.i.d. from
\begin{align*}
P_{X_k|V;\uptheta_r}^{n}&(\underline{x}_k|\underline{v}(w_0),\theta_r)\nonumber\\
&=\prod\limits_{j=1}^n P_{X_k|V;\uptheta_r}(x_{kj}|v_{j}(w_0),\theta_r)\ .
\end{align*}
Index them as $\underline{x}_{k}(w_0,l_{0k})$ with index $l_{0k}\in \big[1,2^{n\hat{R}_{k\theta_r}}\big]$ for $\hat{R}_{k\theta_r}\triangleq I_{\theta_r}(Y_{k};\hat{Y}_k|X_kV)+\epsilon$. 
\item For each $\underline{v}(w_0)$, $\underline{x}_k(w_0,l_{0k})$ and $\theta_r$, randomly and conditionally independently generate $2^{n\hat{R}_{k\theta_r}}$ sequences  $\underline{\hat{y}}_k$ each with probability 
\begin{align*}
&P_{\hat{Y}_k|X_kV;\uptheta_r}^n (\underline{\hat{y}}_k\vert \underline{x}_k(w_0,l_{0k}),\underline{v}(w_0),\theta_r)\nonumber\\
&= \prod\limits_{j=1}^n P_{\hat{Y}_k\vert X_kV;\uptheta_r}(\hat{y}_{kj}\vert x_{kj}(w_0,l_{0k}),v_j(w_0),\theta_r)\ .
\end{align*}
Index them as $\underline{\hat{y}}_k(w_0,l_{0k},l_k)$, where $l_k\in \big[1,2^{n\hat{R}_{k\theta_r}}\big]$ for $\hat{R}_{k\theta_r} \triangleq I_{\theta_r}(Y_{k};\hat{Y}_k|X_kV)+\epsilon$.  
\item As for the second set of codebooks, for each $\theta_r$ randomly and independently generate  $2^{n\hat{R}_{k\theta_r}}$ sequences $\underline{x}_k$ drawn i.i.d. from 
\begin{equation*}
P_{X_k| \uptheta_r }^n(\underline{x}_k| \theta_r)=\prod\limits_{j=1}^n P_{X_k| \uptheta_r}(x_{kj}| \theta_r)\,. 
\end{equation*}
Index them as $\underline{x}_k(l_{0k})$, where $l_{0k}\in \big[1,2^{n\hat{R}_{k\theta_r}}\big]$ for $\hat{R}_{k\theta_r}\triangleq I_{\theta_r}(Y_{k};\hat{Y}_k|X_k)+\epsilon$.  
\item For each $\underline{x}_k(l_{0k})$, randomly and conditionally independently generate $2^{n\hat{R}_k}$ sequences  $\underline{\hat{y}}_k$ each from 
\begin{align*}
P_{\hat{Y}_k|X_k;\uptheta_r}^n &(\underline{\hat{y}}_k\vert \underline{x}_k(l_{0k}),\theta_r)\nonumber\\
&= \prod\limits_{j=1}^n P_{\hat{Y}_k\vert X_k;\uptheta_r}(\hat{y}_{kj}\vert x_{kj}(l_{0k}),\theta_r)\ .
\end{align*}
Index them as $\underline{\hat{y}}_k(l_{0k},l_{k})$, where $l_{k}\in \big[1,2^{n\hat{R}_{k\theta_r} }\big]$ for $\hat{R}_{k\theta_r} \triangleq I_{\theta_r}(Y_{k};\hat{Y}_k|X_k)+\epsilon$.
\end{enumerate}
Note that the rate $\hat{R}_{k\theta_r}$  depends on the relay strategy, i.e., CF or  DF scheme, and so the relay is superimposing the compression index over the DF code. Therefore, the rate $\hat{R}_{k\theta_r}$ varies according to each scenario.
\item For each $\underline{v}(w_0)$, randomly and conditionally independently generate $2^{nR}$ sequences $\underline{x}$ drawn i.i.d. from  
\begin{equation*}
P_{X|V}^n(\underline{x}\vert \underline{v}(w_0))=\prod\limits_{j=1}^n P_{X|V}(x_{j}|v_{j}(w_0))\ .
\end{equation*}
Index them as $\underline{x}(w_0,w)$, where $w\in \left[1,2^{nR}\right]$.

\item Provide the codebooks to every node available except for the collection of codebooks  $\{\underline{x}_k(l_{0k}), \underline{x}_k(w_0,l_{0k})\}$ that cannot be known to the source. 
\end{enumerate}
\subsection*{Encoding:}
\begin{enumerate}[(i)]
\item
In every block $i=[1: B]$, the source sends $w_{i}$ based on $\underline{x}\big(w_{(i-1)},w_i\big)$. Moreover, for blocks $i=[B+1:B+L]$, the source sends the dummy message $w_{i}=1$ known to all nodes.
\item
Since all relays  know $\theta_r$, if $\theta_r\in\mc{D}^{(k)}_{\textrm{DF}}$ for every block $i=[1:B+L]$, the relay $k$ knows $w_{(i-2)}$ by assumption and $w_0=1$. 

Moreover, it searches in the codebook for $\theta_r$ in its first set of codebooks for at least one index $l_{ki}$ with $l_{k0}=1$ such that 
\begin{align*}
\big(&\underline{v}(w_{(i-2)}),\underline{x}_{k}(w_{(i-2)},l_{k(i-1)}),\underline{y}_{k}(i),\nonumber\\
&\underline{\hat{y}}_{k}(w_{(i-2)},l_{k(i-1)},l_{ki})\big)\in \mc{A}^n_\epsilon[VX_kY_{k}\hat{Y}_k|\theta_r]\ .
\end{align*}
The probability of finding such $l_{ki}$ goes to one as $n$ goes to infinity by a choice of the rate $\hat{R}_{k\theta_r}$. Relay $k$ sends $\underline{x}_k(w_{(i-2)},l_{k(i-1)})$ in block $i$ and it repeats $l_{k(B+2)}$ for all blocks $i=[B+3: B+L]$. 
If $\theta_r\notin\mc{D}^{(k)}_{\textrm{DF}}$, the relay $k$ after receiving $\underline{y}_{k}(i)$, searches in the codebook index $\theta_r$ in its second set of codebooks for at least an index $l_{ki}$ with $l_{k0}=1$ such that 
\begin{equation*}
\big(\underline{x}_{k}(l_{k(i-1)}),\underline{y}_{k}(i),\underline{\hat{y}}_{k}(l_{k(i-1)},l_{ki})\big)\in \mc{A}^n_\epsilon[X_kY_{k}\hat{Y}_k|\theta_r]\ .
\end{equation*}
The probability of finding such $l_{ki}$ goes to one as $n$ goes to infinity. The relay $k$ knows from the previous block ${l}_{k(i-1)}$ and it sends $\underline{x}_k(l_{k(i-1)})$. Moreover, relay $k$ repeats $l_{k(B+1)}$ for all blocks $i\in[B+2: B+L]$.
\end{enumerate}

\subsection*{Decoding:} 
If $\theta_r\in\mc{D}_\mc{V}$, then relays $k\in\mc{V}$ use CF scheme while the others relays use DF scheme. It means that the relays with $k\in\mc{V}$ use the codebook for $\theta_r$ in the second set of codebooks and the others relays use the codebook for $\theta_r$ in the first set of codebooks. Now the outage indicator function can be considered as follows:
\begin{equation}
 \mathbf{1}_E\triangleq  \displaystyle\sum_{\mc{V}\subseteq\mc{N}}\mathbf{1}[r \leq I_{\textrm{MNNC}}(\mc{V},\uptheta)\,, \,\uptheta_r\in\mc{D}_\mc{V}]\ .
 \label{outage-event}
\end{equation}
Therefore, if $r$ is less or equal than $I_{\textrm{MNNC}}(\mc{V},\uptheta)$ it can be achieved and the probability of error tends to zero. Hence, the outage probability is calculated easily from \eqref{outage-event} by considering the optimization over all probability distributions and taking the expected value from both sides. 


\section{Outline of the Proof of Proposition~\ref{thm:SCS-NC-MNNC}} \label{proof:SCS-NC-MNNC}

Consider the composite unicast network with parameters $\uptheta=(\uptheta_d,\uptheta_r)$. Transmission is done over $B+L$ blocks. It is assumed that every relay knows $\mc{D}_{\mc{V}}$, i.e., the decisions regions of all others relays. 

\subsection*{Code generation:}
\begin{enumerate} [(i)]
\item The relay $k$ knows $\theta_r$ and for each $\theta_r$ it generates two codebooks $\big(\underline{x}^{(1)}_{k},\underline{x}^{(2)}_{k}\big)$ as follows:
\begin{enumerate}
\item Randomly and independently generate $2^{nR}$ sequences $\underline{x}^{(1)}_{\mc{N}}$ drawn i.i.d. from
\begin{equation*}
P_{X^{(1)}_{\mc{N}}}^{n}\big(\underline{x}^{(1)}_{\mc{N}}\big)=\prod\limits_{j=1}^n P_{X^{(1)}_{\mc{N}}}\big(x^{(1)}_{{\mc{N}}j}\big)\ .
\end{equation*}
Index them as $\underline{x}^{(1)}_{\mc{N}}(r)$ with index $r\in \left[1,2^{nR}\right]$. Since this codebook must be also given to the source it cannot depend on $\theta_r$.

\item Randomly and independently generate  $2^{n\hat{R}_{k\theta_r}}$ sequences $\underline{x}^{(2)}_k$ drawn i.i.d. from 
\begin{equation*}
P_{X^{(2)}_k|\uptheta_r}^n\big(\underline{x}^{(2)}_k | \theta_r\big)=\prod\limits_{j=1}^n P_{X^{(2)}_k|\uptheta_r}\big(x^{(2)}_{kj}| \theta_r\big)\ . 
\end{equation*}
Index them as $\underline{x}^{(2)}_k(r_k)$, where $r_k\in \big[1,2^{n\hat{R}_{k\theta_r}}\big]$ for $\hat{R}_{k\theta_r} \triangleq I_{\uptheta_r}(Y_{k};\hat{Y}_k|X^{(2)}_k)+\epsilon$.
\item For each $\underline{x}^{(2)}_k(r_k)$, randomly and conditionally independently generate $2^{n\hat{R}_{k\theta_r}}$ sequences  $\underline{\hat{y}}_k$ each with probability 
\begin{align*}
P_{\hat{Y}_k|X^{(2)}_k;\uptheta_r}^n& \big(\underline{\hat{y}}_k\vert \underline{x}^{(2)}_k(r_k),\theta_r\big)\nonumber\\
&= \prod\limits_{j=1}^n P_{\hat{Y}_k\vert X^{(2)}_k;\uptheta_r}\big(\hat{y}_{kj}\vert x^{(2)}_{kj}(r_k),\theta_r\big)\ .
\end{align*}
Index them as $\underline{\hat{y}}_k(r_k,\hat{s}_k)$, where $\hat{s}_k\in \big[1,2^{n \hat{R}_{k\theta_r}}\big]$. 
\end{enumerate}
\item For each $\underline{x}^{(1)}_{\mc{N}}(r)$, randomly and conditionally independently generate $2^{nR}$ sequences $\underline{x}$ drawn i.i.d. from  
\begin{equation*}
P_{X|X^{(1)}_{\mc{N}}}^n\big(\underline{x}\vert \underline{x}^{(1)}_{\mc{N}}(r)\big)=\prod\limits_{j=1}^n P_{X|X^{(1)}_{\mc{N}}}\big(x_{j}|x^{(1)}_{{\mc{N}}j}\big)\ .
\end{equation*}
Index them as $\underline{x}(r,w)$, where $w\in \left[1,2^{nR}\right]$. Indeed, this codebook must be also given to the source so it cannot depend on $\theta_r$.
\end{enumerate}

\subsection*{Encoding:}
\begin{enumerate}[(i)]
\item
In every block $i=[1: B]$, the source sends $w_{i}$ based on $\underline{x}\big(w_{(i-1)},w_i\big)$. Moreover, for all blocks $i=[B+1:B+L]$, the source sends the dummy message $w_{i}=1$ known to all nodes.
\item
If $\theta_r\in\mc{D}^{(k)}_{\textrm{DF}}$, for every block $i=[1:B+L]$, the relay $k$ knows $w_{(i-1)}$ by assumption and $w_0=1$ so it sends $\underline{x}^{(1)}_k\big(w_{(i-1)}\big)$. 

If $\theta_r\notin\mc{D}^{(k)}_{\textrm{DF}}$, the relay $k$ after receiving $\underline{y}_{k}(i)$, searches for at least one index $l_{ki}$ with $l_{k0}=1$ such that 
\begin{align}
&\big(\underline{x}^{(2)}_{k}(l_{k(i-1)}),\underline{y}_{k}(i),\underline{\hat{y}}_{k}(l_{k(i-1)},l_{ki})\big)\nonumber\\
&\in \mc{A}^n_\epsilon[X_kY_{k}\hat{Y}_k|\theta_r]\ .
\end{align}
The probability of finding such $l_{ki}$ goes to one as $n$ goes to infinity. The relay $k$ knows from the previous block ${l}_{k(i-1)}$ and it sends $\underline{x}^{(2)}_k(l_{k(i-1)})$. Moreover, relay $k$ repeats $l_{k(B+1)}$ for all blocks $i=[B+2: B+L]$.
\end{enumerate}
\subsection*{Decoding:} 
\begin{enumerate}

\item After transmission of the block  $i=[1:B]$, if $\theta_r\in\mc{D}^{(k)}_{\textrm{DF}}$ the $k$-th relay  decodes the message of block $i$ with the assumption that all messages up to block $i-1$ have been correctly decoded. The relay $k$ searches for the unique index $\hat{w}_{i}\in\mc{M}_n$ such that:
\begin{IEEEeqnarray}{rCl}
& &\left(\underline{x}\big(w_{(i-1)},\hat{w}_{i}\big),\underline{x}^{(1)}_{\mc{N}}\big(w_{(i-1)}\big),\underline{y}_{k}(i)\right)\nonumber\\
\IEEEeqnarraymulticol{3}{r}{\in \mc{A}^n_\epsilon[XX_{\mc{N}}Y_{k}|\theta_r]\ .}
\end{IEEEeqnarray}
The outage event is occurred when
\begin{IEEEeqnarray}{rCl}
r & > I_{\uptheta_r}\left(X;Y_{k}\vert X^{(1)}_{\mc{N}}\right)\ . 
\label{DFerror-1-MNNC}
\end{IEEEeqnarray}
We emphasize that not all $X^{(1)}_{\mc{N}}$ are sending description but only codewords $X^{(1)}_{\mc{V}^c}$ does for $\uptheta_r\in\mc{D}_{\mc{V}}$.
\item
Decoding at the destination is done backwardly. It knows $\uptheta$, $\mc{D}_{\mc{V}}$ and therefore $\mc{V}$, i.e., it is aware of the strategy of each relay (e.g. using DF or CF scheme). Moreover, it chooses $\mc{T}$ to maximize \eqref{NC-MNNC-composite}. It first decodes the last compression indices sent by the relays in $\mc{T}$. It jointly searches for the unique indices $\big(\hat{l}_{k(B+1)}\big)_{k\in\mc{T}}$ such that for all $b=[B+2:B+L]$ the following condition holds:
\begin{IEEEeqnarray*}{rCl}
&\left(\big(\underline{x}_{k}(\hat{l}_{k(B+1)})\big)_{k\in\mc{T}},\underline{x}(1,1),\underline{x}_{\mc{V}^c}(1),\underline{y}(b)\right)\nonumber\\
&\in \mc{A}^n_\epsilon[XX_{\mc{T}}X_{\mc{V}^c}Y|\theta]\ .
\end{IEEEeqnarray*}

After finding the correct index $l_{k(B+1)}$ for all $k\in\mc{T}$ and since $w_{(B+1)}=1$, the destination decodes jointly the message and all compression indices $(w_b,l_{\mc{T}b})$, for each $b=[1:B]$, where $l_{\mc{T}b}=\left(l_{kb}\right)_{k\in\mc{T}}$. Indeed, decoding is performed  backwardly with the assumption that $(w_{b+1},l_{\mc{T}(b+1)})$ have been correctly decoded. The destination finds the unique pair $(\hat{w}_b,\hat{l}_{\mc{T}b})$ such that 
\begin{IEEEeqnarray*}{rCl}
&\Big(\underline{x}(\hat{w}_{b},w_{(b+1)}),\underline{x}_{\mc{V}^c}(\hat{w}_{b}),\underline{y}(b+1),\big(\underline{x}_{k}(\hat{l}_{kb}),\nonumber\\
\IEEEeqnarraymulticol{3}{r}{\underline{\hat{y}}_{k}(\hat{l}_{kb},l_{k(b+1)})\big)_{k\in\mc{T}}\Big)\in \mc{A}^n_\epsilon[XX_{\mc{T}\cup\mc{V}^c}\hat{Y}_{\mc{T}}Y|\theta]\ .}\,\,\,\,
\end{IEEEeqnarray*}
It can be seen from Theorem~\ref{thm:NC-MNNC} that an error occurs if:
\begin{equation}
r > \min_{\mc{S}\subseteq\mc{T}} R_{\mc{T}}(\mc{S},\uptheta)\ ,
\label{CFerror-MNNC}
\end{equation}
where 
\begin{IEEEeqnarray*}{rCl}
R_{\mc{T}}(\mc{S},\uptheta)&\triangleq & I_\uptheta(XX^{(1)}_{\mc{V}^c}X^{(2)}_{\mc{S}};\hat{Y}_{\mc{S}^c}Y|X^{(2)}_{\mc{S}^c})\nonumber\\
\IEEEeqnarraymulticol{3}{r}{-I_\uptheta(Y_{\mc{S}};\hat{Y}_{\mc{S}}|XX^{(2)}_{\mc{T}}X^{(1)}_{\mc{N}}\hat{Y}_{\mc{S}^c}Y)\ .}
\end{IEEEeqnarray*}
Note that $\mc{T}$ is chosen in such a way that the right-hand side achieves its maximum value. From our previous discussion on expression \eqref{MNNC-new}, we know that this set belongs to $\Upsilon(\mc{V})$ and so $Q_{\mc{T}}(\mc{A})\geq 0$ for each $\mc{A}\subseteq\mc{T}$.  
\item Using \eqref{DFerror-1-MNNC} and \eqref{CFerror-MNNC}, the outage indicator function can be defined as
\begin{equation}
\mathbf{1}_E\triangleq \displaystyle\sum_{\mc{V}\subseteq\mc{N}}\mathbf{1}\left[\uptheta_r\in\mc{D}_{\mc{V}} \text{ and } r> I_{\textrm{MNNC}}(\mc{V},\uptheta)\right]\,,
\end{equation}
where 
\begin{IEEEeqnarray*}{rCl}
I_{\textrm{MNNC}}(\mc{V},\uptheta)&\triangleq & \nonumber\\
\IEEEeqnarraymulticol{3}{r}{\max_{\mc{T}\subseteq\mc{V}}\,\min\Big\{\min_{\mc{S}\subseteq\mc{T}} R_{\mc{T}}(\mc{S},\uptheta),\min_{i\in \mc{V}^c}I_{\uptheta_r}(X;Y_{i}|X^{(1)}_{\mc{N}})\Big\}\ . }
\end{IEEEeqnarray*}

As before, the expected value is taken in two steps. For each $\theta_r$, the expected error is calculated with $\prob_{\uptheta|\uptheta_r}$. The relays chose the distribution ${\prod_{j\in \mc{V}}p({x^{(2)}_j})p({\hat{y}_j|x^{(2)}_jy_j})}$ to minimize the conditional expectation for each $\uptheta_r$ and $\mc{D}_{\mc{V}}$. This will lead to the following:
\begin{IEEEeqnarray}{rCl}
\prob[E]_{\uptheta|\uptheta_r}[\mathbf{1}_E] &\triangleq& \displaystyle\sum_{\mc{V}\subseteq\mc{N}}\min_{\prod_{j\in \mc{V}}p({x^{(2)}_j})p({\hat{y}_j|x^{(2)}_jy_j})}
 \nonumber\\
\IEEEeqnarraymulticol{3}{r}{\prob_{\uptheta|\uptheta_r}\big[r > I_{\textrm{MNNC}}(\mc{V},\uptheta)\,,\, \uptheta_r \in \mc{D}_{\mc{V}} \big| \uptheta_r\big] \ .}
\end{IEEEeqnarray}
At the next step, the expected value is taken over $\theta_r$ and is minimized over all decision regions $\mc{D}_{\mc{V}}$ and $p(x,{x^{(1)}_\mc{N}})$ which leads to the following:
\begin{IEEEeqnarray}{rCl}
\bar{\epsilon}(r) &\leq& \min_{p(x,x^{(1)}_{\mc{N}})} \inf_{\left\{\mc{D}_{\mc{V}},\mc{V}\subseteq\mc{N}\right\}\in \Pi\left(\Theta_r,N\right)}\nonumber\\
 \IEEEeqnarraymulticol{3}{r}{\displaystyle\sum_{\mc{V}\subseteq\mc{N}} \prob[E]_{\uptheta_r}  \Big\{ \min_{\prod_{j\in \mc{V}}p({x^{(2)}_j})p({\hat{y}_j|x^{(2)}_jy_j})} }  \nonumber\\
\IEEEeqnarraymulticol{3}{r}{\prob_{\uptheta|\uptheta_r}\big[r > I_{\textrm{MNNC}}(\mc{V},\uptheta), \uptheta_r \in \mc{D}_{\mc{V}} \big| \uptheta_r\big]  \Big\} \ .}
\end{IEEEeqnarray}
At the end, a time sharing random variable $Q$ can be added to the region, however the optimization should be done outside the expectation.
\end{enumerate}

\section*{Acknowledgment}
The authors are grateful to Prof.\ Abbas El Gamal for his valuable advice at the early stage of this work. They are also grateful to Prof. \ Gerhard Kramer for valuable suggestions, the Associate Editor, and to anonymous reviewers for their constructive and helpful comments on the earlier version of the paper, which helped us to improve the manuscript.


\bibliographystyle{IEEEtran}
\bibliography{biblio}


%
\end{document}